\documentclass{aa}
\usepackage{epsfig,graphicx,lscape}
\begin{document}
\title{The VLT-FLAMES survey of massive stars: Surface chemical 
compositions of B-type stars in the Magellanic Clouds\thanks{Based on observations at the European Southern
Observatory in programmes 171.0237 and 073.0234}}  

\author{I. Hunter\inst{1} \and P.L. Dufton\inst{1} \and S.J. Smartt\inst{1} 
\and R.S.I. Ryans\inst{1} \and C.J. Evans\inst{2} \and D.J. Lennon\inst{3,4}
\and C. Trundle\inst{1,4} \and I. Hubeny\inst{5} \and T. Lanz\inst{6}}

\institute{Department of Physics and Astronomy, The Queen's University 
of Belfast, BT7 1NN, Northern Ireland, UK
   \and UK Astronomy Technology Centre, Royal Observatory, Blackford Hill, Edinburgh, EH9 3HJ
   \and The Isaac Newton Group of Telescopes, Apartado de Correos 321, E-38700, Santa Cruz de La Palma, Canary
   Islands, Spain
   \and Instituto de Astrof\'{i}sica de Canarias, 38200 La Laguna, Tenerife, Spain
   \and Steward Observatory, University of Arizona, Tucson, AZ\,85712, USA
   \and Department of Astronomy, University of Maryland, College Park, MD\,20742, USA
}
       
\offprints{I. Hunter,\\ \email{I.Hunter@qub.ac.uk}}

\date{Received; accepted }

\abstract{We present an analysis of high-resolution FLAMES spectra of approximately 
50 early B-type stars in three young clusters at different 
metallicities, NGC\,6611 in the Galaxy, N\,11 in the Large Magellanic Cloud (LMC) and NGC\,346 in the
Small Magellanic Cloud (SMC). Using the {\sc tlusty} non-LTE model atmospheres code, atmospheric 
parameters and photospheric abundances (C, N, O, Mg and Si) of each star have been determined. These results
represent a significant improvement on the number of Magellanic Cloud B-type stars with 
detailed and homogeneous estimates of their atmospheric parameters and chemical compositions. 
The relationships between effective 
temperature and spectral type are discussed for all three metallicity regimes, with the effective temperature
for a given spectral type increasing as one moves to a lower metallicity regime. Additionally the 
difficulties in estimating
the microturbulent velocity and the anomalous values obtained, particularly in the lowest 
metallicity regime, are discussed. Our chemical 
composition estimates are compared with previous studies, both stellar and interstellar with, 
in general, encouraging 
agreement being found. Abundances in the
Magellanic Clouds relative to the Galaxy are discussed and we also present 
our best estimates of the base-line 
chemical composition of the LMC and SMC as derived from 
B-type stars. Additionally we discuss the use of nitrogen as a probe of the
evolutionary history of stars, investigating the roles of rotational mixing, mass-loss, blue loops and binarity
on the observed nitrogen abundances and making comparisons with stellar evolutionary models where possible.
   }
\titlerunning{Abundances of B-type stars in the Magellanic Clouds}
\maketitle{}
%
\section{Introduction}                                         \label{s_intro}

The differing environments of the Magellanic Clouds, compared with our 
own Galaxy and with each other, makes them ideal laboratories for the study of both 
stellar and galactic evolution. Their proximity,
combined with their relatively low extinction has contributed to
the intensity with which they have been studied in recent years (see, for example,
Bouret et al. \cite{bou03}; Korn et al. \cite{kor02}; Garnett \cite{gar99}; 
Westerlund \cite{wes97}). For stellar evolutionary theories it is necessary that the 
present-day chemical composition of the host material in which star-birth occurs
is well understood. The interstellar material can be directly sampled via, 
for example, \ion{H}{ii} regions (Dufour \cite{duf84}; Russell \& Dopita \cite{rus90}; 
Peimbert et al. \cite{pei00}; Kurt \& Dufour \cite{kur98}) but such
studies may be complicated by
depletion onto dust grains. 

Young massive stars allow for an alternative 
means to determine chemical compositions. However, there is evidence (Gies \& Lambert \cite{gie92}; 
Maeder \cite{mae87}; Langer et al. \cite{lan98}; 
Heger \& Langer \cite{heg00}) that mixing of CN-cycled material into the stellar photosphere can occur 
giving rise to N enrichment and to smaller C and O underabundances.
Nevertheless by selecting targets that do not appear to have had material
mixed into their atmospheres, massive stars can provide reliable indicators of the pristine chemical
composition of a recent or ongoing star formation region.

An important challenge for theoretical evolutionary predictions 
is to reliably model the wide range of 
N/H abundances seen in observations of A-type supergiants 
(Venn et al. \cite{ven99}) as well as
in B-type stars, both supergiant and main-sequence 
(see, for example, Korn et al. \cite{kor02}, 
Lennon et al. \cite{len03} and Trundle et al. \cite{tru04}). 
The role of rotational mixing has become increasingly important
particularly for main-sequence objects
(Maeder \& Meynet \cite{mae01}, Meynet \& Maeder \cite{mey02} 
and Lamers et al. \cite{lam01}).
Futhermore, observational evidence suggests that stellar rotational velocity
distributions may depend on metallicity whilst cluster
stars may have greater rotational velocities than those in the field 
(Keller \cite{kel04}, Strom et al. \cite{str05}). As the 
amount of mixing depends on the initial stellar rotational velocity,
it is important that these effects are fully understood.

Metallicity plays an important role in all aspects of massive star evolution. 
For example, at low metallicity the stars are more compact, have lower 
mass-loss rates (Massey \cite{mas03}) and therefore they may lose 
less angular momentum and hence may rotate faster (Meynet \& Maeder 
\cite{mey02}). As discussed above, this
scenario then leads to enhanced mixing between the stellar interior 
and surface. Additionally photospheric mixing is more evident at 
lower metallicity as small absolute
changes in the abundances will be easier to observe and 
hence the Magellanic Clouds provide ideal laboratories within
which metallicity effects upon steller evolution can be tested.

Binarity effects have been less studied, both because these effects 
do not globally affect a given population of stars and also due to 
the wide range of free parameters that are involved in modelling 
a binary stystem. However mass-transfer in binary systems
may be an important contributor to the abundance variations found 
between stars within a given population. Additionally, 
the implications of binarity for other processes such as rotational mixing is not fully understood. 

We have undertaken a high resolution spectroscopic survey of 
young clusters in the Galaxy, LMC and SMC
(see Evans et al. \cite{eva05a}, \cite{eva05b}) and observations have been
obtained for a total of 750 stars, mainly with O and B spectral types. 
Mokiem et al. (\cite{mok05a}, \cite{mok05b})
have derived atmospheric parameters and rotational velocities of the O-type objects in this survey.
In this paper we discuss
B-type stars with narrow absorption line spectra 
in the young Magellanic Cloud clusters, N\,11 
and NGC\,346 along with NGC\,6611, a Galactic cluster with a similar age.

We determine atmospheric parameters and abundances (C, N, O, Mg and Si) of approximately 
50 early B-type stars in these three clusters with the aim of determining the
base-line chemical abundances of each region from a large sample of objects, taking into 
account any chemical enrichments or depletions which may have occured during the 
stellar lifetime. Additionally we examine our results in terms of the stellar evolution
of these objects, through both internal mixing and binary interaction.

In Sect.~\ref{s_obs} we discuss our target selection. In Sect.~\ref{s_atmos} we 
discuss our methodology and present the photospheric abundances derived for
each star in our sample along with estimates of the uncertainites in these
abundances. In Sect.~\ref{s_chem} we discuss the elemental abundances
and in Sect.~\ref{s_comparisons} we compare our results to 
previous stellar and interstellar analyses of 
the three regions. In Sect.~\ref{s_discussion} we discuss both abundance differences between the three clusters
and between stars within the same region, comparing where possible our observations with
current evolutionary theory.
Finally in Sect.~\ref{s_conclusions} we present our best estimates for the 
present-day chemical composition of the LMC and SMC and summarize the evolutionary 
effects observed in the sample.
\section{Observations}                                        \label{s_obs}

The target selection, observational details and data reduction procedures 
for clusters
in our Galaxy and in the Magellanic Clouds have been described 
in Evans et al. (\cite{eva05a}) (hereafter Paper I) and 
Evans et al. (\cite{eva05b}) (hereafter Paper II) respectively.
In summary, the majority of the observations were obtained using the Fibre
Large Array 
Multi-Element Spectrograph (FLAMES), with R$\approx$20000 at the Very Large Telescope (VLT) 
as part of a European Southern Observatory (ESO) Large Programme. These were supplemented
using the Fiber-fed Extended Range Optical
Spectrograph (FEROS), with R$\approx$48000, and the Ultraviolet and Visual 
Echelle Spectrograph
(UVES), with R$\approx$20000.
Seven clusters were observed in the Galaxy and the Small (SMC)
and Large (LMC) Magellanic Clouds and the high resolution FLAMES
spectroscopy covered the wavelength region 
3850-4755\AA\ and the H{$\alpha$} region 6380-6620\AA\ 
 (with more extensive wavelength coverage for the FEROS and UVES observations).
Examples of the quality of the spectra can be found in 
Papers I and II.
In this paper we discuss only stars observed in the youngest cluster 
in each metallicity regime, viz. NGC\,6611 in the Galaxy,
N\,11 in the LMC and NGC\,346 in the SMC.

\begin{table*}[htbp]
\caption{The estimated atmospheric parameters of our sample. Star identifications and spectral 
types have been taken from Paper I and Paper II.
Effective temperatures ($T_{\rm eff})$ are determined
from the ionization balance of \ion{Si}{iii} to \ion{Si}{iv} unless otherwise
noted. The surface gravity, {\it g}, has units of cm\,s$^{-2}$. Note, in Sect.~\ref{s_errors} 
we present alternative microturbulence ($\xi$) values based on the derived Si abundance. The uncertainties in 
these parameters are typically 1\,000\,K for $T_{\rm eff}$, 0.15--0.20\,dex for $\log g$, 3--5\,km\,s$^{-1}$ for $\xi_{\rm Si}$ and
5\,km\,s$^{-1}$ for $v\sin i$ (see Sect.~\ref{s_atmos} for details).}
\label{t_stars+atmos}
\centering
\begin{tabular}{lllcccccl}\hline \hline
Star          & Spectral &  \multicolumn{1}{c}{$T_{\rm eff}$} & $\log g$ & $\xi_{\rm Si}$	 & $v\sin i$ &$\log L/L_{\sun}$&$M_{\rm evol}/M_{\sun}$ &Instrument\\
              &  Type    &  \multicolumn{1}{c}{K}		  &  dex     & km\,s$^{-1}$ & km\,s$^{-1}$ & &&\\
	      
\hline
NGC\,6611-006     & B0 IVp$^*$&  31250        & 4.00     & 7         & 20 & 4.87 &$20^{\rm +3}_{\rm -2}$&FLAMES\\
NGC\,6611-012     & B0.5 V   &  27200         & 3.90     & 6         & 95 & 4.81 &$18^{\rm +2}_{\rm -1}$&FLAMES\\
NGC\,6611-021     & B1 V     &  26250         & 4.25     & 0         & 30 & 4.38 &$13^{\rm +2}_{\rm -1}$&FEROS \\
NGC\,6611-030$^R$ & B1.5 V   &  22500$^1$     & 4.15     & 5         & 10 & 3.63 &$ 8^{\rm +1}_{\rm -1}$&FLAMES\\
NGC\,6611-033     & B1 V     &  25600         & 4.00     & 1         & 25 & 4.20 &$12^{\rm +1}_{\rm -1}$&FLAMES\\
\hline
N\,11-001     & B2 Ia    &  18750$^3$     & 2.50     & 14        & 50 & 5.66 &$38^{\rm +7}_{\rm -3}$&UVES\\
N\,11-002     & B3 Ia    &  15800$^1$     & 2.10     & 12        & 55 & 5.26 &$24^{\rm +3}_{\rm -2}$&UVES\\
N\,11-003     & B1 Ia    &  23200         & 2.75     & 13        & 80 & 5.42 &$30^{\rm +5}_{\rm -3}$&UVES  \\
N\,11-008     & B0.5 Ia  &  25450         & 3.00     & 15        & 75 & 5.39 &$30^{\rm +5}_{\rm -3}$&FLAMES\\
N\,11-009$^{R5}$& B3 Iab   &  15000$^1$     & 2.15     & 17        & 40 & 4.85 &$17^{\rm +2}_{\rm -1}$&FLAMES\\
N\,11-012     & B1 Ia    &  20500         & 2.55     & 14        & 70 & 5.13 &$22^{\rm +2}_{\rm -1}$&FLAMES\\
N\,11-014$^{R5}$& B2 Iab   &  19100$^1$     & 2.55     & 13        & 50 & 5.03 &$19^{\rm +2}_{\rm -1}$&FLAMES\\
N\,11-015     & B0.7 Ib  &  23600         & 2.95     & 11        & 75 & 5.23 &$24^{\rm +2}_{\rm -1}$&UVES  \\
N\,11-016     & B1 Ib    &  21700         & 2.75     & 14        & 60 & 5.13 &$22^{\rm +2}_{\rm -1}$&UVES  \\
N\,11-017$^{R5}$& B2.5 Iab &  16500$^1$     & 2.30     & 17        & 45 & 4.82 &$17^{\rm +2}_{\rm -1}$&FLAMES\\
N\,11-023     & B0.7 Ib  &  24000         & 2.90     & 14        & 70 & 5.09 &$21^{\rm +2}_{\rm -1}$&UVES  \\
N\,11-024     & B1 Ib    &  21600         & 2.80     & 12        & 55 & 4.96 &$18^{\rm +2}_{\rm -1}$&FLAMES\\
N\,11-029     & OC9.7 Ib &  28750         & 3.30     & 11        & 70 & 5.21 &$25^{\rm +3}_{\rm -2}$&FLAMES\\
N\,11-036     & B0.5 Ib  &  23750         & 3.10     & 11        & 55 & 4.95 &$18^{\rm +2}_{\rm -1}$&FLAMES\\
N\,11-037$^R$ & B0 III   &  28100         & 3.25     & 10        & 100& 5.08 &$23^{\rm +2}_{\rm -2}$&FLAMES\\
N\,11-042$^R$ & B0 III   &  29000         & 3.60     & 6         & 30 & 5.05 &$22^{\rm +2}_{\rm -2}$&FLAMES\\
N\,11-047$^R$ & B0 III   &  29200         & 3.65     & 8         & 55 & 5.03 &$21^{\rm +2}_{\rm -2}$&FLAMES\\
N\,11-054     & B1 Ib    &  23500         & 3.05     & 11        & 60 & 4.79 &$16^{\rm +2}_{\rm -1}$&FLAMES\\
N\,11-062$^R$ & B0.2 V   &  30400         & 4.05     & 5         & 25 & 4.95 &$21^{\rm +2}_{\rm -2}$&FLAMES\\
N\,11-069     & B1 III   &  24300         & 3.30     & 10        & 80 & 4.63 &$15^{\rm +1}_{\rm -1}$&UVES  \\
N\,11-072     & B0.2 III &  28800         & 3.75     & 5         & 15 & 4.77 &$18^{\rm +2}_{\rm -2}$&FLAMES\\
N\,11-075$^R$ & B2 III   &  21800$^3$     & 3.35     & 3         & 25 & 4.48 &$12^{\rm +1}_{\rm -1}$&FLAMES\\
N\,11-083$^R$ & B0.5 V   &  29300         & 4.15     & 0         & 20 & 4.71 &$17^{\rm +2}_{\rm -1}$&FLAMES\\
N\,11-100     & B0.5 V   &  29700         & 4.15     & 1         & 30 & 4.68 &$17^{\rm +2}_{\rm -1}$&UVES  \\
N\,11-101     & B0.2 V   &  29800         & 3.95     & 8         & 70 & 4.68 &$17^{\rm +2}_{\rm -2}$&UVES  \\
N\,11-106     & B0 V     &  31200         & 4.00     & 7         & 25 & 4.72 &$18^{\rm +2}_{\rm -2}$&FLAMES\\
N\,11-108     & O9.5 V   &  32150         & 4.10     & 7         & 25 & 4.73 &$19^{\rm +2}_{\rm -2}$&FLAMES\\
N\,11-109     & B0.5 Ib  &  25750         & 3.20     & 14        & 55 & 4.48 &$12^{\rm +2}_{\rm -1}$&FLAMES\\
N\,11-110     & B1 III   &  23100         & 3.25     & 6         & 25 & 4.37 &$12^{\rm +1}_{\rm -1}$&FLAMES\\
N\,11-124$^R$ & B0.5 V   &  28500         & 4.20     & 0         & 45 & 4.47 &$14^{\rm +2}_{\rm -1}$&FLAMES\\
\hline
NGC\,346-012     & B1 Ib    &  24200         & 3.20     & 8         & 30 & 4.77 &$16^{\rm +1}_{\rm -1}$&FLAMES\\
NGC\,346-021     & B1 III   &  25150         & 3.50     & 1         & 15 & 4.61 &$14^{\rm +1}_{\rm -2}$&FLAMES\\
NGC\,346-029$^R$ & B0 V     &  32150         & 4.10     & 0         & 25 & 4.82 &$19^{\rm +2}_{\rm -1}$&FLAMES\\
NGC\,346-037     & B3 III   &  18800$^1$     & 3.20     & 5         & 35 & 4.21 &$10^{\rm +1}_{\rm -1}$&FLAMES\\
NGC\,346-039$^R$ & B0.7 V   &  25800	     & 3.60	& 0	    & 20 & 4.51 &$13^{\rm +1}_{\rm -1}$&FLAMES\\
NGC\,346-040$^R$ & B0.2 V   &  30600	     & 4.00	& 0	    & 20 & 4.67 &$17^{\rm +2}_{\rm -2}$&FLAMES\\
NGC\,346-043     & B0 V     &  33000         & 4.25     & 4         & 10 & 4.71 &$18^{\rm +2}_{\rm -2}$&FLAMES\\
NGC\,346-044     & B1 II    &  23000$^2$     & 3.50     & 0         & 40 & 4.33 &$10^{\rm +2}_{\rm -1}$&FLAMES\\
NGC\,346-056     & B0 V     &  31000         & 3.80     & 1         & 15 & 4.55 &$16^{\rm +2}_{\rm -1}$&FLAMES\\
NGC\,346-062     & B0.2 V   &  29750         & 4.00     & 12        & 25 & 4.45 &$15^{\rm +1}_{\rm -1}$&FLAMES\\
NGC\,346-075$^R$ & B1 V     &  27700         & 4.30     & 0         & 10 & 4.31 &$12^{\rm +1}_{\rm -1}$&FLAMES\\
NGC\,346-094     & B0.7 V   &  28500         & 4.00     & 4         & 40 & 4.28 &$13^{\rm +1}_{\rm -1}$&FLAMES\\
NGC\,346-103     & B0.5 V   &  29500         & 4.00     & 0         & 10 & 4.26 &$13^{\rm +1}_{\rm -1}$&FLAMES\\
NGC\,346-116     & B1 V     &  28250         & 4.10     & 0         & 15 & 4.15 &$12^{\rm +1}_{\rm -1}$&FLAMES\\
\hline
\end{tabular}
\begin{itemize}
\footnotesize
\item[$^R$] Radial velocity variations detected at the 3$\sigma$ level and hence object may be a binary, see Sect.~\ref{s_EW}. $^{R5}$ 
indicates that the radial velocity variation is less than 5\,km\,s$^{-1}$.
\item[$^*$] After the discussion in Paper I, we have since revised the spectral type
of NGC\,6611-006 to B0 IV; its spectrum is intermediate between those of 
$\upsilon$ Ori and HD\,48434 from Walborn \& Fitzpatrick (\cite{wal90}), but with slightly stronger
\ion{He}{ii} $\lambda$4200 than one would expect.
\item[$^1$] $T_{\rm eff}$ estimated from ionization balance of \ion{Si}{ii} to
\ion{Si}{iii}.
\item[$^2$] $T_{\rm eff}$ estimated by placing upper limits on the EW of the
\ion{Si}{ii} and \ion{Si}{iv} lines, see Sect.~\ref{s_teff}.
\item[$^3$] $T_{\rm eff}$ estimates determined from the ionization balance of
\ion{Si}{ii} to \ion{Si}{iii} and \ion{Si}{iii} to \ion{Si}{iv} are in good agreement.
\normalsize
\end{itemize}
\end{table*}

\subsection{Selection of narrow lined stars}                     \label{s_sel}

The spectra of all the stars with a type later than O9 have been examined 
and we have selected stars for
analysis using the criterion that their metal absorption lines were sufficiently
narrow so that they could be easily identified and measured. 
O-type stars were not considered as they lay outside our grid of 
model atmosphere calculations (see Sect. \ref{s_nlte}) and are more appropriately
modelled with unified stellar wind codes, see for example, Mokiem et al. (\cite{mok05a},
\cite{mok05b}).

The spectra were examined for evidence of double-lined binarity, and if
contamination from a secondary object was apparent, they were 
excluded. Additionally stars were 
only selected for analysis if it was possible to reliably determine their 
effective temperature. As the 
clusters are young, many of the stars have early B spectral types and
effective temperatures could be deduced
using the ionization equilibrium  of \ion{Si}{iii} to \ion{Si}{iv}. 
For cooler objects (with effective temperatures less than 18\,000\,K) 
the ionization equilibrium of \ion{Si}{ii} to \ion{Si}{iii} could be used. 
However there was a range of effective temperatures within which both the 
\ion{Si}{ii} and \ion{Si}{iv} lines were too weak to measure and 
the helium spectrum could not be used to 
constrain the temperature. In these cases if the temperature could not be
constrained to better than $\pm$1\,000\,K by putting limits on the strength 
of both the \ion{Si}{ii} and \ion{Si}{iv} lines then the star
was not included in our analysis. The methods used to determine the temperatures 
are further discussed in Sect.~\ref{atm_par}. Due to these selection criteria and the low 
metallicity of the SMC it was only normally possible to analyse stars in NGC\,346 with a
maximum projected rotational velocity of approximately 50\,km\,s$^{-1}$. As the absorption lines 
are stonger in the LMC and Galactic stars, and also as we have several supergiant objects in the LMC sample,
stars with an implied projected rotational 
velocity of up to 100\,km\,s$^{-1}$ could be analysed in NGC\,6611 and N\,11.

An additional selection criterion was whether the spectra were of sufficient 
quality to reliably measure the equivalent widths (EW) for absorption lines from several chemical species.
The \ion{Si}{iii} triplet of lines at 4560\AA\ was observed in two wavelength 
orders leading to independent estimates for their equivalent widths. 
Objects have only been included in this analysis if
agreement between the EW's from the two orders was better than 10\%. 
Table~\ref{t_stars+atmos} lists all the stars
that were selected for analysis based on these criteria. Star identifications have been adopted from 
Paper I and Paper II for the Galactic stars and Magellanic Cloud
stars respectively and alternative identifications along with radial velocity estimates can be found therein.
It should be noted that this analysis represents only a subset of the B-type objects observed in 
the survey as some objects lay outside our strict selection criteria.

\subsection{Equivalent Width Measurements}                         \label{s_EW}

Several exposures were taken for each cluster in each wavelength setting and these are
summarized in Table~\ref{t_exp}. Spectra in each wavelength setting were
cross-correlated and corrected for any resulting velocity shifts. In several wavelength settings the exposures were taken in two sets of three at different dates
and stars were
identified as possible single lined spectroscopic binary objects if the mean radial velocity of each set of exposures differed at the 3$\sigma$ 
level, see Table~\ref{t_stars+atmos}. Of course, especially for small radial velocity variations
binarity is not the only explanation; pulsations or uncertainties in the
wavelength calibrations may also contribute to variations. Additionally, wind effects can lead to a change in the line
profiles and hence spurious radial velocity variations may be detected, especially in the redder wavelength regions. As such we do not include objects as
radial velocity variables in Table~\ref{t_stars+atmos} if changes in the shape of the line
profile were also observed. Generally, velocity shifts could be identified to an accuracy of better 
than 5\,km\,s$^{-1}$, except for the brightest targets in N\,11 where velocity shifts of less than 
2\,km\,s$^{-1}$ could be detected. For the targets not observed with FLAMES only a single exposure
was available and hence it was not possible to examine these objects for evidence of binarity.

The individual exposures in each wavelength region were 
then combined using {\sc IRAF}\footnote{{\sc IRAF} is distributed by the National
Optical Astronomy Observatories, which are operated by the Association of Universities
for Research in Astronomy, Inc., under agreement with the National Science Foundation}
routines which also removed most cosmic 
ray events. The combined FLAMES spectra had signal to noise (S/N) ratios per pixel ranging from 60
to in excess of 400 with the majority of the NGC\,6611 and NGC\,346 stars having S/N ratios
greater than 100 and the majority of the N\,11 targets having S/N ratios greater than 200. The
single spectrum in this sample which was obtained with FEROS had a S/N ratio of 80, while those from 
UVES had S/N ratios ranging from 60-110.

The combined spectra were normalised within the spectral analysis 
package {\sc dipso} (Howarth et al. \cite{how94}) and the 
line fitting program {\sc elf} was used to measure the
EW of the absorption lines by fitting Gaussian profiles to the line and 
using a low order polynominal to represent the continuum. As discussed above, 
we estimated the error in the EW of well observed unblended \ion{Si}{iii} features to 
be better than 10\%. For weak or blended features an error of 20\% may 
be more appropriate. EW measurements of the absorption lines in the spectra of
each star listed in Table~\ref{t_stars+atmos} are 
given in Table~3, Table~4 and Table~5 (only available online at the CDS) for NGC\,6611, N\,11 and NGC\,346
respectively. In these tables we also list
the ionic species, the rest wavelength
and the derived abundance of each absorption line. Lines with measured equivalent widths and no
corresponding abundance estimate were cases where reliable theoretical calculations were not
available.

\begin{table}[htbp]
\caption{Number of exposures taken by the FLAMES spectrograph in each 
wavelength setting. See Papers I and II for further details.}
\label{t_exp}
\centering
\begin{tabular}{lcccc}\hline \hline
Setting    & Central & \multicolumn{3}{c}{Number of exposures} \\
           &Wavelength (\AA)& NGC\,6611 & N\,11 & NGC\,346\\
\hline
\\
HR02 & 3958          & 2         & 6     & 6       \\
HR03 & 4124          & 2         & 6     & 6       \\
HR04 & 4297          & 2         & 6     & 6       \\
HR05 & 4471          & 4         & 6     & 6       \\
HR06 & 4656          & 4         & 6     & 8       \\
HR14 & 6515          & 4         & 6     & 9       \\
\\
\hline
\end{tabular}
\setcounter{table}{6}
\end{table}

\section{Analysis}                              \label{s_atmos}

\subsection{Non-LTE atmosphere calculations}                      \label{s_nlte}

Non-LTE model atmosphere grids generated using {\sc tlusty} and {\sc synspec} 
(Hubeny \cite{hub88}; Hubeny \& Lanz \cite{hub95}; Hubeny et al. \cite{hub98})
were used throughout this analysis to derive atmospheric parameters and 
chemical abundances. An overview of the methods can be found in Hunter et al.
(\cite{hun05}) while a more detailed discussion of the methods and grids can be 
found in Ryans et al. (\cite{rya03}) and Dufton 
et al. (\cite{duf05})\footnote{See also http://star.pst.qub.ac.uk}. The adopted
atomic data is listed in Table~6 (only available online at the CDS). In this table
we list the species, wavelength, $\log gf$ value and also indicate if the line is
considered as part of a blend of lines in the {\sc TLUSTY} code, see Dufton et al., Allende Prieto et al. \cite{all03}
and Lanz \& Hubeny \cite{lan03} for details.

Briefly, four model atmosphere grids have been calculated for
metallicities correponding to a Galactic metallicity of [Fe/H]=7.5 dex, 
and metallicities of 7.2, 6.8 and 6.4 dex to represent the LMC, SMC and lower
metallicity material respectively (see for example, Gies \& Lambert \cite{gie92}, 
Luck \& Lambert \cite{luc92}, Bouret et al. \cite{bou03} and Lehner \cite{leh02}). 
For each of these grids non-LTE models have
been calculated for effective temperatures ranging from 12\,000K to 35\,000K, in
steps of no more than 2\,500K, surface gravities ranging from 4.5\,dex down to the
Eddington limit, in steps of no more than 0.25\,dex and for microturbulences of
0, 5, 10, 20, and 30\,km\,s$^{-1}$. Assuming that the light elements (C, N, O, Mg
and Si) have a negligible effect on line-blanketing and the structure of the
stellar atmosphere, models were then generated with
the light element abundance varied by +0.8, +0.4, -0.4 and -0.8\,dex about their
normal abundance at each point on the {\sc tlusty} grid. A helium abundance of 
11.0\,dex has been adopted throughout these grids and the implication of this with respect to
our sample of stars is discussed in Sect.~\ref{s_helium}.
Theoretical spectra and EW's were then calculated
based on these models. Photospheric abundances for approximately 200 absorption lines at any 
particular set of 
atmospheric parameters covered by the grid 
can then be calculated by interpolation between the models via simple IDL routines.
The reliability of the interpolation
technique has been tested by Ryans et al. (\cite{rya03}) and they
report that no significant errors arise from it.

\subsection{Stellar Atmospheric Parameters}                      \label{atm_par}

A static stellar atmosphere is normally characterised by four parameters, viz. 
the effective temperature, surface gravity, microturbulence and metallicity.
These parameters are interdependent and hence it is necessary to use 
an iterative method to estimate them. The metallicity (iron content) 
was assumed to be constant for each cluster, i.e. Galactic metallicity 
([Fe/H] = 7.5 dex) for NGC\,6611, and reduced by 0.3\,dex  and 0.7\,dex for
N\,11 and NGC\,346 respectively. Hunter et al. (\cite{hun05}) and Dufton et
al. (\cite{duf05}) have found that these assumptions do not normally 
cause a significant
uncertainty in the estimation of the 
atmospheric parameters or derived abundances and this is further discussed in
Sect.~\ref{s_errors}.
By careful selection of the initial parameters (based on the spectral types
from Paper I or estimates given in Dufton et al. \cite{duf06}), 
relatively few iterations were necessary to derive the effective temperature, 
surface gravity and microturbulence. The atmospheric parameters that were 
adopted for each star are given in Table~\ref{t_stars+atmos}.

\subsubsection{Effective Temperature}            \label{s_teff}

\begin{figure}[hbtp]
\centering
\epsfig{file=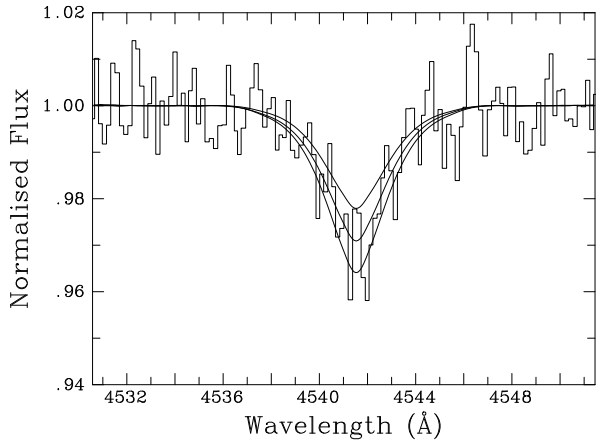, height=90mm, angle=-90}
\caption[]{Observed spectra for the \ion{He}{ii} 4542\AA\ line
in NGC\,346-040. Theoretical spectra for
temperatures of 30\,500\,K, 31\,000\,K and 31\,500\,K (upper, middle and lower
smooth curves respectively) are also plotted. The effective temperature deduced from the 
\ion{Si}{iii} to \ion{Si}{iv} ionization equilibrium is 30\,600\,K.}
\label{f_He_temp_fit}
\end{figure}

Given the age of the clusters, the majority of the targets have early
B spectral types and therefore the ionization equilibrium of \ion{Si}{iii} 
to \ion{Si}{iv} could be used to estimate the effective temperature ($T_{\rm eff}$),
see for example, Kilian et al. (\cite{kil91}). Temperatures were rounded to the nearest
50\,K which resulted in an imbalance of up 
to 0.04\,dex between the abundances of 
two Si ionization states.
For the hottest objects, the \ion{He}{ii} spectrum provided an 
additional criterion. The values derived using the two methods were 
in good agreement, with differences normally being less than 1\,000\,K.
While this agreement is encouraging, in most cases the estimates
derived from the \ion{He}{ii} lines were systematically higher than 
those from the silicon lines. For example, in Fig.~\ref{f_He_temp_fit} we plot the
\ion{He}{ii} line for a typical case (NGC\,346-040) where the temperature
implied by the \ion{He}{ii} line spectra is
several hundred Kelvin higher than that estimated from the silicon lines.

Additionally atmospheric parameters 
of several of our Magellanic Cloud objects have been independently estimated by
Mokiem et al. (\cite{mok05a}, \cite{mok05b}) using the non-LTE code {\sc fastwind}
(Puls et al. \cite{pul05}). In their analysis they adopt an alternative method of fitting 
the hydrogen and helium spectra
to estimate effective temperatures and surface gravities and in all cases higher values were
derived compared to the results given here. As values estimated for
these atmospheric parameters are directly correlated, the higher gravity 
probably simply arises from a higher effective temperature estimate. 
We have derived effective temperatures using a similar method (i.e. fitting the 
\ion{He}{ii} line at 4542\AA) and our results are listed in Table~\ref{tmokiem} 
and compared to those of Mokiem et al. with the two sets of estimates now 
being in better agreement. We therefore
attribute the differences between our adopted atmospheric parameters and those of 
Mokiem et al. to
the choice of methodology. However, Mokiem et al. (\cite{mok05}) have noted that their genetic
algorithm method of determining the atmospheric parameters systematically derives higher gravities than
those estimated `by-eye' and hence, given the correlation between the atmospheric parameters, this
may contribute to the difference in the temperature estimates.

Dufton et al. (\cite{duf06}) have derived an effective temperature of 34\,000\,K for NGC\,6611-006 using the
{\sc fastwind} code to model the \ion{He}{ii} spectrum. Using the \ion{He}{ii} 4542\AA\ line we obtain an 
effective temperature of 33\,000\,K, 1\,750\,K higher than 
that given in Table~\ref{t_stars+atmos} and in reasonable agreement
with Dufton et al.
As the \ion{He}{ii} line
can only be used to estimate the temperature of a small proportion of our sample, we
have elected to adopt the estimates from the \ion{Si}{iii} to \ion{Si}{iv} ionization equilibrium 
wherever possible in order to maintain consistency throughout this analysis.

\begin{table*}[htbp]
\centering
\caption{Comparison of the atmospheric parameters listed in Table~\ref{t_stars+atmos} with those derived by 
Mokiem et al. (\cite{mok05a}, \cite{mok05b}). For these stars, we also estimate the 
effective temperature from the \ion{He}{ii} line
at 4542\AA\ and re-iterate the surface gravity where necessary. Errors in our estimate 
of the effective temperature from
the \ion{He}{ii} spectra are typically about 1\,500\,K.}
\label{tmokiem}
\begin{tabular}{lcccccc}\hline \hline
Star         & \multicolumn{2}{c}{From Table~\ref{t_stars+atmos} } & \multicolumn{2}{c}{Mokiem et al.} 
& \multicolumn{2}{c}{\ion{He}{ii} 4542\AA\ line}\\
             & $T_{\rm eff}$ & $\log g$                       & $T_{\rm eff}$ & $\log g$  & $T_{\rm eff}$ & $\log g$\\
	     \hline           
N\,11-008    & 25450         & 3.00                           & 26000         & 3.00      & 25500         & 3.00\\
N\,11-029    & 28750         & 3.30                           & 29400         & 3.25      & 29000         & 3.30\\
N\,11-036    & 23750         & 3.10                           & 26300         & 3.30      & 25500         & 3.25\\
N\,11-042    & 29000         & 3.60                           & 30200         & 3.70      & 29500         & 3.60\\
N\,11-072    & 28800         & 3.75                           & 30800         & 3.80      & 30500         & 3.75\\
NGC\,346-012 & 24200         & 3.20                           & 26300         & 3.35      & 25500         & 3.25\\
\\
\hline
\end{tabular}
\end{table*}

For our coolest objects, the ionization equilibrium of \ion{Si}{ii} to 
\ion{Si}{iii} could be used to determine the effective temperature. 
For our Magellanic Cloud targets, the resulting silicon abundances (see
Table \ref{t_abund}) appeared to be consistent with those estimated for our 
hotter targets. For the one Galactic object (NGC\,6611-033),
the abundance appeared to be underestimated although this may be
due to uncertainties in the microturbulence as discussed in
Sect.~\ref{s_xi} and Sect.~\ref{s_errors}

Three stars in our sample, N\,11-001, N\,11-014 and N\,11-075, reveal measureable Si lines
from all three ionization stages and this allows us to compare the effective
temperature derived from the ionization balance of \ion{Si}{ii}/\ion{Si}{iii} to
that derived from \ion{Si}{iii}/\ion{Si}{iv}. In the case of N\,11-014 
inspection of the models reveals that the results for
\ion{Si}{iv}
become unstable at its estimated surface gravity, which is close to the Eddington
limit. These instablilites do not appear to be as serious for \ion{Si}{ii} or
\ion{Si}{iii} spectra and hence
we have elected to use these ions to estimate the effective temperature of
this star. For N\,11-001 and N\,11-075, we find that the estimated effective temperatures
from the two ionization equilibria agree to within 200\,K, highlighted by the
almost identical abundance derived for Si from the three ionization stages (see
Table~\ref{t_abund}). This agreement is encouraging as it again indicates that
additional errors should not be present when comparing stars where the temperatures
have been derived using different ionization stages of silicon.

In some cases (especially at low metallicity) neither the \ion{Si}{ii} or 
the \ion{Si}{iv} lines could be observed. By placing upper limits on their 
equivalent widths, lower and upper limits to the effective temperature 
could be estimated (see for example, Trundle et al. \cite{tru04}). As discussed
in Sect.~\ref{s_sel}, these differed by  
less than 2\,000\,K and their mean was adopted. Carbon was the only other element
for which two ionziation stages were  
available, but given the very simple \ion{C}{iii} model atom that was 
used in the {\sc tlusty} grid, the \ion{C}{ii}/\ion{C}{iii} ionization
equilibrium is probably not a reliable temperature estimator. 
We discuss the modelling of the
carbon spectra further in Sect.~\ref{s_carbon}.

On the basis of the quality of the observed spectra and the agreement 
between the different criteria adopted, we believe that the typical 
uncertainty in our effective temperature estimates should be of the 
order of $\pm$1\,000K. 

\subsubsection{Logarithmic Surface Gravity}          \label{s_logg}

The logarithmic surface gravity ($\log g$) of each star was estimated by fitting the observed 
hydrogen Balmer lines with theoretical profiles. Automated procedures 
have been developed to fit model spectra in our {\sc tlusty} model 
atmosphere grid to the observed spectra, with contour maps displaying the
region of best fit. If an effective temperature estimate was available, 
(e.g. from the methods described above), it was a simple matter to estimate 
the gravity. This could then be used to re-iterate for our effective temperature
estimate. For several of the supergiants in N\,11 (N\,11-001, N\,11-003, N\,11-008, N\,11-016)
our gravity estimate lay near to the edge of our
grid and our atmospheric parameters and abundance estimates should be treated 
with some caution.

\begin{figure*}[hbtp]
\centering
\epsfig{file=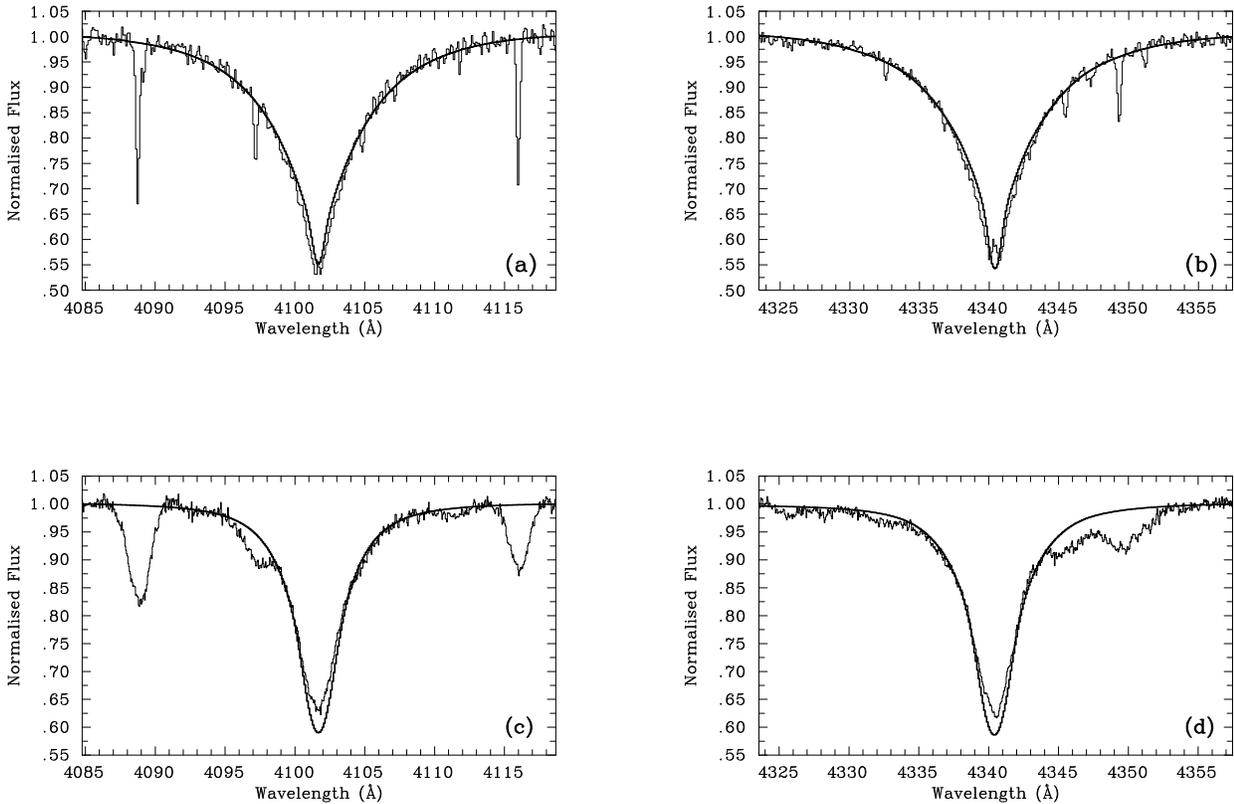, height=180mm, angle=-90}
\caption[]{(a) and (b) show the quality of the model fit onto the observed H$\delta$ and H$\gamma$ lines
of NGC\,346-043 (a slowly rotating B0\,V with a $v\sin i$ of 10\,km\,s$^{-1}$). 
Similarly, (c) and (d) show the fitting of the same hydrogen
lines for N\,11-037 (a fast rotating B0\,III star with a $v\sin i$ of 100\,km\,s$^{-1}$). 
The smooth curves represent the models. Note, the model spectra shown
only includes data for the H-lines and excludes those due to metals.} 
\label{f_logg_fits}
\end{figure*}

In order to quantify the uncertainty in our gravity estimate arising from 
both our normalisation and fitting procedures, we have derived the surface 
gravity from both the H$_{\rm \delta}$ and H$_{\rm \gamma}$ lines. 
Normalisation of the hydrogen lines for the stars in NGC\,6611 was
complicated by the relatively strong metal absorption lines in the wings of the
H~lines but this was less of a problem for the lower
metallicity stars in N\,11 and NGC\,346. Nevertheless,
the surface gravity estimates from the two hydrogen lines were in 
agreement to typically
better than 0.1\,dex for high gravity stars and 0.05\,dex for the supergiants.
Fig.~\ref{f_logg_fits}
shows the agreement of the gravity estimate between the H$\delta$ and the H$\gamma$ line
for a main-sequence and a giant star. The 
uncertainty in the effective temperature estimate would
typically contribute an additional uncertainty of 0.1\,dex.

\subsubsection{Microturbulence}    \label{s_xi}

The microturbulence is normally derived by removing the dependence of the 
abundance estimates for lines of a specific ion on line strength. For B-type 
stars, the \ion{O}{ii} ion 
is often considered (for example, Sim\'{o}n-D\'{i}az et al. \cite{sim06}, Hunter et al. \cite{hun05}, Gies \& Lambert \cite{gie92}, 
and Daflon et al.
\cite{daf04}) as its rich spectrum should improve its 
reliablity. However its use is complicated by the lines arising from different
multiplets,
making any estimate susceptible to errors in the adopted atomic data or in 
the magnitude of non-LTE effects. In order to
remove these uncertainties,  
a single \ion{O}{ii} multiplet can be used, but then one is reduced to 
using relatively few lines. While this was possible for the sharpest lined 
NGC\,6611 stars, it was not always feasible in the lower metallicity 
environments of N\,11 and NGC\,346 as the weakest line in any multiplet 
often could not be identified. In order to maintain consistency throughout the
analysis we have instead initially estimated the microturbulence ($\xi_{\rm Si}$) from 
the \ion{Si}{iii} triplet of lines at 4552--4574\AA\ as these lines could be seen 
in all our spectra. As these lines are from the same multiplet, errors arising
from the oscillator 
strengths and departure co-efficients will be negligible and this method has
been used previously by, for example,  
Dufton et al. (\cite{duf05}) and Vrancken et al. (\cite{vra97}, \cite{vra00}).

As the derived microturbulence depends on the adopted equivalent widths, 
we have investigated the effect of changing these by their estimated 
errors (10\% in the case of strong lines, and by 5\,m\AA\ for lines 
with EW's of less than 50\,m\AA). Typically such errors lead to an 
uncertainty in the microturbulence estimate of approximately 3\,km\,s$^{-1}$, 
although for stars with higher microturbulences (where the derived 
abundances are less sensitive to microturbulence) an error of
5\,km\,s$^{-1}$ is probably more realistic.

\begin{figure}
\centering
\epsfig{file=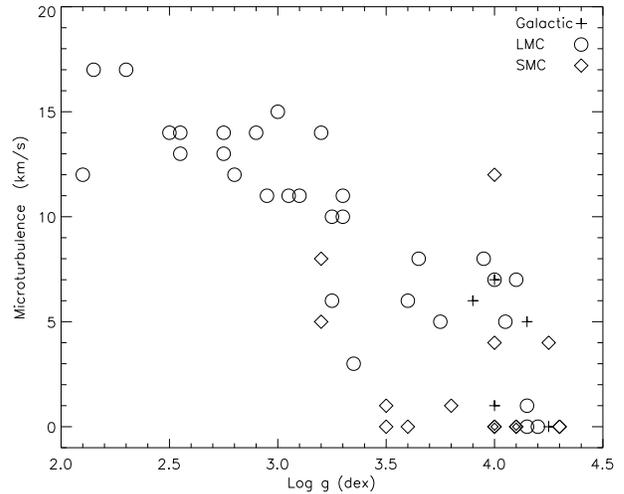, height=70mm, angle=-0}
\caption[]{Microturbulence against surface gravity for all stars listed in
Table~\ref{t_stars+atmos}. The points plotted as diamonds inside diamonds 
represent two stars lying at the same position on the graph.}
\label{f_xivslogg}
\end{figure}

Previous studies (such as Daflon et al. \cite{daf04}) have shown that the
microturbulence  
generally increases as surface gravity decreases and in 
Fig.~\ref{f_xivslogg} we plot our estimates against surface gravity for
all the stars listed in Table~\ref{t_stars+atmos}. The largest sample of 
stars comes from the LMC and these stars also 
have the widest range of gravities. From Fig.~\ref{f_xivslogg} it can 
be seen that the LMC stars show a strong trend of increasing microturbulence 
as surface gravity decreases for surface gravities less than 3.3\,dex. 
However, at gravities greater than 3.3\,dex there is a
much greater scatter, and little evidence of any correlation with gravity; hence 
Fig.~\ref{f_xivslogg}
could be interpreted as a bi-modal distribution. Only five near main-sequence stars were 
analysed in NGC\,6611 and hence it is not possible to identify any trend. 
The 14 targets in NGC\,346 cover a relatively wide 
range of gravities (from 3.2\,dex to 4.3\,dex), but no trend is apparent which is
in agreement with the higher gravity LMC stars.

Examination of the location of the SMC stars in Fig.~\ref{f_xivslogg} 
reveals two important results. Firstly, seven SMC stars were assigned 
a microturbulence of 0\,km\,s$^{-1}$. Furthermore, in some cases adopting 
this value was not sufficent to remove the dependence of the abundance 
upon line strength, in the sense that the weakest line of the multiplet 
still gave the highest abundance. If we adjusted the EW's of the silicon 
lines by their assumed errors, a zero microturbulence could be obtained. However
given that this discrepancy occured in four out of the seven SMC stars
assigned a zero microturbulence, this is not a convincing
explanation. Rather, there may be other physical processes
occuring which affect the curve-of-growth of a given multiplet 
but which may be dominated (or at least masked) by the microturbulence 
parameter at higher metallicity.

Secondly, examination of Fig.~\ref{f_xivslogg} reveals stars 
in NGC\,346 with surface gravities of 4.0\,dex and similar
effective temperatures of approximately 30\,000\,K, two of which have 
microturbulences of 0\,km\,s$^{-1}$ (NGC\,346-103 and NGC\,346-040)
whilst one of the others has a microturbulence of 12\,km\,s$^{-1}$ (NGC\,346-062). 
In order to determine 
if this is a real effect, the EW of the \ion{Si}{iii} lines have again 
been adjusted by their assumed errors in order to derive the lowest 
possible microturbulence for the case of NGC\,346-062 (5\,km\,s$^{-1}$) 
and higher microturbulences for the other two cases (5\,km\,s$^{-1}$ 
and 1\,km\,s$^{-1}$ for NGC\,346-103 and NGC\,346-040 respectively).
However, if we adopt these microturbulences the spread in silicon 
abundance estimates between the three stars increases. Assuming that all three stars
should have the same silicon abundance, would indicate that our original 
microturbulence estimates given in Table~\ref{t_stars+atmos} are more 
reliable at least for abundance determinations. Such differences in the 
microturbulence estimate for stars with similar effective temperatures
and gravities and situated in the same cluster (and hence, with similar metallicity 
and age) is worrying. This again could imply that there are other physical 
processes within the stellar photosphere that are affecting the shape of the curve-of-growth 
but are not fully included in our models. Such differences in microturbulence 
estimates are not found to such an extent in the higher metallicity 
environments of our N\,11 and NGC\,6611 sample.

The choice of microturbulence is more critical when estimating 
abundances from stronger lines and hence will be most important for the 
Galactic cluster, NGC\,6611. In the lowest metallicity region, NGC\,346, 
the microturbulence may be more uncertain for the reasons discussed above, 
but the derived abundances will be less
dependent upon our estimate.

\subsubsection{Microturbulence estimated from other species}

\begin{figure}
\centering
\epsfig{file=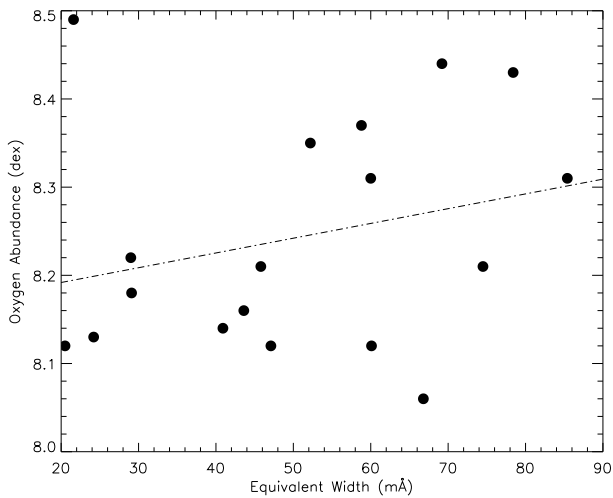, height=70mm, angle=-0}
\caption[]{\ion{O}{ii} abundance against line-strength for each \ion{O}{ii} line observed
in the spectra of N\,11-075. The oxygen abundance has been derived using the
parameters given in Table~\ref{t_stars+atmos}. Blends of lines have not been included in
this figure.} 
\label{f_micro}
\end{figure}

Although we 
have adopted the microturbulence from the silicon lines, we have 
investigated how using alternative species would change our adopted values. 
As discussed above, two methods are available to us, viz. we can use all 
the lines of a species or only lines from a single multiplet. Given the 
wealth of \ion{O}{ii} lines observed in our spectra it would normally be expected that any 
derived microturbulence parameter could be more reliably deduced than when using 
a single multiplet. However as discussed above this
is complicated by the uncertainties arising from the adopted atomic data and non-LTE
departure coefficients (see Sim\'{o}n-D\'{i}az et al. \cite{sim06}
for a detailed discussion). Normally 
including all the oxygen lines (apart from those that are close blends) 
gives significantly higher microturbulences. Indeed, Vrancken et al.
(\cite{vra97}, \cite{vra00}) report differences of up to 9\,km\,s$^{-1}$
between 
the microturbulence derived from the \ion{O}{ii} lines and that derived from the
\ion{Si}{iii} lines. As a typical example, 
the oxygen abundances from the \ion{O}{ii} spectrum in N\,11-075 
have been plotted as a function of line strength in Fig.~\ref{f_micro}, 
adopting the microturbulence from Table~\ref{t_stars+atmos}. 
The linear least squares fit implies that the microturbulence should 
be increased in order to remove the apparent dependence of abundance 
upon line strength. However, given that the uncertainty due to the scatter in this figure 
is greater than the gradient of the best-fitting line, the
credibility of any derived microturbulence must be questionable. Nevertheless,
we find that a microturbulence of 6\,km\,s$^{-1}$ would remove the dependence 
but would not significantly reduce the scatter.  Adopting this larger microturbulence
reduces the oxygen and silicon abundances by 0.1\,dex and 0.3\,dex respectively but has little 
effect on the other abundances.

We have also investigated the effect of using a single multiplet of a species
to derive the microturbulence for N\,11-075. Ideally this would be the best approach for
each individual species but in several cases (e.g. \ion{Mg}{ii}, \ion{C}{ii}) there are
insufficient lines from any single multiplet. Nevertheless, for the 
sake of this comparison we have estimated the microturbulence 
from both an oxygen and a nitrogen multiplet for N\,11-075. The multiplet of \ion{O}{ii} 
lines at 4072\AA, 4076\AA\ and 4079\AA\  
is the most suitable oxygen multiplet as it has the largest range of line strengths 
of those observed. Using this, we find a 
microturbulence of 2\,km\,s$^{-1}$, which is 4\,km\,s$^{-1}$ smaller than
the values found from using all the \ion{O}{ii} lines and in reasonable agreement with 
the value of 3\,km\,s$^{-1}$ deduced from the \ion{Si}{iii} multiplet. The only multiplet of 
N lines available to us are the lines at 4601\AA, 4614\AA\ and 
4630\AA\ and this multiplet has a relatively small spread of EW's. This
multiplet gives an estimated microturbulence of 2\,km\,s$^{-1}$, which again is in good agreement
with the microturbulence derived using the \ion{O}{ii} and \ion{Si}{iii} multiplets. Given this agreement
it seems reasonable to adopt the microturbulence estimate from the \ion{Si}{iii} multiplet which is seen
in all of our stellar spectra.

\subsection{Luminosity}

Luminosities for each star are given in Table~\ref{t_stars+atmos}. For our NGC\,6611 targets these 
have been adopted from Dufton et al. (\cite{duf06}). However, if we adopt
the same methodology as Dufton et al. to calculate the reddening towards the Magellanic Cloud
stars, (i.e. using observed and intrinsic colours)
this leads to unphysical negative values in some cases. Given the low extinction towards the 
Magellanic Clouds it appeared to be more appropriate to adopt a single value for N\,11 and NGC\,346.

For NGC\,346 we adopt an $E(B-V)$ value of 0.09 (Massey et al. \cite{mas95}) and
$A_{\rm V}$\,=\,2.72$E(B-V)$ (Bouchet et al. \cite{bou85})
with a distance modulus of 18.91 (Hilditch et al. \cite{hil05}) and the bolometric corrections from Vacca et al. (\cite{vac96}) and
Bolona (\cite{bol94}). 
Note that adopting a standard Galactic law of $A_{\rm V}$\,=\,3.1$E(B-V)$
changes the estimated luminosities by less than 0.03\,dex. 

N\,11 is complicated by containing at least two
distinct regions of star formation, LH\,9 and LH\,10, and Parker et al. (\cite{par92}) 
give $E(B-V)$
values of 0.05 and 0.17 respectively. Ideally, one would like to associate each star 
in our sample with either
LH\,9 or LH\,10, but this is further complicated as a significant fraction of the 
sample may
be field stars. As such we have adopted the $E(B-V)$ of 0.13 of 
Massey et al. (\cite{mas95}) for our LMC 
targets. The standard law of 
$A_{\rm V}$\,=\,3.1$E(B-V)$ together with a distance modulus of 18.56 (Gieren et al. \cite{gie05}) was used in
estimating the luminosities.

\subsection{Masses}

Evolutionary masses are given in Table~\ref{t_stars+atmos} and have been deduced by 
interpolating between the 
evolutionary tracks of Meynet et al. (\cite{mey94}) along with those of Schaller et al. (\cite{sch92}), 
Schaerer et al. (\cite{sch93}) and Charbonnel et al. (\cite{cha93}) for NGC\,6611, N\,11 and NGC\,346 
respectively.
The uncertainties are calculated assuming
an error of 0.1\,dex in $\log L/L_{\sun}$ with the estimated error of 1\,000\,K in $T_{\rm eff}$ 
generally having a negligible 
effect. Non-rotating tracks have been used although as shown by 
Maeder \& Meynet (\cite{mae01}) inclusion
of rotation does not significantly affect the derived masses. Additionally it 
was possible to calculate spectroscopic masses based on the atmospheric 
parameters and luminosities. However, taking uncertainties of 0.2\,dex in $\log g$, 
1\,000\,K in $T_{\rm eff}$
and 0.1\,dex in $\log L/L_{\sun}$ we calculate an uncertainty in the spectroscopic 
mass of approximately 66\%, with the 
uncertainty in the estimated surface gravity dominating this error. 
Spectroscopic mass estimates are
therefore not particularly useful, but within their errors they are 
consistent with evolutionary masses in the majority of cases.

\subsection{FASTWIND analysis of N\,11-001 and N\,11-002} \label{s_fastwind}

Mass-loss effects in supergiants may be significant, especially for the most massive stars and hence
the plane-parallel static atmosphere that TLUSTY assumes may not be valid.
Therefore we have additionally analysed two of the massive supergiants in our sample, 
N\,11-001 and N\,11-002,
using the unified model atmosphere 
code FASTWIND and the resulting atmospheric parameters are listed in Table~\ref{t_fastwind}.

The atmospheric parameters deduced using the two codes are in reasonable agreement. For N\,11-001, 
the effective temperature disagrees by 1250\,K, the gravity by 0.3\,dex and the microturbulence
by 2\,km\,s$^{-1}$ whilst for N\,11-002, both codes
yield identical estimates. The mass-loss determined from the FASTWIND analyses are 
4.5$\times$10$^{-7}$ and 2.5$\times$10$^{-7}$\,M$_{\sun}$yr$^{-1}$
for N\,11-001 and N\,11-002 respectively. The
agreement between TLUSTY and FASTWIND is expected as Dufton et al.
(\cite{duf05}) used both codes to derive parameters and abundances for SMC
supergiants and found good agreement between the two sets of estimates. As such we can be confident
that no additional uncertainties arise from using TLUSTY to analyze the more evolved objects in this
sample and hence TLUSTY parameters have been adopted throughout for consistency.

\begin{table}[htbp]
\centering
\caption{FASTWIND parameters of N\,11-001 and N\,11-002.}
\label{t_fastwind}
\begin{tabular}{lcc}\hline \hline
                               & N\,11-001 &N\,11-002 \\
\hline
$T_{\rm eff}$(K)               & 17500     & 15800 \\
$\log g$  (dex)                & 2.20      & 2.10  \\
$\xi_{\rm Si}$ (km\,s\,$^{-1}$)& 12        & 12    \\
\\
\hline
\end{tabular}
\end{table}

\subsection{Projected Rotational Velocity}       \label{s_vsini}

The stellar projected rotational velocity ($v\sin i$) has been estimated
using the \ion{Si}{iii}
triplet as these lines are intrinsically narrow and were observed throughout
our sample. For each star, theoretical
spectra with atmospheric parameters similar to those listed in
Table~\ref{t_stars+atmos} were selected and their EW's 
scaled to those observed. These theoretical lines were then convolved with the appropriate instrumental
broadening profiles. IDL
procedures have been developed which rotationally broaden this theorectial line
profile over a specified range of $v\sin i$
values and these broadened profiles were compared to the observed lines. A 
chi-squared minimisation test was then performed and the $v\sin i$
value which gave the best fit to the
observed absorption line profile was returned. This procedure was carried out for
each of the \ion{Si}{iii} lines in the 4560\AA\ multiplet and the agreement
between the three lines was usually better than 5\,km\,s$^{-1}$. From
Table~\ref{t_stars+atmos}, it can be seen that the stars in N\,11 have a larger
distribution of $v\sin i$ values than in the other clusters but this is simply a
consequence of N\,11 containing a proportionally larger number of high
luminosity stars. It should be noted that we have not considered other possible broadening
mechanisms such as macroturbulence (see for example Ryans 
et al. \cite{rya02}) due to the difficulty of distinguishing such broadening
from the rotational broadening through profile fitting. As such the
estimates given in Table~\ref{t_stars+atmos} should be treated as upper limits to the actual
projected rotational velocity especially for the higher
luminosity targets. The $v\sin i$ values quoted in Table~\ref{t_stars+atmos}
have been rounded to the nearest 5\,km\,s$^{-1}$ and an uncertainty of 5\,km\,s$^{-1}$ in these values
can be considered as appropriate.

\subsection{Non-LTE photospheric abundances}                    \label{s_abund}
 
The stellar photospheric abundances of carbon, nitrogen, oxygen, magnesium and 
silicon have been estimated using the non-LTE {\sc tlusty} model 
atmosphere grid and the atmospheric parameters listed in 
Table~\ref{t_stars+atmos}. The abundance derived from each absorption line in the
spectra of each star are given in Tables~3-5 (online only; see Sect.~\ref{s_EW}) with the average abundance 
estimates for each star summarized in Table~\ref{t_abund}.

\subsubsection{Errors in photospheric abundances} \label{s_errors}

The uncertainties given in Table~\ref{t_abund} include both the random 
uncertainties arising from, for example, observational errors and errors in 
individual oscillator strengths, together with the
systematic uncertainties arising from the adopted atmospheric parameters. 
The random uncertainty was taken to be the standard error in the mean, which is the
standard deviation of the abundances 
derived from each line of a given species divided by the square root of the 
number of lines observed for that species. In cases where only one line 
of a particular species was observed (e.g.
\ion{Mg}{ii}), the random uncertainty was taken to be the standard
deviation of the abundances derived for our best observed 
species, \ion{O}{ii}. The random uncertainty will include both measurement errors and random errors in the
atomic data. The systematic uncertainties were estimated by changing each of the atmospheric parameters in turn by their 
associated uncertainty (discussed in the previous sections) and the 
derived abundances were then compared to the values listed in 
Table~\ref{t_abund} to give an estimate of the error. The random and systematic uncertainties were
then summed in quadrature to give the uncertainties listed in 
Table~\ref{t_abund}. However, we note that we have not accounted for the correlation between effective temperature
and surface gravity in the calculation of the systematic errors and as

\begin{landscape}
\pagestyle{empty}
\addtolength{\topmargin}{50mm} 
\addtolength{\textheight}{75mm}
\addtolength{\oddsidemargin}{0mm} 
\addtolength{\evensidemargin}{-10mm}
\addtolength{\textwidth}{20mm}
\begin{table*}
\caption[]{Absolute abundance estimates, together with their estimated uncertainties. The 
quantities in parenthesis indicate the number of lines observed of that species. 
Abundances are presented on the scale 12+log[X/H].}
\begin{center}
\small
\label{t_abund}
\begin{tabular}{lcccccccccccccc}
\hline
\hline
Star&\multicolumn{2}{c}{C II}&\multicolumn{2}{c}{N II} &\multicolumn{2}{c}{O II}&\multicolumn{2}{c}{Mg II}
&\multicolumn{2}{c}{Si II}&\multicolumn{2}{c}{Si III}&\multicolumn{2}{c}{Si IV} \\
\hline
\\
NGC6611-006 &7.85 $\pm$ 0.24&(3)& \,~7.59 $\pm$ 0.13&(3)&8.52 $\pm$ 0.17&(28)&7.38 $\pm$ 0.22&(1)&		 &   &7.46 $\pm$ 0.26&(3)&7.47 $\pm$ 0.32&(2)\\
NGC6611-012 &7.89 $\pm$ 0.23&(1)& \,~7.48 $\pm$ 0.22&(1)&8.50 $\pm$ 0.12&(23)&7.24 $\pm$ 0.21&(1)&		 &   &7.33 $\pm$ 0.25&(3)&7.30 $\pm$ 0.48&(1)\\
NGC6611-021 &7.82 $\pm$ 0.19&(2)& \,~7.51 $\pm$ 0.11&(3)&8.60 $\pm$ 0.19&(24)&7.24 $\pm$ 0.22&(1)&		 &   &7.40 $\pm$ 0.31&(3)&7.41 $\pm$ 0.52&(1)\\
NGC6611-030 &7.94 $\pm$ 0.17&(2)& \,~7.58 $\pm$ 0.19&(5)&8.44 $\pm$ 0.31&(29)&7.15 $\pm$ 0.22&(1)&7.15 $\pm$ 0.22&(2)&7.17 $\pm$ 0.31&(3)&		 &   \\
NGC6611-033 &8.09 $\pm$ 0.19&(2)& \,~7.82 $\pm$ 0.15&(3)&8.69 $\pm$ 0.20&(29)&7.49 $\pm$ 0.21&(1)&		 &   &7.75 $\pm$ 0.33&(3)&7.75 $\pm$ 0.48&(1)\\
\\
    N11-001 &7.29 $\pm$ 0.16&(2)& \,~8.20 $\pm$ 0.23&(5)&8.23 $\pm$ 0.30&(14)&7.12 $\pm$ 0.26&(1)&7.22 $\pm$ 0.27&(2)&7.20 $\pm$ 0.39&(3)&7.23 $\pm$ 0.72&(1)\\
    N11-002 &7.66 $\pm$ 0.20&(2)& \,~8.14 $\pm$ 0.31&(6)&8.42 $\pm$ 0.44&(15)&7.18 $\pm$ 0.33&(1)&7.47 $\pm$ 0.33&(2)&7.44 $\pm$ 0.47&(3)&		 &   \\
    N11-003 &7.34 $\pm$ 0.23&(1)& \,~7.09 $\pm$ 0.26&(1)&8.34 $\pm$ 0.11&(17)&7.07 $\pm$ 0.24&(1)&		 &   &7.17 $\pm$ 0.22&(3)&7.19 $\pm$ 0.60&(1)\\
    N11-008 &7.45 $\pm$ 0.09&(3)& \,~7.86 $\pm$ 0.20&(3)&8.27 $\pm$ 0.16&(15)&7.12 $\pm$ 0.23&(1)&		 &   &7.22 $\pm$ 0.26&(3)&7.22 $\pm$ 0.57&(1)\\
    N11-009 &7.55 $\pm$ 0.23&(2)& \,~7.74 $\pm$ 0.30&(5)&8.38 $\pm$ 0.40&(19)&6.95 $\pm$ 0.25&(1)&7.18 $\pm$ 0.24&(2)&7.17 $\pm$ 0.41&(3)&		 &   \\
    N11-012 &7.24 $\pm$ 0.26&(1)& \,~7.69 $\pm$ 0.08&(4)&8.39 $\pm$ 0.17&(19)&7.02 $\pm$ 0.31&(1)&		 &   &7.10 $\pm$ 0.28&(3)&7.10 $\pm$ 0.67&(1)\\
    N11-014 &7.59 $\pm$ 0.17&(2)& \,~7.86 $\pm$ 0.17&(5)&8.26 $\pm$ 0.28&(26)&7.15 $\pm$ 0.25&(1)&7.13 $\pm$ 0.28&(2)&7.12 $\pm$ 0.39&(3)&6.73 $\pm$ 0.70&(1)\\
    N11-015 &7.45 $\pm$ 0.30&(1)& \,~7.14 $\pm$ 0.30&(1)&8.36 $\pm$ 0.11&(20)&7.01 $\pm$ 0.30&(1)&		 &   &7.21 $\pm$ 0.25&(3)&7.23 $\pm$ 0.62&(1)\\
    N11-016 &7.52 $\pm$ 0.23&(1)& \,~7.86 $\pm$ 0.09&(4)&8.27 $\pm$ 0.17&(17)&7.25 $\pm$ 0.26&(1)&		 &   &7.05 $\pm$ 0.26&(3)&7.07 $\pm$ 0.55&(1)\\
    N11-017 &7.49 $\pm$ 0.26&(1)& \,~7.86 $\pm$ 0.28&(6)&8.32 $\pm$ 0.37&(23)&6.98 $\pm$ 0.23&(1)&7.12 $\pm$ 0.22&(2)&7.15 $\pm$ 0.39&(3)&		 &   \\
    N11-023 &7.45 $\pm$ 0.21&(1)& \,~7.16 $\pm$ 0.24&(1)&8.39 $\pm$ 0.16&(18)&7.00 $\pm$ 0.22&(1)&		 &   &7.13 $\pm$ 0.24&(3)&7.12 $\pm$ 0.58&(1)\\
    N11-024 &7.48 $\pm$ 0.17&(2)& \,~7.85 $\pm$ 0.10&(4)&8.32 $\pm$ 0.20&(22)&7.14 $\pm$ 0.23&(1)&		 &   &7.15 $\pm$ 0.31&(3)&7.15 $\pm$ 0.58&(1)\\
    N11-029 &7.58 $\pm$ 0.40&(1)& \,~7.11 $\pm$ 0.39&(1)&8.31 $\pm$ 0.32&(13)&6.95 $\pm$ 0.33&(1)&		 &   &7.25 $\pm$ 0.40&(3)&7.26 $\pm$ 0.55&(1)\\
    N11-036 &7.32 $\pm$ 0.13&(2)& \,~7.76 $\pm$ 0.12&(4)&8.33 $\pm$ 0.09&(26)&7.03 $\pm$ 0.20&(1)&		 &   &7.17 $\pm$ 0.24&(3)&7.16 $\pm$ 0.59&(1)\\
    N11-037 &7.57 $\pm$ 0.20&(1)& $<$7.18 $\pm$ 0.24&(1)&8.19 $\pm$ 0.20&( 7)&7.02 $\pm$ 0.13&(1)&		 &   &7.23 $\pm$ 0.31&(3)&7.23 $\pm$ 0.52&(1)\\
    N11-042 &7.56 $\pm$ 0.22&(2)& \,~6.92 $\pm$ 0.26&(1)&8.31 $\pm$ 0.18&(21)&7.00 $\pm$ 0.23&(1)&		 &   &7.14 $\pm$ 0.23&(3)&7.16 $\pm$ 0.52&(1)\\
    N11-047 &7.67 $\pm$ 0.26&(1)& $<$6.88 $\pm$ 0.25&(1)&8.24 $\pm$ 0.16&(17)&7.00 $\pm$ 0.24&(1)&		 &   &7.20 $\pm$ 0.22&(3)&7.19 $\pm$ 0.49&(1)\\
    N11-054 &7.51 $\pm$ 0.15&(2)& \,~6.86 $\pm$ 0.13&(2)&8.42 $\pm$ 0.12&(26)&6.97 $\pm$ 0.20&(1)&		 &   &7.10 $\pm$ 0.23&(3)&7.09 $\pm$ 0.58&(1)\\
    N11-062 &7.43 $\pm$ 0.22&(1)& \,~7.16 $\pm$ 0.17&(2)&8.25 $\pm$ 0.14&(26)&6.99 $\pm$ 0.20&(1)&		 &   &7.16 $\pm$ 0.22&(3)&7.18 $\pm$ 0.47&(1)\\
    N11-069 &7.63 $\pm$ 0.25&(1)& \,~6.95 $\pm$ 0.24&(1)&8.47 $\pm$ 0.15&(26)&7.07 $\pm$ 0.24&(1)&		 &   &7.25 $\pm$ 0.26&(3)&7.23 $\pm$ 0.59&(1)\\
    N11-072 &7.46 $\pm$ 0.14&(3)& \,~7.38 $\pm$ 0.08&(3)&8.36 $\pm$ 0.15&(25)&7.12 $\pm$ 0.20&(1)&		 &   &7.21 $\pm$ 0.24&(3)&7.21 $\pm$ 0.40&(3)\\
    N11-075 &7.55 $\pm$ 0.13&(1)& \,~8.11 $\pm$ 0.26&(5)&8.24 $\pm$ 0.30&(25)&7.24 $\pm$ 0.20&(1)&7.32 $\pm$ 0.23&(2)&7.36 $\pm$ 0.38&(3)&7.33 $\pm$ 0.59&(1)\\
    N11-083 &7.53 $\pm$ 0.17&(2)& \,~6.86 $\pm$ 0.20&(1)&8.33 $\pm$ 0.10&(27)&7.00 $\pm$ 0.19&(1)&		 &   &7.06 $\pm$ 0.22&(3)&7.06 $\pm$ 0.45&(1)\\
    N11-100 &7.45 $\pm$ 0.25&(1)& \,~7.68 $\pm$ 0.19&(2)&8.38 $\pm$ 0.11&(26)&7.15 $\pm$ 0.25&(1)&		 &   &7.44 $\pm$ 0.26&(3)&7.45 $\pm$ 0.47&(1)\\
    N11-101 &7.74 $\pm$ 0.22&(1)& $<$7.09 $\pm$ 0.22&(1)&8.32 $\pm$ 0.12& 21)&7.21 $\pm$ 0.20&(1)&		 &   &7.16 $\pm$ 0.17&(3)&7.17 $\pm$ 0.46&(1)\\
    N11-106 &7.50 $\pm$ 0.28&(1)& \,~7.12 $\pm$ 0.28&(1)&8.31 $\pm$ 0.17&(22)&7.17 $\pm$ 0.24&(1)&		 &   &7.10 $\pm$ 0.23&(3)&7.10 $\pm$ 0.31&(2)\\
    N11-108 &7.66 $\pm$ 0.30&(1)& \,~7.20 $\pm$ 0.32&(1)&8.23 $\pm$ 0.20&(17)&7.12 $\pm$ 0.23&(1)&		 &   &7.08 $\pm$ 0.26&(3)&7.08 $\pm$ 0.40&(1)\\
    N11-109 &7.39 $\pm$ 0.16&(2)& \,~7.22 $\pm$ 0.23&(1)&8.25 $\pm$ 0.12&(17)&6.83 $\pm$ 0.20&(1)&		 &   &7.01 $\pm$ 0.22&(3)&7.02 $\pm$ 0.51&(1)\\
    N11-110 &7.50 $\pm$ 0.20&(1)& \,~7.41 $\pm$ 0.08&(3)&8.54 $\pm$ 0.26&(26)&7.08 $\pm$ 0.20&(1)&		 &   &7.35 $\pm$ 0.36&(3)&7.36 $\pm$ 0.58&(1)\\
    N11-124 &7.56 $\pm$ 0.16&(1)& \,~7.25 $\pm$ 0.17&(1)&8.12 $\pm$ 0.09&(19)&6.97 $\pm$ 0.15&(1)&		 &   &6.94 $\pm$ 0.21&(3)&6.97 $\pm$ 0.43&(1)\\
\\
\hline
\end{tabular}
\normalsize
\end{center}
\end{table*}
\end{landscape}

\begin{landscape}
\pagestyle{empty}
\addtolength{\topmargin}{50mm} 
\addtolength{\textheight}{75mm}
\addtolength{\oddsidemargin}{10mm} 
\addtolength{\evensidemargin}{0mm}
\addtolength{\textwidth}{20mm}
\begin{table*}
\addtocounter{table}{-1}
\caption[]{--continued.}
\begin{center}
\small
\begin{tabular}{lcccccccccccccc}
\hline
\hline
Star&\multicolumn{2}{c}{C II}&\multicolumn{2}{c}{N II} &\multicolumn{2}{c}{O II}&\multicolumn{2}{c}{Mg II}
&\multicolumn{2}{c}{Si II}&\multicolumn{2}{c}{Si III}&\multicolumn{2}{c}{Si IV} \\
\hline
\\
 NGC346-012 &7.10 $\pm$ 0.09&(4)& \,~6.93 $\pm$ 0.13&(2)&8.15 $\pm$ 0.10&(27)&6.71 $\pm$ 0.16&(1)&		 &   &6.89 $\pm$ 0.19&(3)&6.91 $\pm$ 0.55&(2)\\
 NGC346-021 &7.38 $\pm$ 0.12&(4)& \,~6.85 $\pm$ 0.11&(2)&8.24 $\pm$ 0.18&(28)&6.80 $\pm$ 0.17&(1)&		 &   &6.95 $\pm$ 0.26&(3)&6.96 $\pm$ 0.51&(2)\\
 NGC346-029 &7.17 $\pm$ 0.29&(1)& $<$6.99 $\pm$ 0.29&(1)&8.02 $\pm$ 0.24&(18)&6.69 $\pm$ 0.21&(1)&		 &   &6.69 $\pm$ 0.24&(3)&6.70 $\pm$ 0.39&(1)\\
 NGC346-037 &7.04 $\pm$ 0.12&(2)& \,~7.51 $\pm$ 0.28&(4)&7.89 $\pm$ 0.39&( 8)&6.56 $\pm$ 0.17&(1)&6.64 $\pm$ 0.17&(2)&6.68 $\pm$ 0.36&(3)&		 &   \\
 NGC346-039 &7.34 $\pm$ 0.12&(4)& $<$6.61 $\pm$ 0.18&(1)&8.37 $\pm$ 0.15&(28)&6.78 $\pm$ 0.18&(1)&		 &   &7.07 $\pm$ 0.28&(3)&7.07 $\pm$ 0.52&(2)\\
 NGC346-040 &7.11 $\pm$ 0.22&(1)& $<$6.88 $\pm$ 0.22&(1)&7.95 $\pm$ 0.15&(17)&6.39 $\pm$ 0.20&(1)&		 &   &6.56 $\pm$ 0.19&(3)&6.57 $\pm$ 0.32&(3)\\
 NGC346-043 &7.24 $\pm$ 0.32&(1)& $<$6.73 $\pm$ 0.35&(1)&7.97 $\pm$ 0.22&(19)&6.81 $\pm$ 0.26&(1)&		 &   &6.57 $\pm$ 0.24&(3)&6.55 $\pm$ 0.20&(3)\\  
 NGC346-044 &7.30 $\pm$ 0.13&(2)& $<$6.99 $\pm$ 0.15&(2)&8.28 $\pm$ 0.30&(21)&6.73 $\pm$ 0.20&(1)&		 &   &7.15 $\pm$ 0.38&(3)&		 &   \\
 NGC346-056 &6.99 $\pm$ 0.25&(1)& \,~7.40 $\pm$ 0.18&(3)&8.00 $\pm$ 0.26&(17)&6.81 $\pm$ 0.20&(1)&		 &   &6.77 $\pm$ 0.25&(3)&6.74 $\pm$ 0.31&(3)\\
 NGC346-062 &7.18 $\pm$ 0.19&(1)& \,~7.27 $\pm$ 0.12&(3)&7.87 $\pm$ 0.08&(21)&6.72 $\pm$ 0.18&(1)&		 &   &6.61 $\pm$ 0.15&(3)&6.63 $\pm$ 0.32&(3)\\
 NGC346-075 &7.48 $\pm$ 0.13&(4)& $<$6.43 $\pm$ 0.15&(1)&8.08 $\pm$ 0.12&(27)&6.91 $\pm$ 0.15&(1)&		 &   &6.94 $\pm$ 0.22&(3)&6.94 $\pm$ 0.43&(1)\\
 NGC346-094 &7.34 $\pm$ 0.14&(2)& \,~7.37 $\pm$ 0.20&(1)&8.17 $\pm$ 0.12&(18)&6.77 $\pm$ 0.17&(1)&		 &   &6.92 $\pm$ 0.21&(3)&6.92 $\pm$ 0.45&(1)\\
 NGC346-103 &7.02 $\pm$ 0.17&(1)& \,~7.60 $\pm$ 0.13&(5)&7.99 $\pm$ 0.09&(21)&6.83 $\pm$ 0.17&(1)&		 &   &6.87 $\pm$ 0.21&(3)&6.87 $\pm$ 0.36&(3)\\
 NGC346-116 &7.27 $\pm$ 0.10&(3)& \,~6.93 $\pm$ 0.18&(1)&8.13 $\pm$ 0.09&(24)&6.70 $\pm$ 0.17&(1)&		 &   &6.81 $\pm$ 0.20&(3)&6.81 $\pm$ 0.44&(2)\\
\\
\hline
\end{tabular}
\normalsize
\end{center}
\end{table*}
\end{landscape}

\begin{landscape}
\pagestyle{empty}
\addtolength{\topmargin}{80mm} 
\addtolength{\textheight}{105mm}
\begin{figure*}
\centering
\epsfig{file=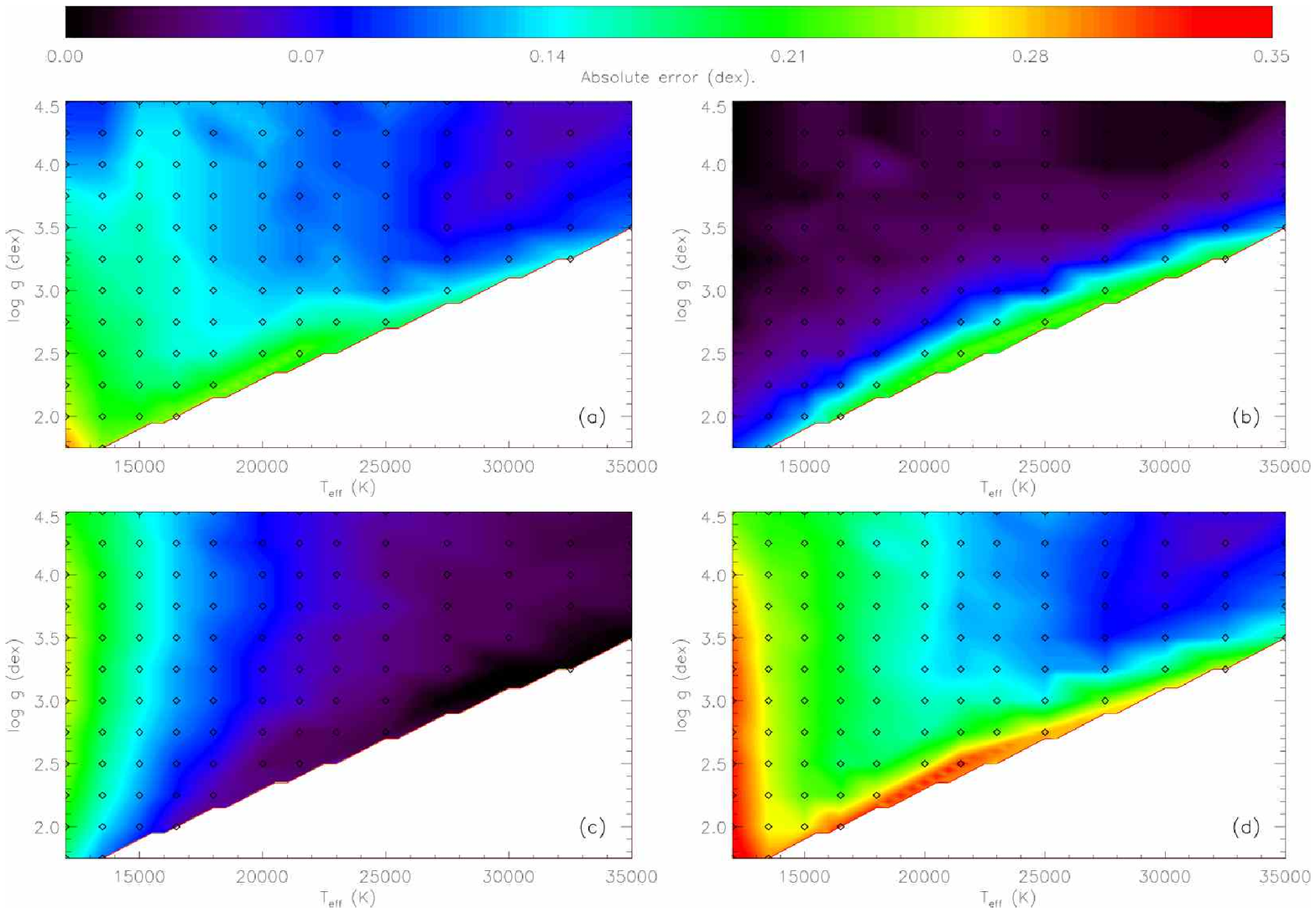, height=160mm, angle=-0}
\caption[]{Contour maps displaying the effect of the errors in atmospheric
parameters upon the derived abundances of the \ion{Mg}{ii} 4481\AA\,line 
at all points on the {\sc tlusty} grid at a LMC metallicity calculated at a
microturbulence 
of 5\,km\,s$^{-1}$ for (a) an error in $T_{\rm eff}$ of 1\,000\,K, (b) an error in
$\log g$ of 0.2\,dex and (c) an error in $\xi$ of 3\,km\,s$^{-1}$. Panel (d) shows the
combined error calculated as the errors from (a), (b) and (c) summed in
quadrature.}
\label{f_Mgerrors}
\end{figure*}
\end{landscape}

\begin{landscape}
\pagestyle{empty}
\addtolength{\topmargin}{80mm} 
\addtolength{\textheight}{105mm}
\begin{figure*}
\centering
\epsfig{file=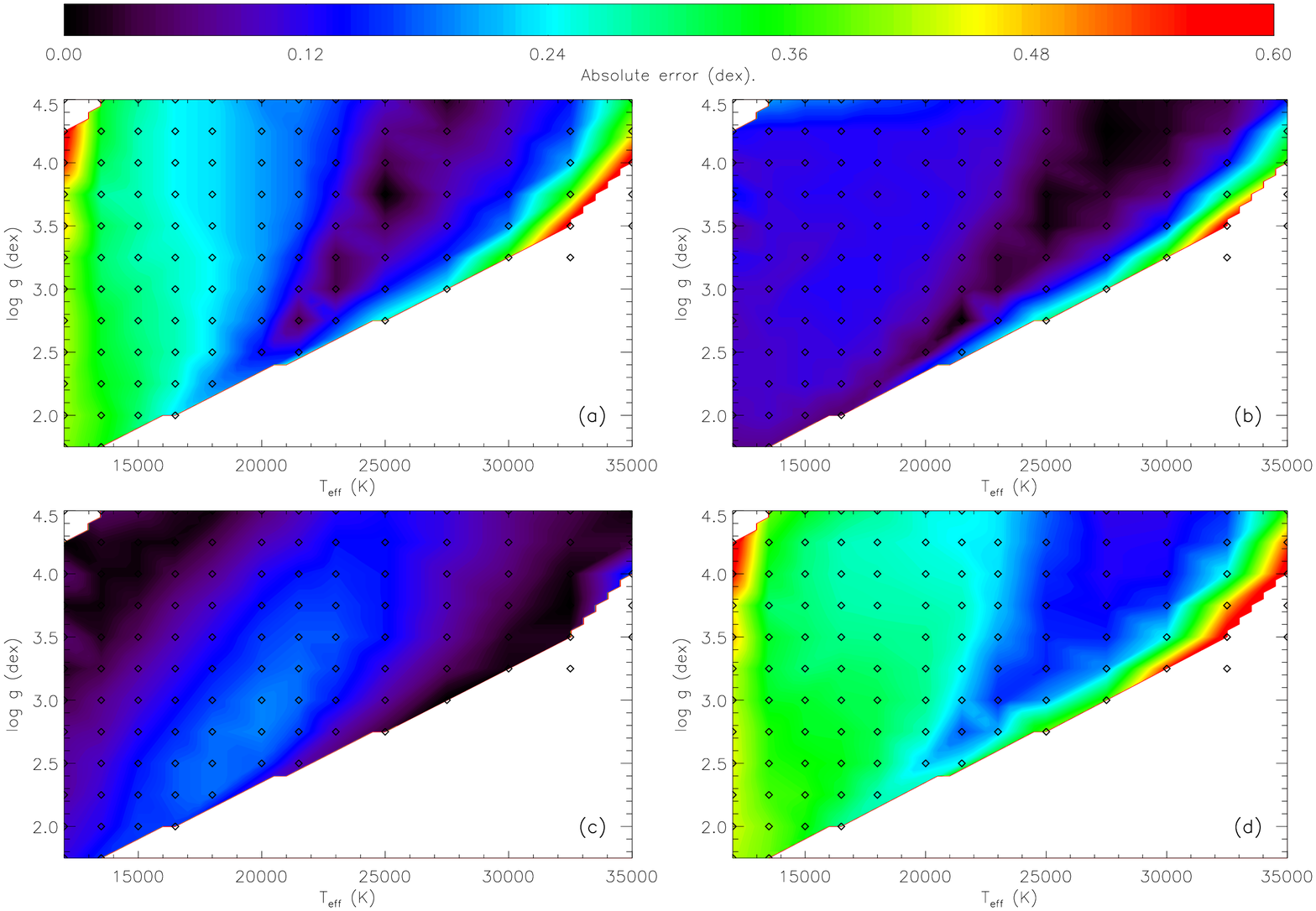, height=160mm, angle=-0}
\caption[]{Contour maps displaying the effect of the errors in atmospheric
parameters upon the derived abundances of the \ion{N}{ii} 3995\AA\,line 
at all points on the {\sc tlusty} grid at a LMC metallicity calculated at a
microturbulence 
of 5\,km\,s$^{-1}$ for (a) an error in $T_{\rm eff}$ of 1\,000\,K, (b) an error in
$\log g$ of 0.2\,dex and (c) an error in $\xi$ of 3\,km\,s$^{-1}$. Panel (d) shows the
combined error calculated as the errors from (a), (b) and (c) summed in
quadrature. Grid points for which the EW of the model line is less than 1\,m\AA\,
have not been included in the contour mapping.}
\label{f_Nerrors}
\end{figure*}
\setcounter{figure}{31}
\end{landscape}

\begin{table}
\caption{Dependence of atmospheric parameters and abundances upon the assumed
metallicity of a cluster for N\,11-062.}
\label{t_metallicity}
\centering
\begin{tabular}{lccc}\hline \hline
               & \multicolumn{3}{c}{Metallicity Regime} \\
               & Galactic & LMC & SMC\\
\hline
\\
$T_{\rm eff}$  & 30100 & 30400 & 31000 \\
$\log g$       &  4.05 &  4.05 &  4.10 \\
$\xi_{\rm Si}$ &     5 &     5 &     5 \\
\ion{C}{ii}    &  7.37 &  7.43 &  7.51 \\
\ion{N}{ii}    &  7.15 &  7.16 &  7.22 \\
\ion{O}{ii}    &  8.25 &  8.25 &  8.27 \\
\ion{Mg}{ii}   &  6.97 &  6.99 &  7.00 \\
\ion{Si}{iii}  &  7.16 &  7.16 &  7.21 \\
\ion{Si}{iv}   &  7.15 &  7.18 &  7.20 \\
\\
\hline
\end{tabular}
\end{table}

\noindent  such we may have 
overestimated the actual error in the abundances from the adopted atmospheric parameters.

To illustrate how the errors given in Table~\ref{t_abund} depend on atmospheric
parameters at all points on our {\sc tlusty} grid, in Fig.~\ref{f_Mgerrors} and
Fig~\ref{f_Nerrors} we plot the 
contribution of the uncertainty in each of the atmospheric 
parameters to the uncertainty in the derived photospheric abundances for two
typical lines, \ion{Mg}{ii} 4481\AA\,and \ion{N}{ii} 3995\AA. These maps
were
calculated at LMC metallicity and at a microturbulence of 5\,km\,s$^{-1}$. 
Of course 
it should be noted that the systematic errors will not affect each line 
of a particular species to the same extent, with for example, the predictions for a strong
line being more dependent on the microturbulence 
estimate than for
a weaker line. Figures.~7-31, only included online,
show similar error contour maps for the following lines - \ion{C}{ii}\,4267\AA, 
\ion{C}{iii}\,4647\AA, \ion{N}{ii}\,3995\AA, \ion{O}{ii}\,4075\AA, \ion{O}{ii}\,4132\AA, 
\ion{Si}{ii}\,4128\AA, \ion{Si}{iii}\,4567\AA\ and \ion{Si}{iv}\,4116\AA - at Galactic, 
LMC and SMC metallicities.

In order to test the dependence of the atmospheric parameters and photospheric
abundances upon the metallicity (iron content) adopted for the model atmosphere, we
have analysed a single LMC target using both the Galactic and SMC grids and the resulting 
atmospheric parameters and abundances are given in Table~\ref{t_metallicity}. It 
can be seen that uncertainties in the atmospheric parameters arising from the metallicity 
adopted are well within those discussed in the above sections. Indeed, changing from one
metallicity grid to another does not effect our derived 
microturbulence. The uncertainty
in the abundances from the choice of metallicity grid is less than 0.1\,dex for all 
species and less than 0.05\,dex for \ion{O}{ii}, \ion{Mg}{ii}, \ion{Si}{iii} and \ion{Si}{iv}.
This approach to estimating the error associated with the assumption of the base metallicity 
may be  conservative in that, although the Fe abundance in any particular cluster may vary, it
is unlikely to be as large as the +0.3\ or -0.5\,dex assumed above.

The derived Si abundance in
each star is sensitive to the adopted atmospheric parameters, especially
effective temperature and microturbulence. This is consistent with their use in 
estimating these parameters and is highlighted by the larger
errors for the different species of silicon in Table~\ref{t_abund}. Additionally 
the spread of silicon abundances for stars in the same
cluster is relatively large compared to that of the other species with the exception of 
nitrogen. The Si abundance
should not be affected by
chemical processing in the stellar interior (unlike nitrogen) and hence
we would not expect to see this variation. Most of
the error quoted in Table~\ref{t_abund} arises from the uncertainty in microturbulence due to the \ion{Si}{iii}
lines being amongst the strongest observed in the spectra. To
test if the spread in Si abundance estimates represents a real variation between stars in the same cluster or is an
effect of microturbulence we have performed the following procedure. The average Si abundance of
each cluster was calculated and then, holding $T_{\rm eff}$ and $\log g$ constant, the microturbulence was
adjusted for every star so that the derived \ion{Si}{iii} abundance from that star was the same as the cluster
average. In Table~\ref{t_vtadjust} the adjusted microturbulence $\xi_{\rm av}$ and corresponding 
abundances for each species are listed. For comparison purposes the originally adopted miroturbulence $\xi_{\rm Si}$
is also given. For the majority of our sample the difference between $\xi_{\rm av}$ and 
$\xi_{\rm Si}$ is within our estimated error and the abundances given in Table~\ref{t_vtadjust} are generally
within the uncertainties of those given in Table~\ref{t_abund}.

It is important to note that this procedure has not been possible for all stars. Two LMC stars,
N\,11-083 and N\,11-124, have Si abundances lower than the cluster 
average. However they also have
microturbulence 
values ($\xi_{\rm Si}$) of 0\,km\,s$^{-1}$ and hence it is not possible 
to increase their stellar Si abundances. This problem also occurs in two SMC stars,
NGC\,346-029 and NGC\,346-040. The
abundances listed in Table~\ref{t_vtadjust} are therefore the same as those given in
Table~\ref{t_abund} for these four stars. NGC\,346-043 has a microturbulence of 4\,km\,s$^{-1}$ 
and even reducing 
the microturbulence to 0\,km\,s$^{-1}$ does not increase the \ion{Si}{iii} abundance to the 
cluster average and the estimates
given in Table~\ref{t_vtadjust} are for a microturbulence of 0\,km\,s$^{-1}$.
Nevertheless, this procedure has been possible for the vast majority of our sample, and
most of the cases where it fails are the problematic cases where a real value for the microturbulence cannot 
be determined from the
Si lines. Although we no longer achieve ionization balance for several of the stars in our 
sample the different
Si abundance estimates are consistent with an error of 1000\,K in $T_{\rm eff}$. For 
NGC\,6611, where we previously had the largest spread of Si abundances for stars in the same
cluster, using the above procedure has eliminated both this spread and also noticeably 
improved the
agreement of the other abundances. In the following sections we adopt the abundances listed in
Table~\ref{t_vtadjust} but we note that our principle conclusions would be effectively 
unchanged if
we had adopted the abundance estimates in Table~\ref{t_abund}.

\begin{landscape}
\newpage
\pagestyle{empty}
\addtolength{\topmargin}{50mm} 
\addtolength{\textheight}{75mm}
\addtolength{\oddsidemargin}{10mm} 
\addtolength{\evensidemargin}{0mm}
\addtolength{\textwidth}{20mm}
\begin{table*}
\caption[]{Absolute abundance estimates derived assuming the $T_{\rm eff}$ and 
$\log g$ given in Table~\ref{t_stars+atmos} 
with the microturbulence adjusted in order to derive the same Si abundance for each 
star in each cluster where possible. The 
uncertainties are estimated by the same method as those in Table~\ref{t_abund}. The corrected carbon 
abundances (see Sect.~\ref{s_carbon}) are given in brackets.}
\begin{center}
\small
\label{t_vtadjust}
\begin{tabular}{lccccccccc}
\hline
\hline
Star       &$\xi_{\rm av}$&$\xi_{\rm Si}$&C II&             N II&            O II           &Mg II           &Si II         &Si III         &Si IV \\
\hline
\\
NGC6611-006&	  8	  &    7	 &7.85 $\pm$ 0.24 (7.96)&\,~7.58 $\pm$ 0.12&8.50 $\pm$ 0.16&7.36 $\pm$ 0.22&               &7.41 $\pm$ 0.24&7.39 $\pm$ 0.30\\
NGC6611-012&	  5	  &    6	 &7.92 $\pm$ 0.25 (8.26)&\,~7.50 $\pm$ 0.24&8.55 $\pm$ 0.14&7.26 $\pm$ 0.23&               &7.45 $\pm$ 0.30&7.41 $\pm$ 0.51\\
NGC6611-021&	  0	  &    0	 &7.82 $\pm$ 0.19 (7.99)&\,~7.51 $\pm$ 0.11&8.60 $\pm$ 0.19&7.24 $\pm$ 0.22&               &7.40 $\pm$ 0.31&7.41 $\pm$ 0.52\\
NGC6611-030&	  1	  &    5	 &8.09 $\pm$ 0.18 (8.26)&\,~7.69 $\pm$ 0.21&8.57 $\pm$ 0.32&7.32 $\pm$ 0.25&7.34 $\pm$ 0.24&7.46 $\pm$ 0.32&               \\
NGC6611-033&	  4	  &    1	 &7.99 $\pm$ 0.16 (8.16)&\,~7.70 $\pm$ 0.12&8.54 $\pm$ 0.18&7.38 $\pm$ 0.19&               &7.43 $\pm$ 0.30&7.53 $\pm$ 0.50\\
\\
    N11-001&	 14	  &   14	 &7.29 $\pm$ 0.16 (7.46)&\,~8.20 $\pm$ 0.23&8.23 $\pm$ 0.30&7.12 $\pm$ 0.26&7.22 $\pm$ 0.27&7.20 $\pm$ 0.39&7.23 $\pm$ 0.72\\
    N11-002&	 18	  &   12	 &7.55 $\pm$ 0.16 (7.72)&\,~8.00 $\pm$ 0.28&8.32 $\pm$ 0.40&7.07 $\pm$ 0.29&7.32 $\pm$ 0.28&7.18 $\pm$ 0.44&               \\
    N11-003&	 13	  &   13	 &7.34 $\pm$ 0.23 (7.68)&\,~7.09 $\pm$ 0.26&8.34 $\pm$ 0.11&7.07 $\pm$ 0.24&               &7.17 $\pm$ 0.22&7.19 $\pm$ 0.60\\
    N11-008&	 16	  &   15	 &7.45 $\pm$ 0.09 (7.56)&\,~7.84 $\pm$ 0.20&8.25 $\pm$ 0.16&7.12 $\pm$ 0.24&               &7.19 $\pm$ 0.25&7.15 $\pm$ 0.55\\
    N11-009&	 17	  &   17	 &7.55 $\pm$ 0.23 (7.72)&\,~7.74 $\pm$ 0.30&8.38 $\pm$ 0.40&6.95 $\pm$ 0.25&7.18 $\pm$ 0.24&7.17 $\pm$ 0.41&               \\
    N11-012&	 13	  &   14	 &7.24 $\pm$ 0.26 (7.58)&\,~7.71 $\pm$ 0.08&8.42 $\pm$ 0.18&7.02 $\pm$ 0.31&               &7.15 $\pm$ 0.30&7.14 $\pm$ 0.67\\
    N11-014&	 12	  &   13	 &7.60 $\pm$ 0.17 (7.77)&\,~7.89 $\pm$ 0.19&8.27 $\pm$ 0.29&7.16 $\pm$ 0.25&7.14 $\pm$ 0.27&7.17 $\pm$ 0.40&6.76 $\pm$ 0.71\\
    N11-015&	 11	  &   11	 &7.45 $\pm$ 0.30 (7.79)&\,~7.14 $\pm$ 0.30&8.36 $\pm$ 0.11&7.01 $\pm$ 0.30&               &7.21 $\pm$ 0.25&7.23 $\pm$ 0.62\\
    N11-016&	 12	  &   14	 &7.55 $\pm$ 0.25 (7.89)&\,~7.90 $\pm$ 0.10&8.31 $\pm$ 0.18&7.27 $\pm$ 0.26&               &7.17 $\pm$ 0.30&7.15 $\pm$ 0.58\\
    N11-017&	 16	  &   17	 &7.51 $\pm$ 0.26 (7.85)&\,~7.89 $\pm$ 0.28&8.33 $\pm$ 0.37&7.00 $\pm$ 0.25&7.14 $\pm$ 0.22&7.19 $\pm$ 0.39&               \\
    N11-023&	 13	  &   14	 &7.46 $\pm$ 0.21 (7.80)&\,~7.16 $\pm$ 0.24&8.41 $\pm$ 0.17&7.00 $\pm$ 0.22&               &7.17 $\pm$ 0.25&7.18 $\pm$ 0.61\\
    N11-024&	 12	  &   12	 &7.48 $\pm$ 0.17 (7.65)&\,~7.85 $\pm$ 0.10&8.32 $\pm$ 0.20&7.14 $\pm$ 0.23&               &7.15 $\pm$ 0.31&7.15 $\pm$ 0.58\\
    N11-029&	 15	  &   11	 &7.57 $\pm$ 0.39 (7.91)&\,~7.10 $\pm$ 0.38&8.28 $\pm$ 0.31&6.93 $\pm$ 0.32&               &7.17 $\pm$ 0.36&6.93 $\pm$ 0.43\\
    N11-036&	 11	  &   11	 &7.32 $\pm$ 0.13 (7.49)&\,~7.76 $\pm$ 0.12&8.33 $\pm$ 0.09&7.03 $\pm$ 0.20&               &7.17 $\pm$ 0.24&7.16 $\pm$ 0.59\\
    N11-037&	 12	  &   10	 &7.56 $\pm$ 0.20 (7.90)&$<$7.17 $\pm$ 0.23&8.18 $\pm$ 0.20&7.01 $\pm$ 0.13&               &7.18 $\pm$ 0.30&7.03 $\pm$ 0.44\\
    N11-042&	  5	  &    6	 &7.56 $\pm$ 0.21 (7.73)&\,~6.92 $\pm$ 0.26&8.34 $\pm$ 0.18&7.00 $\pm$ 0.23&               &7.21 $\pm$ 0.24&7.32 $\pm$ 0.55\\
    N11-047&	  8	  &    8	 &7.67 $\pm$ 0.26 (8.01)&$<$6.88 $\pm$ 0.25&8.24 $\pm$ 0.16&7.00 $\pm$ 0.24&               &7.20 $\pm$ 0.22&7.19 $\pm$ 0.49\\
    N11-054&	 10	  &   11	 &7.52 $\pm$ 0.16 (7.69)&\,~6.86 $\pm$ 0.13&8.45 $\pm$ 0.13&6.98 $\pm$ 0.21&               &7.16 $\pm$ 0.26&7.14 $\pm$ 0.60\\
    N11-062&	  5	  &    5	 &7.43 $\pm$ 0.22 (7.77)&\,~7.16 $\pm$ 0.17&8.25 $\pm$ 0.14&6.99 $\pm$ 0.20&               &7.16 $\pm$ 0.22&7.18 $\pm$ 0.47\\
    N11-069&	 11	  &   10	 &7.62 $\pm$ 0.25 (7.96)&\,~6.94 $\pm$ 0.22&8.44 $\pm$ 0.14&7.06 $\pm$ 0.24&               &7.18 $\pm$ 0.24&7.18 $\pm$ 0.56\\
    N11-072&	  5	  &    5	 &7.46 $\pm$ 0.14 (7.57)&\,~7.38 $\pm$ 0.08&8.36 $\pm$ 0.15&7.12 $\pm$ 0.20&               &7.21 $\pm$ 0.24&7.21 $\pm$ 0.40\\
    N11-075&	  5	  &    3	 &7.52 $\pm$ 0.13 (7.86)&\,~8.00 $\pm$ 0.25&8.18 $\pm$ 0.29&7.16 $\pm$ 0.17&7.23 $\pm$ 0.21&7.17 $\pm$ 0.37&7.23 $\pm$ 0.59\\
    N11-083&	  -	  &    0	 &7.53 $\pm$ 0.17 (7.70)&\,~6.86 $\pm$ 0.20&8.33 $\pm$ 0.10&7.00 $\pm$ 0.19&               &7.06 $\pm$ 0.22&7.06 $\pm$ 0.45\\
    N11-100&	  4	  &    1	 &7.44 $\pm$ 0.23 (7.78)&\,~7.62 $\pm$ 0.17&8.30 $\pm$ 0.10&7.10 $\pm$ 0.22&               &7.20 $\pm$ 0.23&7.19 $\pm$ 0.48\\
    N11-101&	  8	  &    8	 &7.74 $\pm$ 0.22 (8.08)&$<$7.09 $\pm$ 0.22&8.32 $\pm$ 0.12&7.21 $\pm$ 0.20&               &7.16 $\pm$ 0.17&7.17 $\pm$ 0.46\\
    N11-106&	  5	  &    7	 &7.50 $\pm$ 0.28 (7.84)&\,~7.13 $\pm$ 0.28&8.35 $\pm$ 0.17&7.19 $\pm$ 0.24&               &7.19 $\pm$ 0.24&7.25 $\pm$ 0.34\\
    N11-108&	  4	  &    7	 &7.67 $\pm$ 0.30 (8.01)&\,~7.21 $\pm$ 0.32&8.27 $\pm$ 0.20&7.14 $\pm$ 0.23&               &7.17 $\pm$ 0.25&7.43 $\pm$ 0.46\\
    N11-109&	 10	  &   14	 &7.41 $\pm$ 0.16 (7.58)&\,~7.24 $\pm$ 0.23&8.32 $\pm$ 0.14&6.84 $\pm$ 0.20&               &7.18 $\pm$ 0.24&7.32 $\pm$ 0.58\\
    N11-110&	  7	  &    6	 &7.49 $\pm$ 0.18 (7.83)&\,~7.39 $\pm$ 0.07&8.48 $\pm$ 0.25&7.06 $\pm$ 0.18&               &7.23 $\pm$ 0.32&7.28 $\pm$ 0.57\\
    N11-124&	  -	  &    0	 &7.56 $\pm$ 0.16 (7.90)&\,~7.25 $\pm$ 0.17&8.12 $\pm$ 0.09&6.97 $\pm$ 0.15&               &6.94 $\pm$ 0.21&6.97 $\pm$ 0.43\\
\\
\hline
\end{tabular}
\normalsize
\end{center}
\end{table*}
\end{landscape}

\begin{landscape}
\newpage
\pagestyle{empty}
\addtolength{\topmargin}{50mm} 
\addtolength{\textheight}{75mm}
\addtolength{\oddsidemargin}{0mm} 
\addtolength{\evensidemargin}{-10mm}
\addtolength{\textwidth}{20mm}
\begin{table*}
\addtocounter{table}{-1}
\caption[]{--continued}
\begin{center}
\small
\begin{tabular}{lccccccccc}
\hline
\hline
Star       &$\xi_{\rm av}$&$\xi_{\rm Si}$&C II           &N II              &O II           &Mg II          &Si II          &Si III          &Si IV          \\
\hline
\\
 NGC346-012&	  9	  &    8	 &7.10 $\pm$ 0.08 (7.18)&\,~6.93 $\pm$ 0.13&8.13 $\pm$ 0.08&6.70 $\pm$ 0.15&	            &6.82 $\pm$ 0.16&6.85 $\pm$ 0.54\\
 NGC346-021&	  3	  &    1	 &7.36 $\pm$ 0.12 (7.45)&\,~6.84 $\pm$ 0.11&8.16 $\pm$ 0.16&6.76 $\pm$ 0.15&	            &6.78 $\pm$ 0.22&6.83 $\pm$ 0.48\\
 NGC346-029&	  -	  &    0	 &7.17 $\pm$ 0.29 (7.51)&$<$6.99 $\pm$ 0.29&8.02 $\pm$ 0.24&6.69 $\pm$ 0.21&	            &6.69 $\pm$ 0.24&6.70 $\pm$ 0.39\\
 NGC346-037&	  3	  &    5	 &7.06 $\pm$ 0.12 (7.23)&\,~7.55 $\pm$ 0.29&7.94 $\pm$ 0.39&6.62 $\pm$ 0.19&6.72 $\pm$ 0.18&6.79 $\pm$ 0.37&               \\
 NGC346-039&	  3	  &    0	 &7.32 $\pm$ 0.11 (7.40)&$<$6.59 $\pm$ 0.15&8.24 $\pm$ 0.12&6.73 $\pm$ 0.15&	            &6.79 $\pm$ 0.22&6.84 $\pm$ 0.50\\
 NGC346-040&	  -	  &    0	 &7.11 $\pm$ 0.22 (7.45)&$<$6.88 $\pm$ 0.22&7.95 $\pm$ 0.15&6.39 $\pm$ 0.20&	            &6.56 $\pm$ 0.19&6.57 $\pm$ 0.32\\
 NGC346-043&	  -	  &    4	 &7.21 $\pm$ 0.33 (7.55)&$<$6.75 $\pm$ 0.34&8.00 $\pm$ 0.22&6.83 $\pm$ 0.27&	            &6.69 $\pm$ 0.23&6.78 $\pm$ 0.26\\  
 NGC346-044&	  4	  &    0	 &7.27 $\pm$ 0.10 (7.44)&$<$6.94 $\pm$ 0.12&8.13 $\pm$ 0.27&6.66 $\pm$ 0.16&               &6.76 $\pm$ 0.30&               \\
 NGC346-056&	  1	  &    1	 &6.99 $\pm$ 0.25 (7.33)&\,~7.40 $\pm$ 0.18&8.00 $\pm$ 0.26&6.81 $\pm$ 0.20&	            &6.77 $\pm$ 0.25&6.74 $\pm$ 0.31\\
 NGC346-062&	  6	  &   12	 &7.15 $\pm$ 0.18 (7.49)&\,~7.28 $\pm$ 0.11&7.91 $\pm$ 0.08&6.75 $\pm$ 0.17&	            &6.82 $\pm$ 0.18&6.93 $\pm$ 0.36\\
 NGC346-075&	  2	  &    0	 &7.47 $\pm$ 0.12 (7.55)&$<$6.42 $\pm$ 0.15&8.03 $\pm$ 0.10&6.88 $\pm$ 0.15&	            &6.79 $\pm$ 0.19&6.83 $\pm$ 0.41\\
 NGC346-094&	  6	  &    4	 &7.33 $\pm$ 0.13 (7.50)&\,~7.34 $\pm$ 0.18&8.11 $\pm$ 0.10&6.75 $\pm$ 0.16&	            &6.80 $\pm$ 0.19&6.78 $\pm$ 0.43\\
 NGC346-103&	  1	  &    0	 &7.02 $\pm$ 0.16 (7.36)&\,~7.58 $\pm$ 0.13&7.96 $\pm$ 0.08&6.82 $\pm$ 0.16&	            &6.80 $\pm$ 0.19&6.81 $\pm$ 0.35\\
 NGC346-116&	  0	  &    0	 &7.27 $\pm$ 0.10 (7.38)&\,~6.93 $\pm$ 0.18&8.13 $\pm$ 0.09&6.70 $\pm$ 0.17&	            &6.81 $\pm$ 0.20&6.81 $\pm$ 0.44\\
\\
\hline
\end{tabular}
\normalsize
\end{center}
\end{table*}
\end{landscape}

\begin{table*}[htbp]
\caption{Average abundances for NGC\,6611, N\,11 and NGC\,346 from 
this analysis and CNO abundances from \ion{H}{ii} regions in these clusters. The
mean corrected carbon abundances are given in brackets.
\ion{H}{ii} region abundances are taken from Rodr\'{i}guez (\cite{rod99}), 
Shaver et al. (\cite{sha83})
and Pe\~{n}a et al.(\cite{pen00}) for NGC\,6611, Tsamis et al. (\cite{tsa03}) and 
Kurt \& Dufour (\cite{kur98}) for N\,11 and
Kurt \& Dufour (\cite{kur98}), Peimbert et al. (\cite{pei00}) and Shaver et al. (\cite{sha83}) 
for NGC\,346. Errors in \ion{H}{ii} 
region abundances are typically estimated to be 0.1-0.2\,dex. Reyes (\cite{rey99}) estimate the SMC carbon abundace
from several \ion{H}{ii} regions to be 7.39\,dex, see Sect.~\ref{s_HII_regions}. Solar abundances are given for 
comparison and are taken from Asplund et al. (\cite{asp05}). }
\label{t_averages}
\centering
\begin{tabular}{lccccccc}\hline \hline
   & \multicolumn{2}{c}{NGC\,6611} & \multicolumn{2}{c}{N\,11} & \multicolumn{2}{c}{NGC\,346} & Solar\\
   & This&\ion{H}{ii}& This&\ion{H}{ii}& This&\ion{H}{ii} & Abundances\\
   &Paper&Region    &Paper&Region    &Paper&Region    &\\
\hline
\\
C  & 7.95 $\pm$ 0.11 (8.13)&8.23 & 7.48 $\pm$ 0.14 (7.73)&7.81& 7.23 $\pm$ 0.15 (7.37)&7.17&8.39 $\pm$ 0.05\\
N  & 7.59 $\pm$ 0.10       &7.64 & 7.54 $\pm$ 0.40	 &6.88& 7.17 $\pm$ 0.29       &6.50&7.78 $\pm$ 0.06\\
O  & 8.55 $\pm$ 0.04       &8.56 & 8.33 $\pm$ 0.08	 &8.41& 8.06 $\pm$ 0.10       &8.11&8.66 $\pm$ 0.05\\
Mg & 7.32 $\pm$ 0.06       &     & 7.06 $\pm$ 0.09	 &    & 6.74 $\pm$ 0.07       &    &7.53 $\pm$ 0.09\\
Si & 7.41 $\pm$ 0.05       &     & 7.19 $\pm$ 0.07	 &    & 6.79 $\pm$ 0.05       &    &7.51 $\pm$ 0.04\\
\\
\hline
\end{tabular}
\end{table*}

\subsubsection{Correlations of abundances with atmospheric parameters} \label{s_correlations}

We have searched for any dependence in our abundances with the stellar atmospheric abundances 
for the Magellanic Cloud clusters using the values given in both Tables~\ref{t_abund} and \ref{t_vtadjust}.
For carbon, oxygen, magnesium and silicon
there is no evidence of any significant correlation. For example, in N\,11 a linear least squares fit
suggests that 
the oxygen abundance decreases by less than 0.1\,dex over a range in gravity
from 2.0\,dex to 4.5\,dex, while the 2$\sigma$ errors
are greater than the gradients in all cases. Additionally, if we adopt the 
abundances given in Table~\ref{t_vtadjust} and do not
include the stars where we where unable to 
determine $\xi_{\rm av}$, the gradients of the best-fitting lines in the majority of cases are less than their
1$\sigma$ errors. We have not investigated
any dependence of the nitrogen abundance with the atmospheric parameters due to the scatter in the
nitrogen abundances and also because we expect to see a correlation with surface gravity due to
evolutionary effects (see Sect.~\ref{s_evolve_effects}).

\section{Chemical composition of the three clusters}       \label{s_chem} 

In Table~\ref{t_averages}, the average C, N, O, Mg and Si abundances
for each cluster are presented. These averages have
been calculated from the stellar abundances listed in Table~\ref{t_vtadjust}
and are weighted by the quoted uncertainties. The Si abundances are the weighted 
average of the \ion{Si}{ii}, \ion{Si}{iii} and \ion{Si}{iv} abundances. The 
five stars without a $\xi_{\rm av}$ value in Table~\ref{t_vtadjust}
have not been included in the estimate of these averages (see Sect.~\ref{s_errors}) nor are upper
limits to the nitrogen abundance included in the nitrogen average. The
quoted errors are the 1$\sigma$ standard deviation in abundances derived from each star 
analysed in the cluster.

\subsection{Helium} \label{s_helium}

\begin{figure}
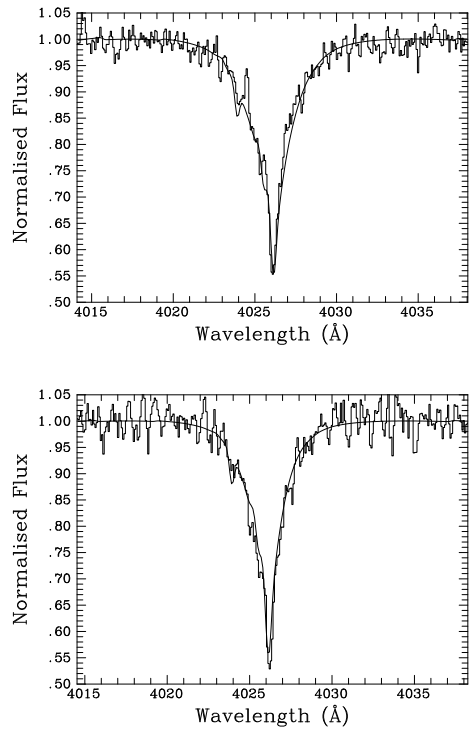

\centering
\begin{tabular}{c}
\epsfig{file=6148f32a.eps, height=70mm, angle=-90}\\
\epsfig{file=6148f32b.eps, height=70mm, angle=-90}\\
\end{tabular}
\caption[]{Fits of model spectra with a He abundance of 11.0\,dex (smooth curve) to the observed \ion{He}{i} 4026\AA\ line in 
NGC\,346-075 and NGC\,346-103, upper and lower plots respectively. These objects have nitrogen abundances which differ by over 1.0\,dex.} 
\label{f_He_abund_fits}
\end{figure}

In our analysis we have not explicitly derived helium abundances
but instead have assumed a nominal value (11.0\,dex) throughout. To test the validity 
of this assumption we have fitted theoretical models at the appropriate 
atmospheric parameters to the observed \ion{He}{i} line at 4026\AA. 
It should be noted
that due to the strength of this line the theoretical profiles are
dependent on the adopted microturbulence. 
Other weaker He lines are available, such as the 4169\AA\ and the 
4437\AA\ lines but these lines are not well observed in all of our spectra.
Within the uncertainties in our atmospheric parameters we find 
excellent agreement between theory and observation for
the majority of stars in our sample. In Fig.~\ref{f_He_abund_fits} we show
the quality of the fit for two SMC objects, NGC\,346-075 and NGC\,346-103, which have
the lowest and highest estimated SMC nitrogen abundance estimates respectively.

The  
few cases where discrepancies are found are all supergiant stars having relatively 
large values of microturbulence.
In these cases the discrepancy can 
be attributed to the senstivity of the 
theoretical profiles to this parameter and indeed
the weaker He lines (where available) are in excellent agreement with theory. 
As such, within the uncertainties in the atmospheric parameters, 
we believe that we are justified in assuming normal helium
abundances throughout our analysis but small variations cannot be ruled out.

\subsection{Carbon}                                            \label{s_carbon} 

In Tables~\ref{t_abund} and \ref{t_vtadjust}, the carbon abundances have been determined solely from \ion{C}{ii}  
lines, despite the spectra of our hotter targets containing measurable \ion{C}{iii} lines.  
The latter have not been included in our abundance analysis because our {\sc tlusty}  
model grid used a relatively simple \ion{C}{iii}  
ion. Our choice of model ions was discussed in 
Dufton et al. (\cite{duf05}) and the models were obtained from 
Lanz \& Hubeny (\cite{lan03})
and Allende-Prieto et al. (\cite{all03}). Although
a grid including a more  
sophisticated \ion{C}{iii} ion could have been generated, this was not
considered worthwhile given the relatively few stars containing \ion{C}{iii} lines.

In previous studies (Sigut \cite{sig96} and Lennon et al. \cite{len03}) it has been found that 
estimates of the carbon abundance from the 
\ion{C}{ii} line at 4267\AA\ gives lower abundances than estimates from other \ion{C}{ii} lines.
Recently, Nieva \& Przybilla (\cite{nie06}) have constructed a comprehensive \ion{C}{ii} non-LTE model
ion which removes this discrepancy. However, given the relatively simple model ion used in our {\sc tlusty} 
code, we have adopted the
correction of +0.34\,dex reported by Lennon et al. (\cite{len03}) to the  
abundances estimated from the 4267\AA\ line. Indeed, as discussed in Sect.~\ref{s_HII_regions} this  
improves the agreement of our stellar carbon abundances with previous  
interstellar studies. Both uncorrected and corrected carbon abundances are  
given in Tables~\ref{t_vtadjust} and \ref{t_averages}, 
with the error estimates for the latter being either identical or similar to those
for the uncorrected values.

\subsection{Nitrogen}                                         \label{s_nitrogen}

In several stars, especially at the lowest metallicity studied, it was not possible to 
observe any nitrogen lines. In these cases an upper limit to
the EW of the nitrogen line at 3995\AA\ was estimated by adding a Gaussian profile (with
the same width as the other observed metal absorption lines) to the spectra. The strength of the
Gaussian was then varied until it became obvious in the spectral noise. The size of the 
apparent 
absorption line was then measured and the resulting EW was taken as an upper limit to 
the EW of a nitrogen line. This upper limit was then used to derive an upper limit to the 
nitrogen abundance in the star.

In the spectra where we observe more than one nitrogen 
line the derived nitrogen abundances of each line agree well. However, there is
a very large spread in the nitrogen abundance between stars in
each of the Magellanic Cloud clusters that is not replicated in our NGC\,6611 sample.
As discussed in Sect.~\ref{s_evolve_effects}, we believe that this is due to evolutionary 
effects during the stellar lifetimes and is not seen in NGC\,6611 due to its higher metallicity and
lack of supergiant targets.

\subsection{Oxygen}                                           \label{s_oxygen}

As can be seen from
Table~\ref{t_vtadjust} the oxygen abundances derived for individual stars in each of the
three clusters are in good agreement. This suggests that the oxygen abundance is effectively 
constant within each cluster
or at least, any variations being too small to be detected. It should be noted that 
as seen in Fig.~\ref{f_micro} there is quite a large spread between the oxygen 
abundance derived from different absorption lines of the same star. We believe that
these differences are due to uncertainties in the atomic data and the theoretical 
calculations, but given the large
number of oxygen lines observed, the mean
oxygen abundances should be robust.

\subsection{Magnesium}                                     \label{s_magnesium}

The magnesium abundances are based solely upon the 4481\AA\ line and
very good agreement between the estimates for stars in the same cluster
is seen. This agreement is encouraging as we do not expect to see variations in
Mg as it should be unaffected by the nucleosynthetic processes that affect our CNO abundances.

\subsection{Silicon}                                        \label{s_silicon}

As discussed above the Si abundances are especially dependent on 
the adopted atmospheric parameters and hence we base the parameters upon these lines.
Because of the methodology the Si abundances given in Table~\ref{t_vtadjust} are essentially 
constant. Nevertheless, from Table~\ref{t_abund} it can be seen that the majority of 
the uncorrected Si abundances in each cluster are in good agreement within the uncertainties.

\section{Comparisons with previous work} \label{s_comparisons}

\subsection{Stellar studies}

\subsubsection{NGC6611}

\begin{table*}[hbtp]
\caption[]{Comparison of NGC\,6611 stars with those analysed by
Kilian-Montenbruck et al. (\cite{kil94}) (Kil94), who
have used non-LTE line 
formation calculations with LTE models.}
\label{t_6611compare}
\centering
\begin{tabular}{lcccccccc}
\hline
\hline
 & \multicolumn{2}{c}{NGC\,6611-006} & \multicolumn{2}{c}{NGC\,6611-012} &
\multicolumn{2}{c}{NGC\,6611-021} &\multicolumn{2}{c}{NGC\,6611-033} \\
                        &This &Kil94&This &Kil94&This &Kil94&This &Kil94\\
                        &Paper&     &Paper&     &Paper&     &Paper&     \\
\hline
\\
$T_{\rm eff}$ (K)       &31250&32600&27200&29400&26250&29400&25600&28600\\
$log g$ (dex)           &4.00 &4.17 &3.90 &4.17 &4.25 &4.39 &4.00 &4.21 \\
$\xi$ (km\,s\,$^{-1}$)    &8    &8    &5    &10   &0    &0    &4    &5    \\
$v\sin i$ (km\,s\,$^{-1}$)&20   &29   &95   &86   &30   &38   &25   &41   \\
Carbon                  &7.85 &8.28 &7.92 &8.23 &7.82 &8.23 &7.99 &8.41 \\
Nitrogen                &7.58 &7.78 &7.50 &7.72 &7.51 &7.89 &7.70 &8.02 \\
Oxygen                  &8.50 &8.69 &8.55 &8.39 &8.60 &8.62 &8.54 &8.52 \\
Magnesium               &7.36 &7.21 &7.26 &7.39 &7.24 &7.32 &7.38 &7.51 \\
Silicon                 &7.41 &7.63 &7.45 &7.11 &7.40 &7.21 &7.43 &7.01 \\
\\
\hline
\end{tabular}
\end{table*}

Four of the five stars analysed in NGC\,6611 have previously been analysed by
Kilian-Montenbruck et al. (\cite{kil94}) and in Table~\ref{t_6611compare} the two analyses 
are compared. It can been seen that Kilian-Montenbruck et al. derive
systematically higher effective temperatures and hence higher gravities, which may be due to their adoption of LTE models with
only partial line blanketing. In one case, NGC\,6611-021,
this difference is over 3\,000\,K. Although our {\sc tlusty} model predicts that the
\ion{He}{ii} lines would be present at this higher temperature, they would probably be hidden by the noise
in the observational spectra and hence cannot be used to constrain the temperature estimate.
However, the predicted \ion{Si}{iv} line at 4116\AA\ would be 50\% stronger at this higher temperature
and a change of this magnitude would be incompatible with the observed spectra. Although there are some differences 
between the derived abundances, given the differences in the atmospheric parameters, 
these are not significant. Additionally we note that the scatter in abundances derived here is
less than that previously seen. For example we see a range of less than 0.1\,dex in our oxygen abundance compared with
a range of 0.3\,dex in the O abundances of Kilian-Montenbruck et al.
As
we have used both non-LTE line formation calculations and non-LTE models (rather than LTE models adopted by
Kilian-Montenbruck et al.),
we believe the results presented here are an improvement 
on those currently available in the literature.

\subsubsection{N11}

A search of the literature revealed that only one of our N\,11 target stars had previously 
been analysed, N\,11-100, by Rolleston et al. (\cite{rol02}). In Table~\ref{t_11compare} we present a 
comparision of the two analyses. Given that Rolleston et al. adopted LTE techinques,
the atmospheric parameters compare well and the abundances differ by at most
0.3\,dex between the two analyses. 

\subsubsection{NGC346}

Two stars in NGC\,346 have previously been analysed, viz. NGC\,346-029 by
Rolleston et al. (\cite{rol93}) and NGC\,346-043 by Hunter et al.
(\cite{hun05}). In Table~\ref{t_346compare} we present a 
comparision of the two analyses.
We find good agreement between our analysis of NGC\,346-029 and that of 
Rolleston et al., who used LTE methods and observational data with a
S/N ratio of 35, compared to a S/N ratio
of 140 for the FLAMES data. Hunter et al. have
used 
effectively the same methods as discussed here to analyse UVES spectra of NGC\,346-043
and the agreement between the two analyses is encouraging, with all the
parameters and abundance estimates consistent within the uncertainties.

\begin{table}[hbtp]
\caption[]{Comparison of our derived parameters and abundances of N\,11-100 with 
those derived by Rolleston et al. (\cite{rol02}) (R02).}
\label{t_11compare}
\centering
\begin{tabular}{lcc}
\hline
\hline
 & \multicolumn{2}{c}{N\,11-100}  \\
                        &This      &  R02     \\
                        &Paper     &          \\
\hline
$T_{\rm eff}$ (K)       &29700     &29500     \\
$log g$ (dex)           &4.15      &4.10       \\
$\xi$ (km\,s\,$^{-1}$)    &4         &6         \\
$v\sin i$ (km\,s\,$^{-1}$)&30        &30        \\
Carbon                  &$~\,$7.44 &  $<$7.64 \\
Nitrogen                &$~\,$7.62 &$~\,$7.86 \\
Oxygen                  &$~\,$8.30 &$~\,$8.28 \\
Magnesium               &$~\,$7.10 &$~\,$6.81 \\
Silicon                 &$~\,$7.20 &$~\,$7.21 \\
\\
\hline
\end{tabular}
\end{table}

\begin{table}[hbtp]
\caption[]{Comparison of our derived parameters and abundances for NGC\,346-029
and NGC\,346-043 with those estimated by
Rolleston et al. (\cite{rol93}) (R93) and Hunter et al. (\cite{hun05}) (H05) respectively.
Note that the abundances given in this table are from Table~\ref{t_abund} rather than
Table~\ref{t_vtadjust} as the method of adjusting the microturbulence to derive 
the average Si cluster abundance for these stars was not possible, see Sect.\ref{s_errors}.}
\label{t_346compare}
\centering
\begin{tabular}{lcccc}
\hline
\hline
 & \multicolumn{2}{c}{NGC\,346-029} & \multicolumn{2}{c}{NGC\,346-043} \\
                        &This      &  R93     &This     &  H05\\
                        &Paper     &          &Paper    &     \\
\hline
$T_{\rm eff}$ (K)       &32150     &30500     &33000    &32500\\
$log g$ (dex)           &4.10      &4.0       &4.25     &4.25 \\
$\xi$ (km\,s\,$^{-1}$)    &0         &5         &4        &5   \\
$v\sin i$ (km\,s\,$^{-1}$)&25        &28        &10       &8   \\
Carbon                  &$~\,$7.17 &$~\,$6.80 &$~\,$7.24&$~\,$7.45 \\
Nitrogen                &$~\,$6.99 &  $<$7.20 &  $<$6.73&  $<$6.73 \\
Oxygen                  &$~\,$8.02 &$~\,$8.00 &$~\,$7.97&$~\,$7.82 \\
Magnesium               &$~\,$6.69 &$~\,$7.10 &$~\,$6.81&$~\,$6.77 \\
Silicon                 &$~\,$6.69 &$~\,$6.50 &$~\,$6.56&$~\,$6.42 \\
\\
\hline
\end{tabular}
\end{table}

\subsubsection{Other stellar studies}

Venn (\cite{ven99}) has obtained SMC abundances from A-type supergiants and these
are generally in good agreement with the NGC\,346 abundances presented here.
Venn 
also observed a similar variation of the nitrogen abundances from different stars. 
Additionally Dufton et al. (\cite{duf05}), Trundle \& Lennon (\cite{tru05}) and
Trundle et al. (\cite{tru04}) have determined chemical abundances for approximately 
30 SMC supergiants. Their C, O, Mg and Si 
abundances are in excellent agreement (within 0.1\,dex)
with those presented in Table~\ref{t_averages}. They report higher N abundances 
than those derived here
but this may be due to the more evolved 
nature of their sample compared to that presented here. Bouret et al. (\cite{bou03})
and Heap et al. (\cite{hea06}) have derived abundances from SMC O-type stars. 
Their O and Si abundances are in excellent agreement with those in Table~\ref{t_averages}
while they find a similar spread in the nitrogen abundances. Their average 
C abundance is 0.2\,dex higher than that given in Table~\ref{t_averages}
but given the uncertainties in our estimate this is unlikely to be significant. In Sect.~\ref{s_evolve_effects}, we
discuss further the implications of these studies for stellar evolutionary models.

Korn et al. (\cite{kor05}) have obtained
LMC abundances from targets in NGC\,2004 and the abundances which we derive for 
N\,11 are within their uncertainties 
for all elements except carbon. As discussed in Sect.~\ref{s_carbon} 
this may be at least partially due to problems with our derived carbon abundances.

\subsection{\ion{H}{ii} regions} \label{s_HII_regions}

In Table~\ref{t_averages} 
we present CNO abundances for the \ion{H}{ii} regions NGC\,6611, N\,11 and NGC\,346 
taken from the literature. Unfortunately it was not possible to find one source which covered
all three clusters and so the abundance estimates utilise
a variety of methods.

It is particularly difficult to obtain carbon abundances from these \ion{H}{ii}
regions and we have found only one recent reference
for each cluster. This is primarily due to the high quality UV spectra that is required. 
In the case of both NGC\,6611 and N\,11 our uncorrected carbon abundances are 
$\sim$0.3\,dex lower than those found from the \ion{H}{ii} region analyses, whilst our corrected values
are in excellent agreement with those from \ion{H}{ii} region studies. However, we find poorer
agreement between the corrected carbon abundance of NGC\,346 and that from Kurt \& Dufour (\cite{kur98}) compared to the uncorrected value,
although within the uncertainties this may not be significant. For example, Reyes (\cite{rey99}) find that the average carbon abundance 
derived from five different \ion{H}{ii} regions in the SMC is 7.39\,dex (compared with 7.20\,dex from Kurt \& Dufour), which is in excellent agreement 
with our corrected value. 

Our estimated nitrogen abundance from B-type stars in NGC\,6611 is in excellent agreement with
those from \ion{H}{ii} regions, in contrast to N\,11 and NGC\,346 where our average nitrogen
abundances are higher than those given in Table~\ref{t_averages}. However as discussed in
Sect.~\ref{s_nitrogen} we see a very large spread in the nitrogen abundances between stars in the
same cluster and we believe that this is due to chemical processing in the stellar 
interior and not due to any
inconsistency in our analysis. For the stars with the lowest nitrogen abundance estimates
in each cluster, there is excellent agreement with \ion{H}{ii} region results.

The average oxygen abundances in each cluster estimated from our sample are in 
excellent agreement with the averages of the \ion{H}{ii} region oxygen abundances
given in Table~\ref{t_averages}. This is expected as, given the relatively large 
oxygen abundance in these clusters, any small changes due to mixing of 
nucleosynthetic processed material to the surface would be difficult to detect.

\section{Discussion}	\label{s_discussion}

\subsection{Atmospheric parameters and abundances}                     \label{s_dis_atmos}

Atmospheric parameters and abundances have been presented for approximately 50 narrow lined
B-type stars covering three metallicity regions, Galactic, LMC and SMC. To date, 
this is the most extensive abundance analysis of these regions using a consistent method of analysis.
Also, the uncertainties in our estimated abundances
have been fully quantified to take into account the errors in the atmospheric parameters.
Although we determine projected rotational velocities it is not
possible for us to examine these distributions as a function of metallicity 
due to the selection effects that we have used. 
In a future paper we intend to 
derive atmospheric parameters and $v\sin i$ values for all the LMC and SMC stars observed in the 
FLAMES survey using similar methods to those described in Dufton et al. (\cite{duf06}).

\subsection{Abundance ratios}

In Table~\ref{t_ratios} the ratios of the Magellanic Cloud cluster abundances to 
the Galactic cluster abundances are presented. Excluding nitrogen, the abundances in
the LMC cluster are on average 0.3\,dex lower than those in the Galactic cluster, and 
the SMC abundances are on average 0.6\,dex lower than the Galactic case which is in good agreement with
the iron abundances adopted in our model atmosphere grids (Sect.~\ref{s_nlte}). However, there is some spread 
in these values. For example, oxygen appears to be the least depleted
element in both the LMC and SMC cluster and nitrogen is the most depleted.

In Table~\ref{t_ratios} we also present the ratios for CNO as derived from the \ion{H}{ii} region 
results given in Table~\ref{t_averages}. It can be seen, in the majority of cases, that 
these are in excellent agreement with those derived in this 
analysis. The only exception being the carbon NGC\,346/NGC\,6611 ratio, although as 
discussed earlier this may be
due to choice of \ion{H}{ii} region analysis. Adopting the \ion{H}{ii} region carbon abundance
from SMC analysis of Reyes (\cite{rey99}) (see Sect.~\ref{s_HII_regions}) results in a ratio of -0.86\,dex which 
is in better agreement with that derived from our stellar analysis. Slightly greater depletions of carbon than for
other elements (except nitrogen) are evident and this has previously been observed by Dufour et al. (\cite{duf82})
from \ion{H}{ii} region analyses. We also note the relatively small oxygen
underabundances implied by the \ion{H}{ii} region studies, which are consistent with our stellar results.

\begin{table}
\caption{Logarithmic ratios of abundances derived from the Magellanic Cloud clusters to those derived
from the Galactic cluster. Note that the ratios for nitrogen use the lowest nitrogen 
abundance we derive from the Magellanic Cloud cluster stars rather than the average nitrogen 
given in Table~\ref{t_averages}. Additionally the ratios for the carbon abundance use the corrected abundance estimates
(compare with ratios of -0.47 and -0.72 for N\,11/NGC\,6611 and NGC\,346/NGC\,6611 respectively from the
uncorrected values). The ratios derived from the
\ion{H}{ii} region abundances in Table~\ref{t_averages} are also given.}
\label{t_ratios}
\centering
\begin{tabular}{lcccc}\hline \hline
   & \multicolumn{2}{c}{N\,11/NGC\,6611} & \multicolumn{2}{c}{NGC\,346/NGC\,6611}\\
   & This & \ion{H}{ii} & This & \ion{H}{ii} \\
   &Paper & Regions     & Paper& Regions\\
\hline
\\
C  & -0.40 $\pm$ 0.17 &-0.42   & -0.76 $\pm$ 0.19 &-1.10\\
N  & -0.73 $\pm$ 0.18 &-0.77   & -1.17 $\pm$ 0.18 &-1.14\\
O  & -0.22 $\pm$ 0.09 &-0.15   & -0.49 $\pm$ 0.11 &-0.45\\
Mg & -0.26 $\pm$ 0.11 &        & -0.58 $\pm$ 0.09 &\\
Si & -0.22 $\pm$ 0.09 &        & -0.62 $\pm$ 0.07 &\\
\\
\hline
\end{tabular}
\end{table}

\subsection{Temperature scales of B-type stars}

\begin{figure*}[htbp]
\centering
\epsfig{file=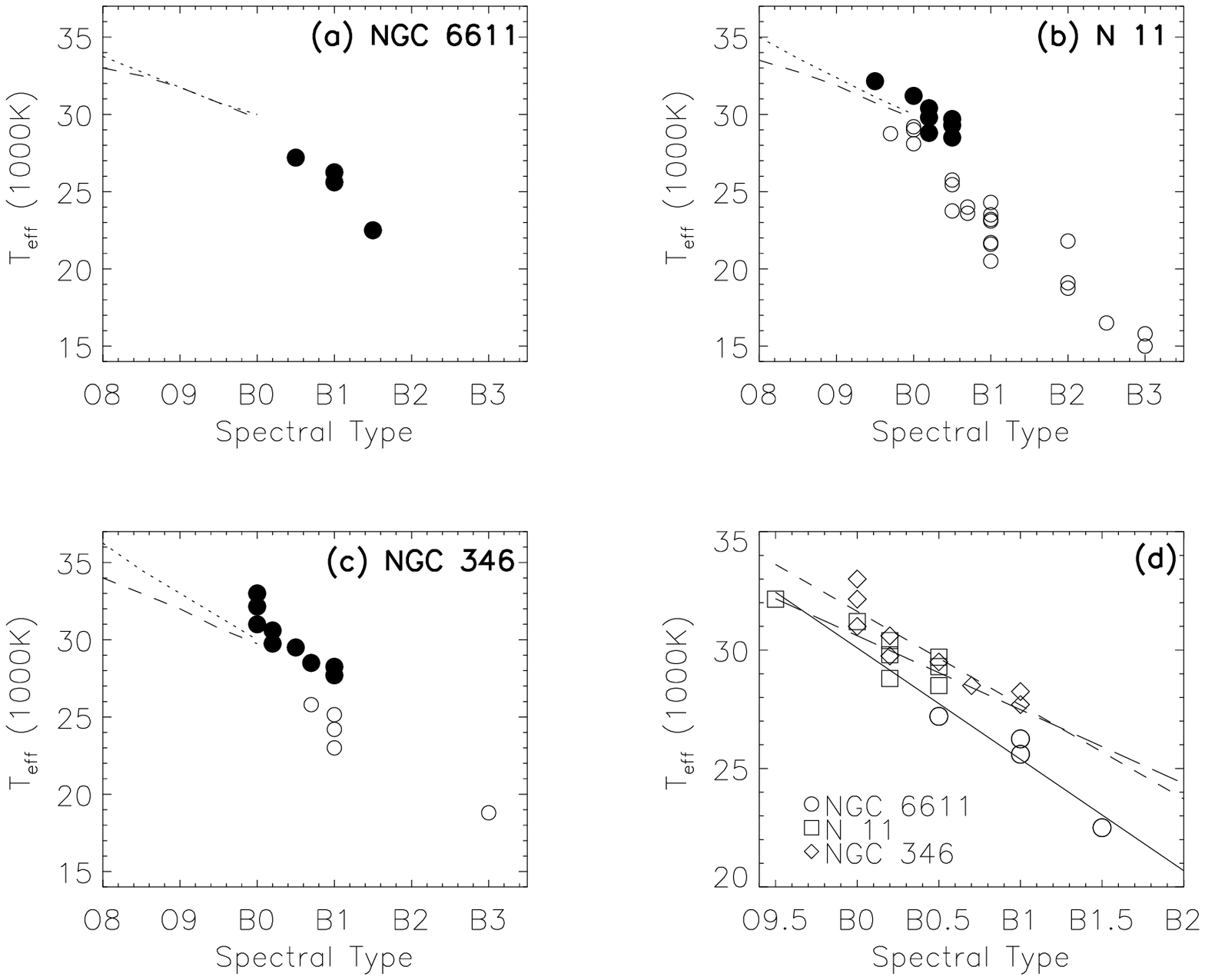, height=140mm, angle=0}
\caption[]{Temperature scales derived from the atmospheric parameters given in 
Table~\ref{t_stars+atmos} for each cluster, plots (a), (b) and (c).
The solid points represent stars with
$\log g$\,$\geq$\,3.70\,dex and open circles represent stars with 
$\log g$\,$<$\,3.70\,dex. The dotted and dashed lines
are the temperature scales derived from the O-star analysis of 
Massey et al. (\cite{mas05}) for luminosity class V \& III stars and luminosity class I 
stars respectively. In (d) we compare the temperature scales from 
the three clusters for stars with $\log g$ values 
$\geq$\,3.70\,dex and the solid, long-dashed and
short-dashed lines represent the best-fitting lines for the NGC\,6611, N\,11
and NGC\,346 stars, respectively. Note that NGC\,6611-006 is not 
included in plots (a) and (d) due to the uncertainty in its
spectral type, see footnote in Table~\ref{t_stars+atmos}. } 
\label{f_temp_scale}
\end{figure*}

In previous studies, see for example Voels et al. (\cite{voe89}) and Massey et al. (\cite{mas05}), it has been
reported that there is a significant difference in temperature scales between different luminosity class 
stars at the same metallicity and also
for stars of the same spectral type at different metallicities. 
In Figs.~\ref{f_temp_scale}~(a), (b) and (c) plots of spectral type against $T_{\rm eff}$  for all 
the stars in Table~\ref{t_stars+atmos} 
are presented for NGC\,6611, N\,11 and NGC\,346 respectively. From Figs.~\ref{f_temp_scale}~(b) and (c)
it can be seen that the temperature derived from the lower gravity stars in the LMC and SMC clusters are
smaller than those from higher gravity stars. Unfortunately for NGC\,6611 we have no stars with 
$\log g$\,$<$\,3.70\,dex. However, 
Crowther et al. (\cite{cro05})
have derived atmospheric parameters for Galactic B-type supergiants and a comparison with our results 
shows again that the lower gravity stars have 
lower temperatures for a given spectral type. This is in contrast to the temperature scale 
derived by  Massey et al.
for O-stars who have previously reported that while the temperature of early O-type stars can be higher 
by up to 6\,000\,K for main-sequence compared to
supergiant stars, the difference decreases as one moves towards later O-types with
no significant difference by B0. However we note that they had 
only three stars 
in their SMC sample (two giants and one supergiant) and no stars 
in their LMC sample
are later than O8. Additionally, in deriving temperature scales, 
Massey et al. group together giant
and main-squence stars as they find little difference between the gravity
estimates for these luminosity classes. However,
their two SMC giant stars later than O8 have gravities of 3.27 and 3.52\,dex and 
hence may not be representative of the main-sequence temperature scale for 
late type O stars. In Figs.~\ref{f_temp_scale}~(a),
(b) and (c) the dotted and dashed lines represent the
temperature scales from the O-star analysis of Massey et al. for luminosity 
class V \& III stars and luminosity class I stars respectively. 

In Fig.~\ref{f_temp_scale}~(d) we compare the temperature scale for 
the (near) main-sequence B-type stars at Galactic, LMC and SMC metallicities. 
Although there are relatively few data points covering
only a small range in spectral type there appears to be a trend of 
increasing temperature as one
moves towards lower metallicities for stars of the same spectral type. Massey et al. (\cite{mas05}) 
report the same effect for both high and low gravity 
early O-type stars although they find little difference by B0. 
In contrast we see a difference of 2000\,K between dwarf stars in the SMC and 
dwarf stars in the Galaxy. Mokiem et al. (\cite{mok05a}) also 
report a similar effect for their O and early B-type stars with the temperature scale
derived from their SMC stars being in good agreement with that 
shown in Fig.~\ref{f_temp_scale}~(d). At present we
are unable to undertake a detailed comparison of the temperature scale for low gravity stars due to a 
lack of giants and supergiants in our Galactic sample.

\subsection{Evolutionary effects} \label{s_evolve_effects}

\begin{figure*}[bhtp]
\centering
\begin{tabular}{c}
\epsfig{file=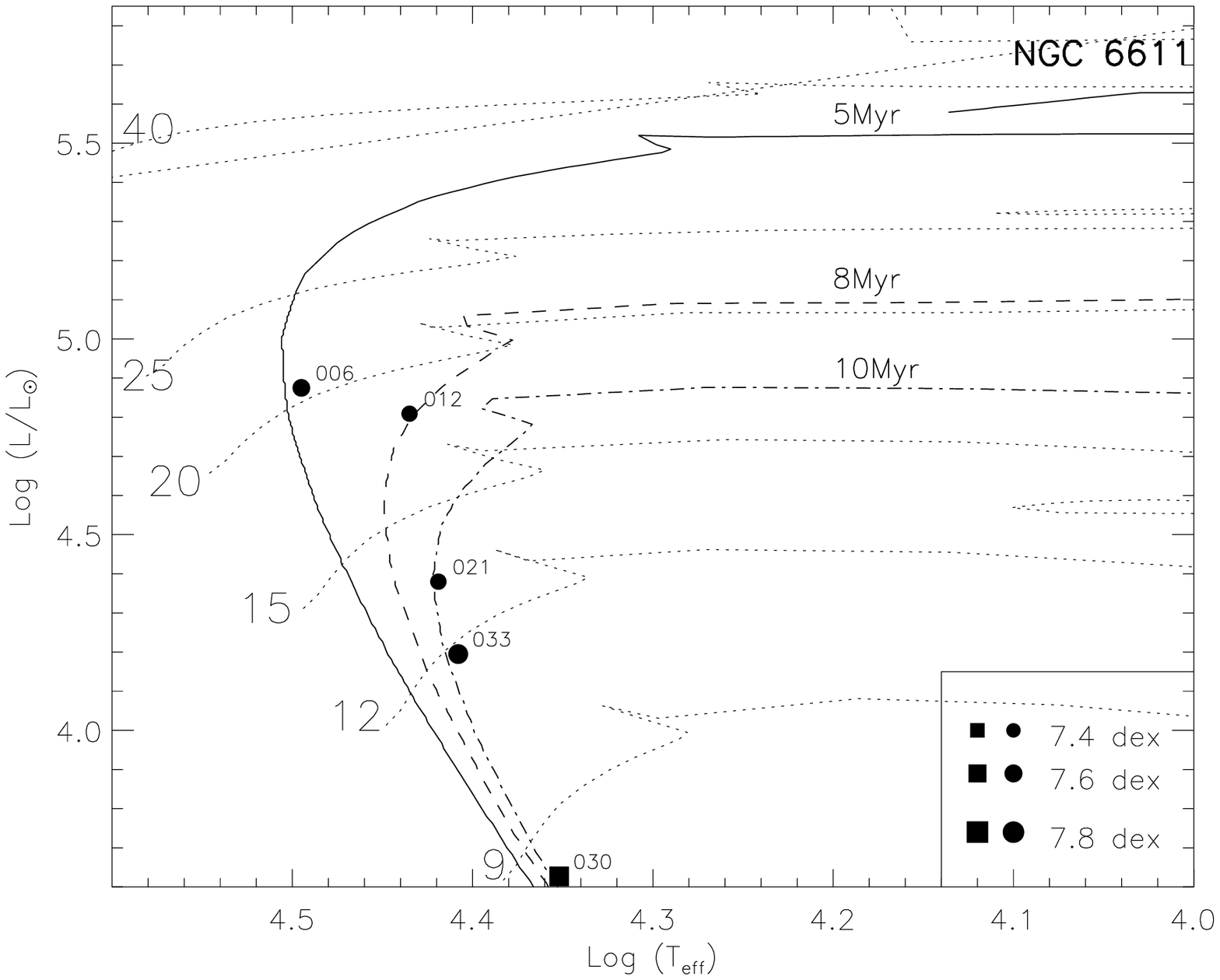, height=115mm, angle=0}\\
(a)\\
\end{tabular}
\caption[]{Hertzsprung-Russell diagrams for NGC\,6611 (a), N\,11 (b) and
NGC\,346 (c). The size of the symbols represent the absolute nitrogen abundances. 
Open symbols represent upper limits to the nitrogen abundances. Squares and circles
represent potential binaries and single stars respectively. Evolutionary tracks have
been obtained from Meynet et al. (\cite{mey94}), Schaller et al. (\cite{sch92}), 
Schaerer et al. (\cite{sch93}), and Charbonnel et al. (\cite{cha93}). Isochrones have been 
calculated from Meynet et al. (\cite{mey93}). In panel (b) stars 100 and 101 overlap.} 
\label{f_HRdiagrams}
\end{figure*}

\addtocounter{figure}{-1}
\begin{figure*}[htbp]
\centering
\begin{tabular}{c}
\epsfig{file=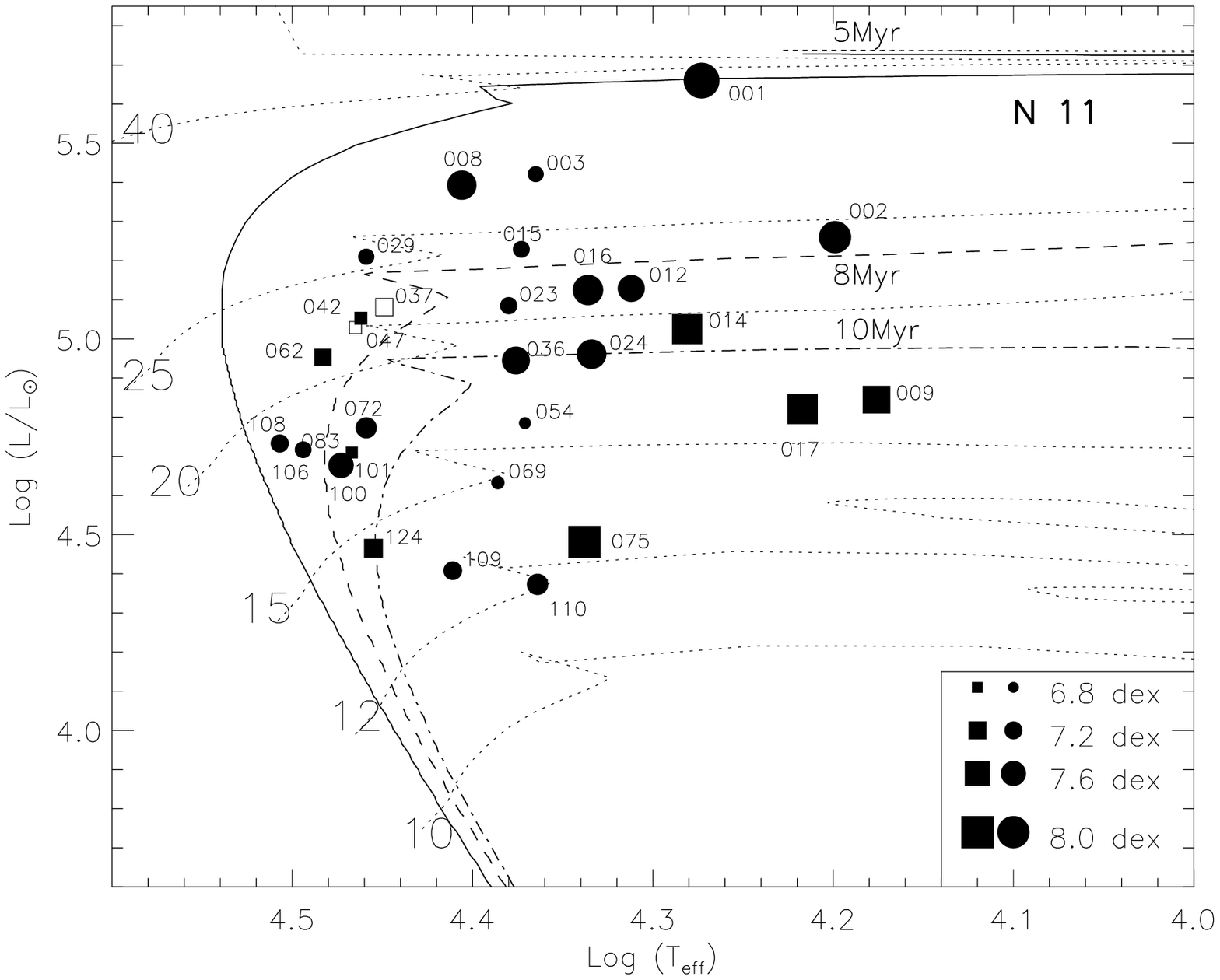, height=115mm, angle=0}\\
(b)\\
\epsfig{file=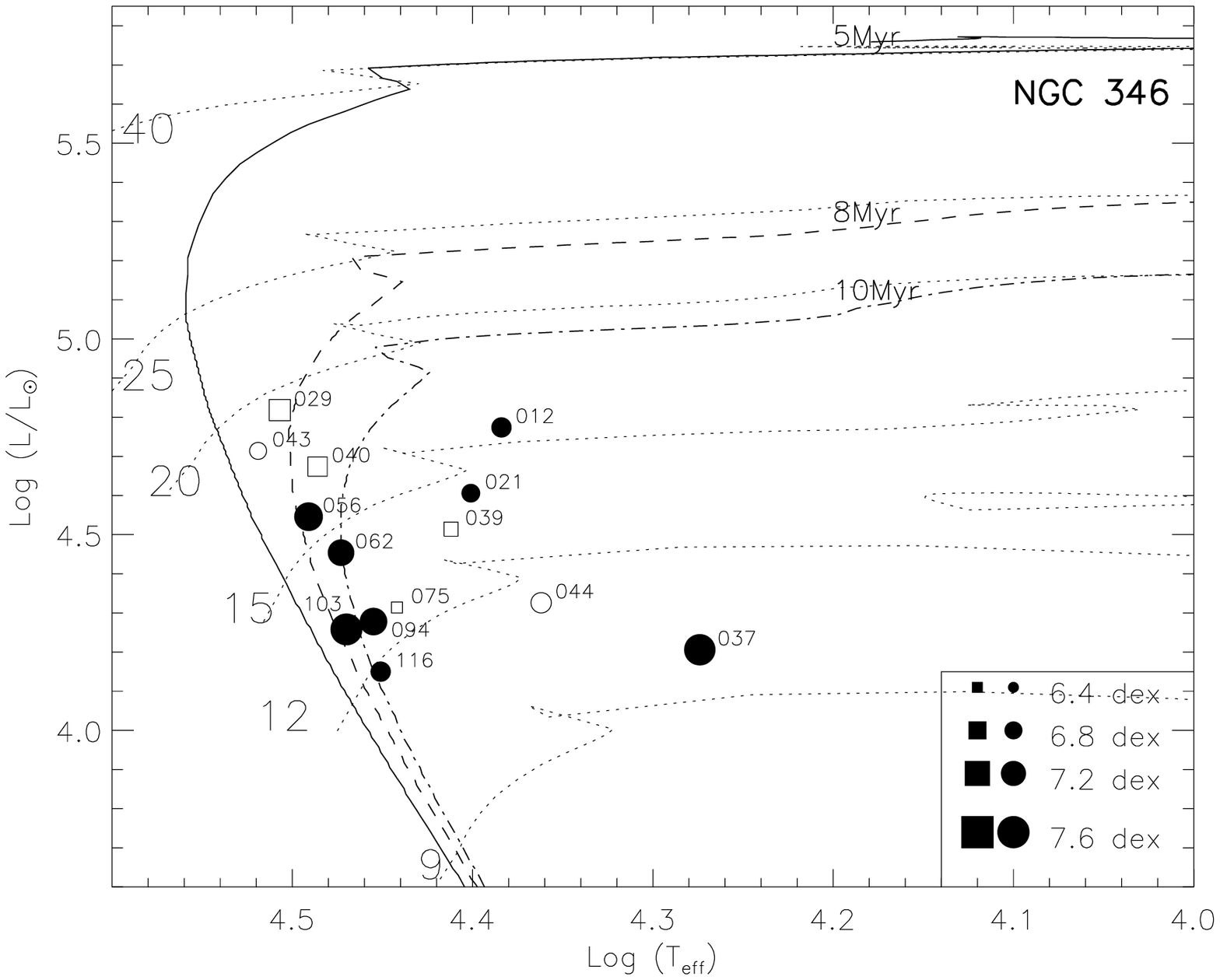, height=115mm, angle=0}\\
(c)\\
\end{tabular}
\caption[]{--continued} 
\end{figure*}

Nitrogen is often used as a probe of the evolutionary history of B-type stars due to its initial low abundance, 
especially in the Magellanic Clouds, (see, for example, Lennon et al. \cite{len03}, Venn \cite{ven99} and
Gies \& Lambert \cite{gie92}). Observed surface enrichments and depletions arise from processes which mix 
core processed material to the photosphere and hence the observed surface abundances can trace evolutionary
effects such as rotational mixing, mass-loss and binarity. In this section the evolutionary status of the objects
in this sample is discussed and compared with current theoretical models where possible. Carbon depletions 
can also be used as a probe of evolutionary history
but given its initial high abundance (compared with nitrogen) it is difficult to detect small abundance
variations. For example, assuming that the sum of the carbon and nitrogen nuclei are conserved, the uncertainities 
in the carbon abundance estimates appear to mask any evolutionary effects for the majority of objects. 
Additionally, no convincing anti-correlation between the carbon and nitrogen abundances is observed, hence
the former are not discussed in this section. 

Figure~\ref{f_HRdiagrams} displays Hertzsprung-Russell (HR) diagrams of our target stars. 
The absolute nitrogen abundance is indicated by the size of the symbol,
open symbols represent upper limits to the nitrogen abundance and square symbols represent stars where
radial velocity variations (potential binary systems) have been detected, see Sect.~\ref{s_EW}. 
From Evans et al. (\cite{eva05b}) it is 
evident that we sample both cluster and field stars and hence the isochrones shown in
Fig.~\ref{f_HRdiagrams} are for information only as the populations are unlikely to be coeval. 

We sample only five main-sequence 
NGC\,6611 stars and these objects
have very similar nitrogen abundances and are in good agreement with the nitrogen abundance
estimated from \ion{H}{ii} regions. Given the lower baseline nitrogen abundance of N\,11 and NGC\,346
it is easier to detect small nitrogen enhancements. For example, if we add the average nitrogen enrichments observed in the
main-sequence objects of N\,11 and NGC\,346 to the base-line nitrogen abundance of NGC\,6611 (by number) we 
obtain enrichments of only 0.10 and 0.12\,dex which is consistent with the scatter seen 
in the nitrogen abundances of our Galactic sample. Hence we limit our following discussion to
the Magellanic Cloud objects.

From Fig.~\ref{f_HRdiagrams} (b) and (c) it is
immediately apparent that in both the LMC and SMC samples, nitrogen variations of over 1.0\,dex are seen.
These variations are present even for the main-sequence objects, with a spread of
0.7 in N\,11 and 1.1\,dex in NGC\,346. 
For the more evolved objects
a range of nitrogen enhancements is also evident. 
In NGC\,346 the evolved objects do not appear to have systematically higher nitrogen enhancements compared
to those on the main-sequence, although the numbers of evolved objects are limited. Additionally in N\,11 
there are non-main-sequence objects with close to baseline nitrogen abundances, for 
example N\,11-054, and yet there are also 
giant objects (such as N\,11-036) showing significant nitrogen enhancement.

In Fig.~\ref{f_Nvslogg_LMC} and Fig.~\ref{f_Nvslogg_SMC} the nitrogen abundance as a function of
$\log g$ is plotted for the LMC and SMC samples (divided into those targets with masses less than and greater than 
20M$_{\sun}$) respectively. Given the smaller sample
size of NGC\,346 and also that all our objects have masses $<$20M$_{\sun}$, to make a direct comparison with the
N\,11 sample we have included other published results for SMC B-type supergiants 
(see Fig.~\ref{f_Nvslogg_SMC} caption
for details). Additionally Venn (\cite{ven99}) has analysed several AF-type supergiants in the SMC and 
these objects are plotted left of the dotted line in Fig.~\ref{f_Nvslogg_SMC} using
updated nitrogen abundances from Venn \& Przybilla (\cite{ven03}). Unfortunately, due to a lack of 
other analyses in the LMC it is not possible to compare our N\,11 sample with similar objects.

Histograms of the stellar nitrogen abundances have also been plotted in
Fig.~\ref{f_LMC_hist} and Fig.~\ref{f_SMC_hist}. Objects for which only upper limits to
the nitrogen abundance have been derived are included as dashed lines on these plots. The dotted line 
indicates the position of the 
baseline nitrogen abundance of each of the Magellanic Clouds (see Sect.~\ref{s_conclusions}). 

In Sect.~\ref{s_evolve_LMC}
and Sect.~\ref{s_evolve_SMC} we now discuss the observed nitrogen enhancements
for LMC and SMC samples respectively. In particular, the observational evidence for rotational mixing, blue loops, 
mass-loss and binarity is discussed and compared with theoretical predictions, where possible, in each 
metallicity regime.

\subsubsection{LMC} \label{s_evolve_LMC}

\begin{figure}[htbp]
\centering
\begin{tabular}{c}
\epsfig{file=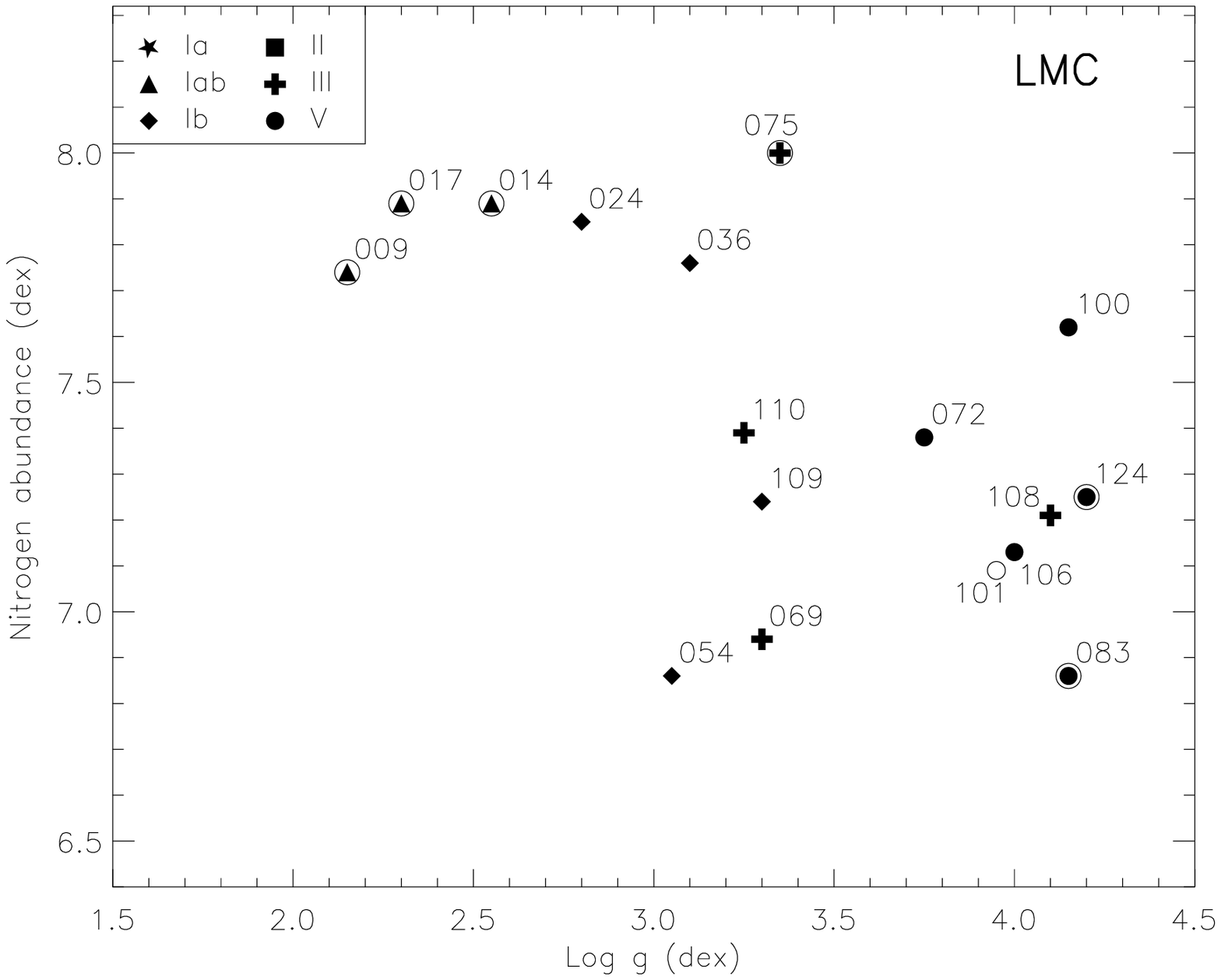, height=70mm, angle=0}\\
(a)\\
\epsfig{file=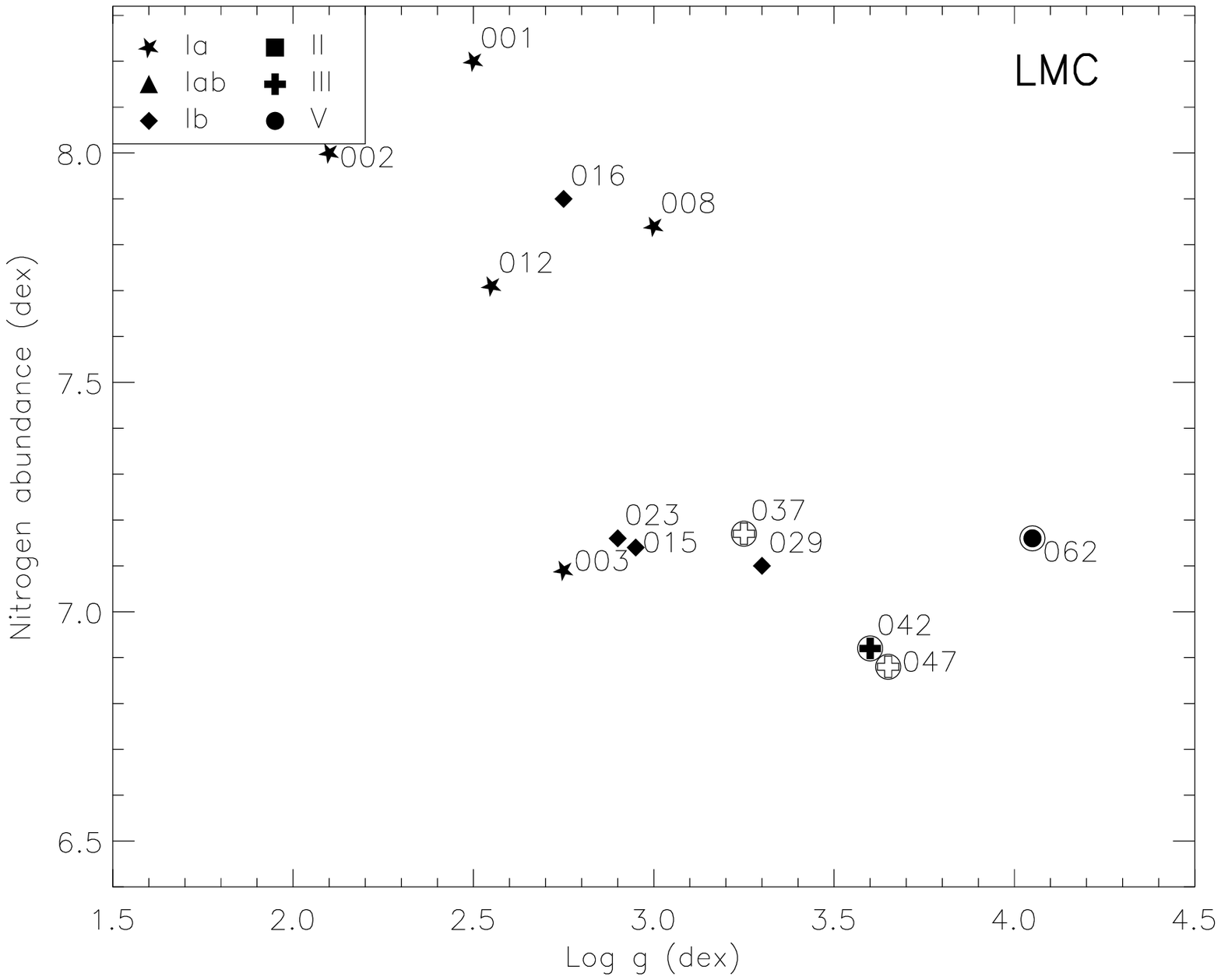, height=70mm, angle=0}\\
(b)\\
\end{tabular}
\caption[]{Variation of the N\,11 stars as a function of 
$\log g$. Open symbols represent stars with upper limits to 
their nitrogen abundance. Symbols which have been circled 
represent stars for which we see evidence of binarity.
Stars with evolutionary masses $<$20M$_{\sun}$ and $\geq$20M$_{\sun}$
are plotted in (a) and (b) respectively.}
\label{f_Nvslogg_LMC}
\end{figure}

\begin{figure}[htbp]
\centering
\begin{tabular}{c}
\epsfig{file=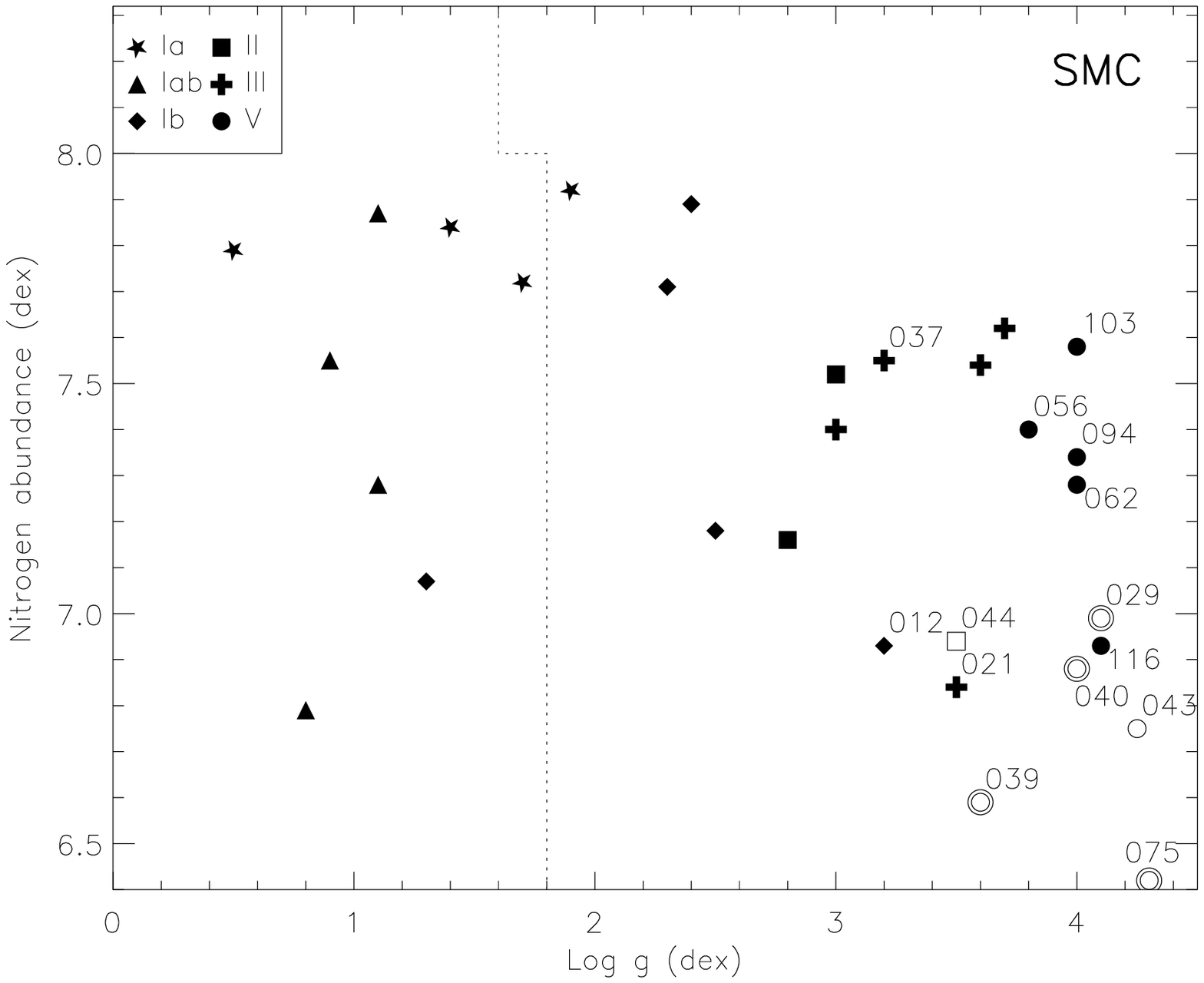, height=70mm, angle=0}\\
(a)\\
\epsfig{file=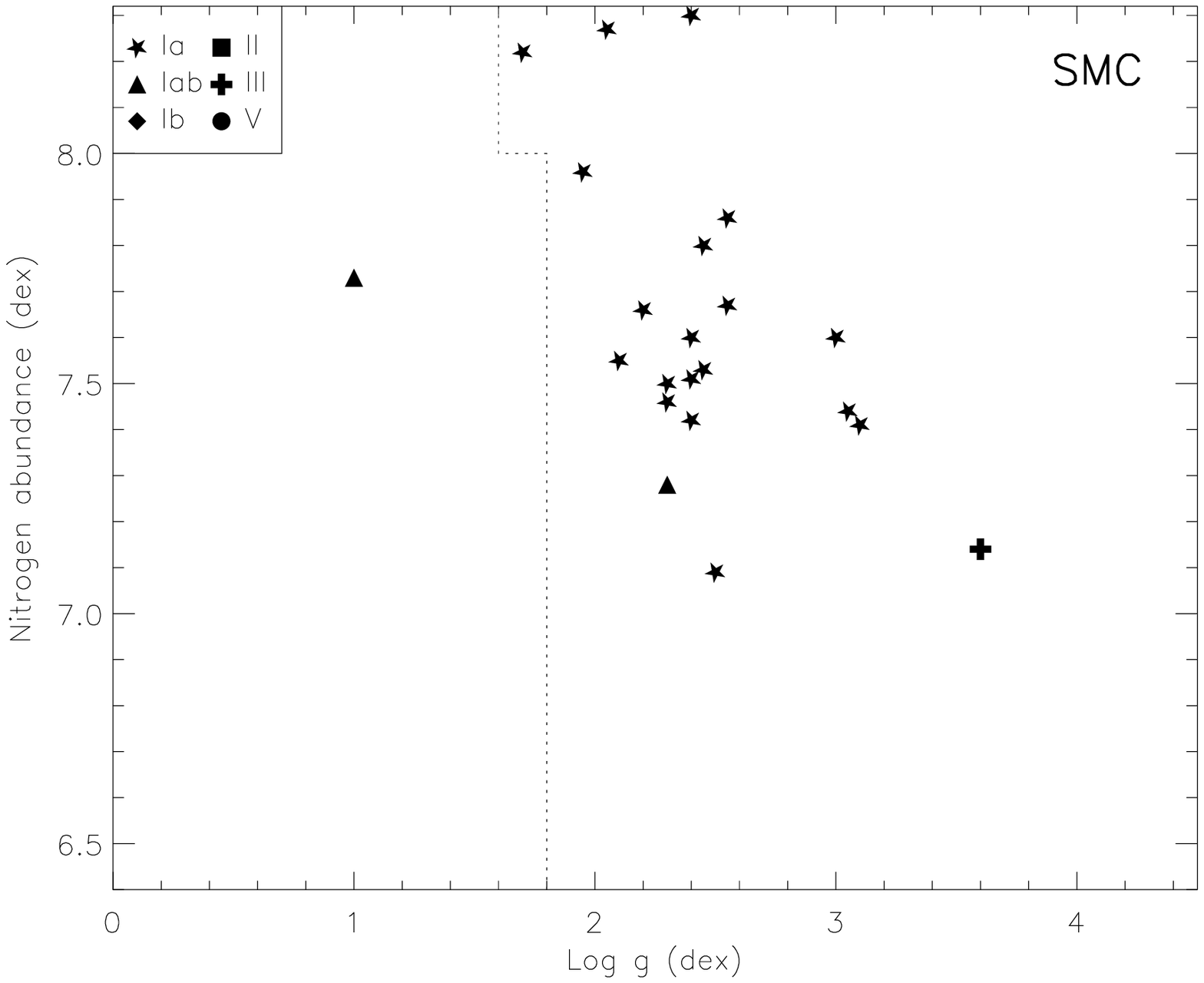, height=70mm, angle=0}\\
(b)\\
\end{tabular}
\caption[]{Variation of the NGC\,346 stars as a function of 
$\log g$. Symbols are equivalent with those in Fig.~\ref{f_Nvslogg_LMC}. In addition to the
B-type stars analysed in this sample, SMC B supergiants and giants (Dufton et al. \cite{duf05},
Trundle et al. \cite{tru04}, Trundle \& Lennon \cite{tru05} and Lennon et al. \cite{len03}) and 
AF supergiants (Venn \cite{ven99} updated with N abundances from Venn \& Przybilla \cite{ven03})
have also been plotted for comparison.
Points left of the dotted line are AF stars, points to the right
are B type stars. The stars analysed in this analysis are labelled.} 
\label{f_Nvslogg_SMC}
\end{figure}

\begin{figure}[htbp]
\centering
\begin{tabular}{c}
\epsfig{file=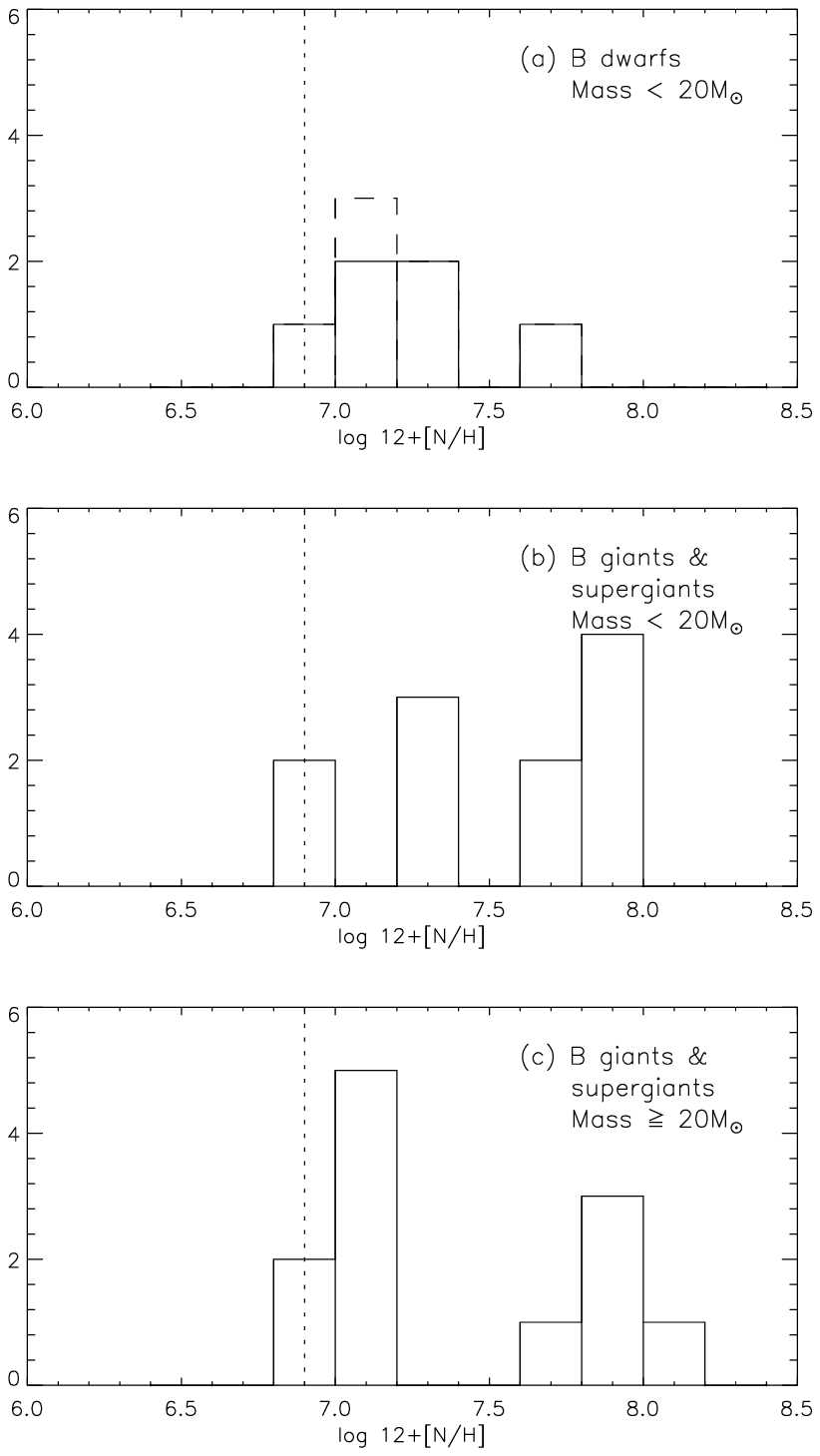, height=115mm, angle=0}\\
\end{tabular}
\caption[]{Histograms showing the spread the in nitrogen abundances of the N\,11 stars. The dashed
line represents upper limits to the abundance. The dotted lined indicates the baseline nitrogen 
abundance of the LMC. Similarly to Fig.~\ref{f_Nvslogg_LMC} the stars are split by mass.}
\label{f_LMC_hist}
\end{figure}

Both Fig.~\ref{f_Nvslogg_LMC} and Fig.~\ref{f_LMC_hist} show 
that significant nitrogen enhancement can occur during the main-sequence
lifetime. The most evolved objects in the sample appear to be further enhanced in nitrogen
than the main-sequence objects and there is some evidence of a bi-modal distribution.
However, there are some Ia supergiants with low nitrogen abundances, 
for example, N\,11-003 which has a nitrogen abundance 0.5\,dex lower than the most enhanced main-sequence object.

Rotational mixing of core processed material into the stellar photosphere can increase the photospheric 
nitrogen abundance and the variable nitrogen abundances that are observed can be interpreted as being due to different
rotation rates. Lamers et al. (\cite{lam01}) have calculated, for LMC metallicity, 
the N/O ratio that would be expected by the end of the main-sequence
due to rotation, as a function of the ratio of the mixing time to the main-sequence
lifetime for masses of 85, 40, 60 and 20M$_{\sun}$. 
For their longest mixing times (where the mixing time to main-sequence lifetime
ratio is 5), a nitrogen abundance of approximately
8.0\,dex (scaled by the oxygen abundance that we determined here) 
would be expected by the end of the main-sequence for an initial mass of 20M$_{\sun}$. 
This is significantly greater than the nitrogen abundances of our objects with such masses. 
Although not all of our LMC objects are at the end of the
main-sequence, the majority of rotational mixing is thought to occur early in the main-sequence lifetime where
rotational rates are highest (Maeder \& Meynet \cite{mae01}) and hence this may indicate that longer mixing
times or less efficient mixing needs to be considered.

A possible explanation for the high nitrogen abundances derived for several of our higher mass evolved objects in N\,11 is that
these stars are on blue loops and the enhancement arises from convective mixing during the
red supergiant phase. Indeed this was the scenario that was invoked to explain the apparently high
nitrogen abundance of the progenitor to the LMC supernova 1987A (N/O=1.6$\pm$0.8\,dex, 
Fransson et al. \cite{fra89}). Lamers et al. (\cite{lam01}) have used
the evolutionary tracks of Meynet et al. (\cite{mey94}) to predict the N/O ratio that would be expected
after convective mixing in the red supergiant phase as a function of the remaining fraction of the mass of the
star at the onset of outer convection, but they do not include any rotational
mixing in this calculation. For negligible mass-loss at an inital mass of 20M$_{\sun}$ their predicted
N/O abundance ratio is approximately 0.5, which again equates to a nitrogen abundance of approximately 8.0\,dex. 
Although the majority of the N\,11 supergiants have nitrogen abundances
of less than 8.0\,dex, within the uncertainties this is a viable explanation for
many of the high nitrogen objects in this sample. However, inspection of Fig.~\ref{f_HRdiagrams}(b) shows
that although blue
loops are indeed predicted for several of the low mass evolutionary tracks, they do not extend far enough 
into the blue or to high enough masses to encompass the objects in this sample. Additionally, given that the
Lamers et al. (\cite{lam01}) predictions of the nitrogen abundance from both rotational mixing (at 20M$_{\sun}$
for their longest mixing time) and convective mixing during the red-supergiant phase (at 20M$_{\sun}$
with no mass-loss) are both 8.0\,dex it is difficult to distinguish these processes.

Mass-loss in massive stars can affect the observed abundances as the outer layers of the
star are removed and the more mixed inner layers are revealed. More massive stars generally have 
higher mass-loss rates. However we note that in both
Fig.~\ref{f_Nvslogg_LMC} and Fig.~\ref{f_LMC_hist} there is little evidence for a trend of increasing nitrogen
enhancement with increasing mass, although the majority of the sample covers a small range in mass.

In Fig.~\ref{f_Nvslogg_LMC} objects for which radial velocity variations have been detected are circled. The circled 
main-sequence objects typically have radial velocity variations on the order of 10-40\,km\,s$^{-1}$ and are likely
to be binary systems. These objects
have near-normal nitrogen abundances and within the uncertainties these 
are similar to those stars where we have not detected evidence of binarity. The evolved radial velocity variable objects
in the sample are all significantly enhanced in nitrogen. It is worth emphasizing that several
of the high nitrogen objects in N\,11 were observed with UVES and as only a single observation was available it was not possible to
examine these objects for velocity variations and so there may be more radial velocity variable objects than
shown in Fig.~\ref{f_Nvslogg_LMC}. Additionally three of the four evolved velocity variable objects have detected 
velocity shifts of less than 5\,km\,s$^{-1}$ and binarity may not be the only explanation (see Sect.~\ref{s_EW}).

However, if these objects are in binary systems they may have undergone a mass-transfer process from which 
we would expect to see an enhancement of nitrogen, although at present there are no detailed theoretical predictions
of abundances in post mass-transfer LMC binary systems. Wellstein et al. (\cite{wel01}) calculate the
masses and orbital parameters of post mass-transfer contact-free binary systems where the remnants are O- and
early B-type stars with a helium star companion. These systems are generally long period systems (greater then 60
days) and the primary objects have orbital velocities of less than 20\,km\,s$^{-1}$. Given the time coverage of
our N\,11 observations (those of the 6515\AA\ region were separated by 35 days) 
and random angles of inclination the velocities shifts of a few km\,s$^{-1}$ which we detect for the most evolved
objects may be compatible with the predictions of Wellstein et al. and so the high nitrogen
abundance could be due to mass-transfer. Wellstein et al. calculate nitrogen enhancement factors of 5.4 and 4.0 
for a representative Case A
(mass-transfer during core hydrogen burning) and Case B (mass-transfer during core helium burning) binary scenario.
However as these calculations are at solar metallicity it is difficult to directly compare them with our LMC abundances.
Follow-up observations of these objects are
necessary in order to determine if these objects truly are binary systems and if so to obtain accurate orbital parameters
to compare with the predictions of Wellstein et al. It is also important to note that the nitrogen enhancements in these 
systems are similar to the apparently single objects and hence it may not be necessary to invoke binarity to explain the
observed abundances.

\subsubsection{SMC}     \label{s_evolve_SMC}

\begin{figure*}[htbp]
\centering
\begin{tabular}{c}
\epsfig{file=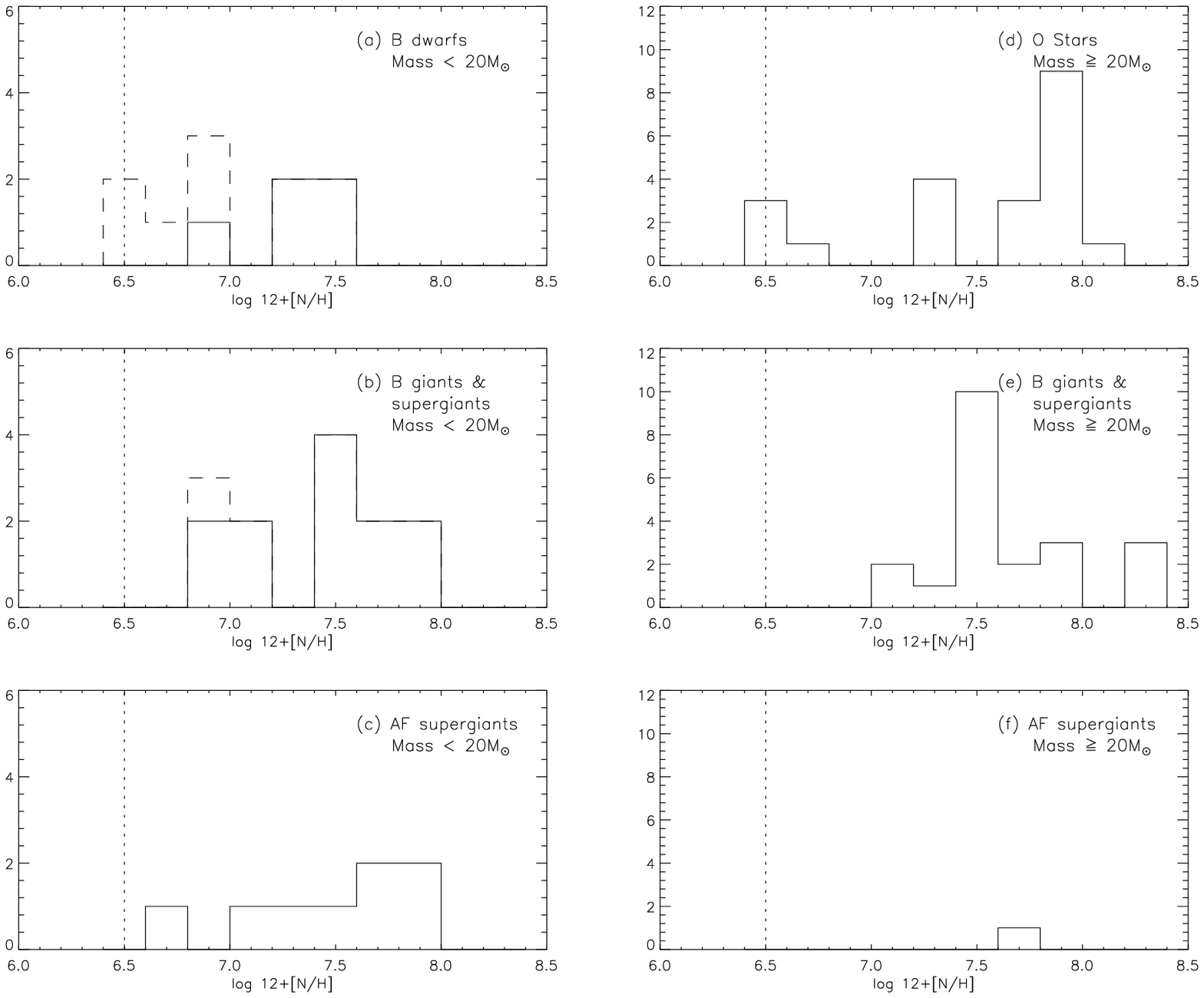, height=115mm, angle=0}\\
\end{tabular}
\caption[]{Histograms showing the spread in the nitrogen abundances of the NGC\,346 stars. The dashed
line represents upper limits to the abundance. The dotted lined indicates the baseline nitrogen 
abundance of the SMC. Similarly to Fig.~\ref{f_Nvslogg_SMC} the stars are split by mass. For comparison SMC
O type stars (Heap et al. \cite{hea06}, Evans et al.\cite{eva04} and Crowther et al.
\cite{cro02}), AF supergiants and additional B supergiants and giants are included (see references in 
Fig.~\ref{f_Nvslogg_SMC}).} 
\label{f_SMC_hist}
\end{figure*}

Both Fig.~\ref{f_Nvslogg_SMC} and Fig.~\ref{f_SMC_hist} show that significant nitrogen 
enhancement occurs on the SMC main-sequence. As in the LMC sample it appears
that the more evolved stars can have higher enhancements than that observed
on the main-sequence and the most massive stars can have higher nitrogen enhancements than their less massive
counterparts. Indeed the similarlity of Fig.~\ref{f_Nvslogg_SMC}(a) and
Fig.~\ref{f_Nvslogg_LMC}(a), when the offset due to baseline nitrogen abundance is accounted for, is 
remarkable and probably shows that the LMC and SMC stars go through similar evolutionary processes. This
appears to suggest that the maximum N abundance detected in both the LMC and SMC are independent of metallicity,
but we should caution that the SMC objects with the highest nitrogen abundances have been obtained from the
literature and hence differences in methodology must be considered.
The A-supergiants in Fig.~\ref{f_Nvslogg_SMC}(a) have similar, although slightly higher, 
enhancements to the main-sequence
objects again indicating that a significant proportion of enhancement occurs during the main-sequence lifetime.
Comparison of Fig.~\ref{f_Nvslogg_SMC}(a) and (b) shows that on the main-sequence a nitrogen enhancement of typically 1.0\,dex occurs
while after the main-sequence an additional 0.4\,dex of enhancement may occur. However, it is important to note that
it is necessary to add only 7.56\,dex of material to a baseline abundance of 6.5\,dex to be able to achieve an abundance of
7.6\,dex, yet it takes a further 7.78\,dex of material to increase the observed nitrogen abundance from
7.6\,dex to 8.0\,dex. This highlights the importance of studying low metallicity environments in order to
observe small amounts of mixing.

\begin{figure}[htbp]
\centering
\begin{tabular}{c}
\epsfig{file=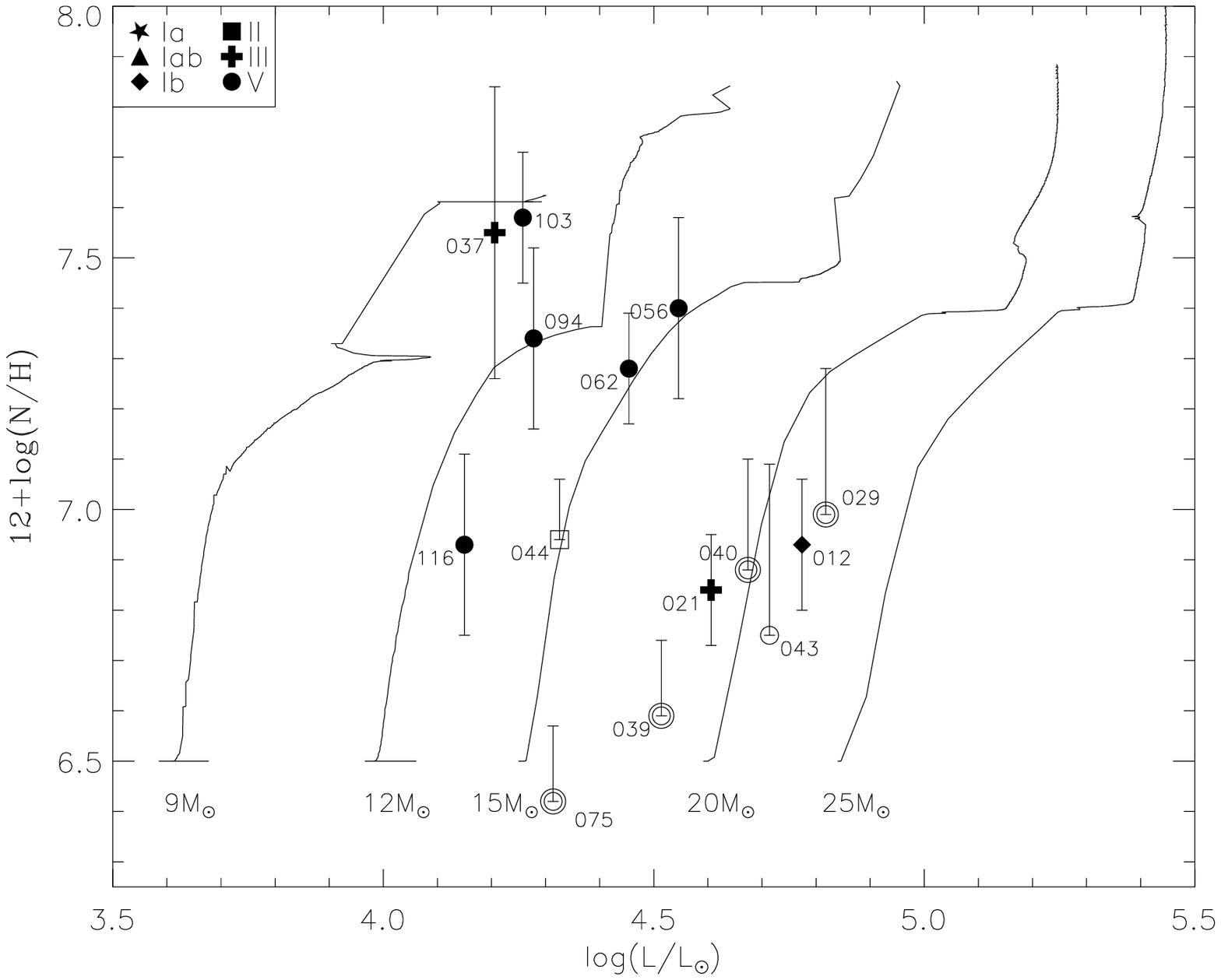, height=70mm, angle=0}\\
\end{tabular}
\caption[]{Photospheric nitrogen abundance as a function of luminosity for the
NGC\,346 stars. Open circles represent stars with upper limits to the nitrogen abundance (lower error
estimates are not shown for these upper limits). 
Stars where evidence of binarity is detected are circled. The rotational evolutionary tracks of
Maeder \& Meynet (\cite{mae01}) which assume an initial rotational velocity
of 300\,km\,s$^{-1}$ are plotted. Note that these tracks have been scaled to a baseline nitrogen
abundance of 6.5\,dex, see text.}
\label{f_SMCrotation}
\end{figure}

Various degrees of rotational mixing on the main-sequence can again explain the variation of nitrogen abundances. 
In Fig.~\ref{f_SMCrotation} the nitrogen abundance of our NGC\,346 sample of stars is plotted as a function
of luminosity and compared to the rotational evolutionary tracks of Maeder \& Meynet (\cite{mae01}), where
an initial rotational velocity ($v$) of 300\,km\,s$^{-1}$ on the zero-age main-sequence (ZAMS) is assumed.
However, as discussed by Lennon et al. (\cite{len03}) it is important to note that 
Maeder \& Meynet have scaled their initial abundances to one fifth solar. While this is 
reasonable for the majority of elements it is not appropriate for nitrogen. Indeed, from \ion{H}{ii} regions 
the base-line nitrogen abundance of the SMC is approximately one twentieth solar. Using similar methods 
to that discussed in Trundle et al. (\cite{tru04})
we have scaled the evolutionary tracks to our base-line nitrogen abundance of 6.5\,dex. This assumes
that the enhancement in nitrogen is independent of the initial nitrogen abundance. While this may not
be strictly true it does allow us to make some comparison between theoretical models and observational
data. 

It should be noted that all the main-sequence dwarf objects, of
less than 20M$_{\sun}$, are close to the beginning of the main-sequence (Fig.~\ref{f_HRdiagrams}(c))
and hence the total amount of mixing could be lower than that expected by the end of the main-sequence. 
However, as indicated by the initial steepness of the evolutionary tracks shown in Fig.~\ref{f_SMCrotation} 
(and Fig.16 in Maeder \& Meynet \cite{mae01}),
significant rotational mixing occurs at the beginning of the main-sequence. 
For example, the 20M$_{\sun}$ rotational tracks (see Maeder \& Meynet) 
show that by an age of 5\,Myr the nitrogen
abundance has been enhanced to 7.15\,dex and it takes almost another 5\,Myr for the enhancement to reach
7.4\,dex. 
It can be seen that several of the more evolved objects have lower nitrogen abundances than that
predicted at the end of the main-sequence by the evolutionary tracks. A simple 
explanation is that these stars had an initial rotational velocity of less than 300\,km\,s$^{-1}$ and therefore have undergone
less rotational mixing. Similarly, 
NGC\,346-103 has a higher nitrogen abundance than predicted for a main-sequence object 
and so may therefore have initially been rotating more rapidly. Indeed, as rotation is thought to extend the
main-sequence lifetime, if the objects in our sample are of similar ages then it is likely that the more
evolved objects are slower rotators than the main-sequence objects and hence should be less rotationally mixed.

Again blue loops can be invoked to explain the higher nitrogen abundances of the more 
evolved SMC objects, which
has been discussed by Lennon et al. (\cite{len03}) to explain the high nitrogen abundances 
of their objects. Indeed, for low-mass single stars, where mass-loss is generally negligible, rotation does
not account for enhancements above 7.5\,dex, while at the end of the blue-loop Maeder \& Meynet (\cite{mae01})
predict a nitrogen abundance of 7.80\,dex for a 12M$_{\sun}$ model. However, as seen in the LMC sample, this
requires that the blue loops extend further into the blue and to higher masses
than currently predicted by theory to explain our observations. 
Smiljanic et al. (\cite{smi06}) have dervied CNO abundances for 
evolved intermediate mass stars (2M$_{\sun}$$<$M$<$15M$_{\sun}$) and also find that extended blue
loops can explain their nitrogen abundances for their most nitrogen rich objects.

Comparison of Fig.~\ref{f_Nvslogg_SMC} (a) and (b) shows that the more massive stars tend towards 
higher nitrogen abundances and this may be an effect of mass-loss. In addition to the tracks shown in 
Fig.~\ref{f_SMCrotation}, Maeder \& Meynet (\cite{mae01}) have
also calculated tracks for 40M$_{\sun}$ and 60M$_{\sun}$ and predict nitrogen abundances of 
approximately 7.6\,dex and 7.9\,dex by the end of the main-sequence respectively. Hence above 25M$_{\sun}$,
mass-loss effects appear to become significant at SMC metallicity and so the high nitrogen abundances of many of
the objects in Fig.~\ref{f_Nvslogg_SMC}(b) may be attributed to this mechanism. However, Heap et al. (\cite{hea06}) have found
that the nitrogen abundances which they derive from O-type stars can be greater than those predicted by Maeder \& Meynet,
which may suggest that the mixing efficiency in their evolutionary models is too low or the assumed rotational velocities are not great enough
for the more massive objects.

The detected binaries in Fig.~\ref{f_Nvslogg_SMC}(a) all have upper limits to their
nitrogen abundances and these limits are generally lower than the nitrogen abundance of apparently single
stars with similar atmospheric parameters. This may indicate that 
binary stars have lower initial rotational velocities than single stars and therefore smaller amounts of nitrogen would
be mixed to the surface via rotational mixing. However, tidal forces may also slow down the rotational velocity of an
object as the rotational velocity may become tidally locked with the orbitial velocity. Recent observational studies
(Huang \& Gies \cite{hua05} and Abt et al. \cite{abt02}) have shown that binary systems experience more spin down than single
stars and this is attributed to tidal interaction causing synchronization. Theoretical studies by Zahn
(\cite{zha77}) have also shown that the timescales for synchronization of close binaries can be much less that the main-sequence
lifetime of the primary star. Additionally Zahn (\cite{zha94}) discusses how tidal interaction reduces the rotational velocity and hence
rotational mixing and relates this to the low lithium depletions that are observed in late-type close binary
systems. It is therefore plausible to similarly attribute the low nitrogen enhancements observed for 
the binary objects in Fig.~\ref{f_Nvslogg_SMC}(a) to a loss of angular momentum (and hence reduced rotational mixing)
through tidal interaction.

\section{Conclusions}                     \label{s_conclusions}

\subsection{Chemical composition of the Magellanic Clouds}

Atmospheric parameters and abundances have been estimated for approximately 
50 stars in three clusters, NGC\,6611 in the Galaxy, N\,11 in the LMC
and NGC\,346 in the SMC. In Table~\ref{t_finalabund} we present our best estimates
for the present-day chemical composition of the LMC and SMC. Our carbon (corrected values), oxygen, magnesium and 
silicon estimates are taken directly from Table~\ref{t_averages} rounded to the
nearest 0.05~dex. Given the agreement 
between our lowest observed nitrogen abundance in N\,11 and NGC\,346 and the nitrogen abundance 
estimated from these \ion{H}{ii} regions we adopt 6.90\,dex and 6.50\,dex in Table~\ref{t_finalabund} as the
pristine nitrogen abundance of the LMC and SMC respectively. 

Given the large number
of stars which have been analysed and the use of a single methodology throughout, with sophisticated non-LTE calculations 
in determining the abundances given in Table~\ref{t_finalabund}, 
we believe that these are the best estimates currently available
for the present-day chemical composition of the Magellanic Clouds derived from 
B-type stars and are suitable for use as the base-line abundances in stellar evolution 
models.

\begin{table}[htbp]
\caption{Present-day chemical composition of the LMC and SMC. Abundances are presented on the scale 12+log[X/H]. For
comparison the solar abundances of Asplund et al. (\cite{asp05}) are also given.}
\label{t_finalabund}
\centering
\begin{tabular}{lccc}\hline \hline
   & \multicolumn{1}{c}{LMC} & \multicolumn{1}{c}{SMC} & \multicolumn{1}{c}{Solar}\\
\hline
\\
C  &                    7.75 & 7.35 & 8.40 \\
N  &                    6.90 & 6.50 & 7.80 \\
O  &                    8.35 & 8.05 & 8.65 \\
Mg &                    7.05 & 6.75 & 7.55 \\
Si &                    7.20 & 6.80 & 7.50 \\
\\
\hline
\end{tabular}
\end{table}

\subsection{Chemical evolution of massive stars}

Given the large
number of stars which we have analysed and the wide range of observed nitrogen abundances, the evolutionary effects
of rotational mixing, mass-loss, blue loops and binarity have been discussed and compared with theory where possible.
Variable rates of rotational mixing and mass-loss can be used to explain the observed photospheric nitrogen enhancements 
of the main-sequence objects. However for objects beyond the main-sequence, particularly low mass objects, these
effects do not account for the observed abundances in many cases. At LMC metallicity Lamers et al. (\cite{lam01}) predictions
of nitrogen abundances by the end of the main-sequence are greater than those observed in almost all cases. At SMC 
metallicity, the models of Maeder \& Meynet (\cite{mae01}) do not predict the high nitrogen abundances observed in many B-type giant
and supergiant objects. The nitrogen abundances predicted by the end of the blue loops at SMC metallicity are 
in good agreement with those
observed in many of the non-main-sequence objects, but these loops do not extend far 
enough into the blue or to high enough masses in current stellar evolutionary calculations to make this a viable explanation. Tidal forces in binary systems may supress
rotational mixing and all our main-sequence binary objects have close to baseline nitrogen abundances. Several evolved
binary objects have high nitrogen enhancements, which may be explained via mass-transfer events. However these abundances
are in agreement with those observed from apparently single stars and hence binary evolution may not
be necessary to account for the observed abundances.

Although this analysis represents one of the largest systematic abundance analyses of B-type stars in the 
Magellanic Clouds, much still remains unclear about the chemical evolution of these objects. In particular, 
as only narrow lined stars have been analysed, our main-sequence sample may be biased towards slowly rotating
stars. Additionally this analysis highlights the need for evolutionary models that adopt the correct initial 
chemical composition of these regions. In a future paper we intend to estimate nitrogen abundances (or limits) for all the
B-type stars (including fast rotators) observed by Evans et al. (\cite{eva05b}), some 300 objects from two LMC and two SMC clusters, in order to place better
constraints on the evolutionary processes that occur in massive stars.

\begin{acknowledgements}

We are grateful to staff from the European Sourthern Observatory at both Paranal and
La Silla for
assistance in obtaining the observational data. We also thank the UK 
Particle Physics and Astronomy Research Council (PPARC) for financial support. IH
acknowledges financial support from the Department of Employment \& Learning Northern Ireland. SJS 
acknowledges the European Heads of Research Councils and Eurpoean Science Foundation EURYI (European 
Young Investigator) Awards scheme, supported by funds from the Participating Organisations of EURYI and the 
EC Sixth Framework Programme. We would also like to thank the referee, Andreas Korn, for valuable 
comments on a previous draft of this paper.

\end{acknowledgements}


\end{document}